\documentclass[1p,sort,final]{elsarticle}
\usepackage{graphicx,amsmath,amssymb,bm}
\usepackage{color}
\usepackage{rotating}
\usepackage{simplewick}
\usepackage{natbib}
\usepackage[breaklinks]{hyperref}
\usepackage{subcaption}

\usepackage[maxfloats=60]{morefloats}

\makeatletter
\providecommand{\doi}[1]{%
  \begingroup
    \let\bibinfo\@secondoftwo
    \urlstyle{rm}%
    \href{http://dx.doi.org/#1}{%
      doi:\discretionary{}{}{}%
      \nolinkurl{#1}%
    }%
  \endgroup
}
\makeatother

\allowdisplaybreaks

\DeclareMathSymbol{\NS}{\mathord}{AMSb}{"4E}

\newcommand{\ket}[1]{\ensuremath{\,|{#1}\rangle}}
\newcommand{\braket}[2]{\ensuremath{\langle{#1}|{#2}\rangle}}
\newcommand{\ketbra}[2]{\ensuremath{|{#1}\rangle}\langle{#2}|}
\newcommand{\matrixe}[3]{\ensuremath{\langle{#1}|\,{#2}\,|{#3}\rangle}}

\newcommand{\dmatrixe}[2]{\matrixe{#1}{#2}{#1}}

\newcommand{\sgn}{\mathrm{sgn}\,}
\newcommand{\tr}{\ensuremath{\mathrm{tr}}}

\newcommand{\expect}[1]{\ensuremath{\langle{#1}\rangle}}

\newcommand{\comm}[2]{\ensuremath{[{#1},{#2}]}}
\newcommand{\acomm}[2]{\ensuremath{ \big\{ {#1}, {#2} \big\} }}

\newcommand{\op}[1]{\ensuremath{#1}}
\newcommand{\adj}[1]{\ensuremath{{{#1}}^{\dag}}}

\renewcommand{\vec}[1]{\ensuremath{\bm{#1}}}

\newcommand{\sixj}[1]{
 \ensuremath{
   \begin{Bmatrix}
     #1
   \end{Bmatrix}
 }
}

\newcommand{\totd}[2]{\ensuremath{ \frac{d {#1}} {d {#2}} }}

\newcommand{\nord}[1]{\ensuremath{\,:\!#1\!:\,}}

\newcommand{\aO}{\ensuremath{\op{a}}}
\newcommand{\hO}{\ensuremath{\op{h}}}
\newcommand{\etaO}{\ensuremath{\op{\eta}}}

\newcommand{\aaO}{\ensuremath{\adj{\op{a}}}}

\newcommand{\AO}{\ensuremath{\op{A}}}
\newcommand{\BO}{\ensuremath{\op{B}}}
\newcommand{\CO}{\ensuremath{\op{C}}}
\newcommand{\HO}{\ensuremath{\op{H}}}
\newcommand{\OO}{\ensuremath{\op{O}}}
\newcommand{\TO}{\ensuremath{\op{T}}}
\newcommand{\UO}{\ensuremath{\op{U}}}

\newcommand{\UUO}{\ensuremath{\adj{\op{U}}}}

\newcommand{\pOV}{\ensuremath{\vec{\op{p}}}}
\newcommand{\qOV}{\ensuremath{\vec{\op{q}}}}
\newcommand{\rOV}{\ensuremath{\vec{\op{r}}}}

\newcommand{\KOV}{\ensuremath{\vec{\op{K}}}}
\newcommand{\POV}{\ensuremath{\vec{\op{P}}}}
\newcommand{\ROV}{\ensuremath{\vec{\op{R}}}}

\newcommand{\fz}{\ensuremath{\overline{f}^{[0]}}}
\newcommand{\ft}{\ensuremath{\overline{f}^{[2]}}}
\newcommand{\Deltaz}{\ensuremath{\overline{\Delta}^{[0]}}}
\newcommand{\Deltat}{\ensuremath{\overline{\Delta}^{[2]}}}

\newcommand{\etao}{\ensuremath{\overline{\eta}^{[1]}}}
\newcommand{\etat}{\ensuremath{\overline{\eta}^{[2]}}}

\newcommand{\Gammao}{\ensuremath{\overline{\Gamma}^{[1]}}}
\newcommand{\Gammat}{\ensuremath{\overline{\Gamma}^{[2]}}}

\newcommand{\Wt}{\ensuremath{\overline{W}^{[2]}}}

\newcommand{\hJ}{\widehat{J}}

\newcommand{\PC}{\ensuremath{\mathcal{P}}}
\newcommand{\OC}{\ensuremath{\mathcal{O}}}
\newcommand{\QC}{\ensuremath{\mathcal{Q}}}

\newcommand{\Hint}{\ensuremath{\HO_\text{int}}}
\newcommand{\Hcm}{\ensuremath{\HO_\text{cm}}}
\newcommand{\Ecm}{\ensuremath{E_\text{cm}}}

\newcommand{\hwBar}{\ensuremath{\hbar\widetilde{\omega}}}
\newcommand{\nn}{\ensuremath{\bar{n}}}

\newcommand{\NNLO}{N${}^2$LO }
\newcommand{\NNNLO}{N${}^3$LO }

\newcommand{\aHO}{\ensuremath{a_{\text{HO}}}}
\newcommand{\eMax}{\ensuremath{e_{\text{max}}}}

\newcommand{\EMax}{\ensuremath{E_{3\text{max}}}}
\newcommand{\Nmax}{\ensuremath{N_\text{max}}}

\newcommand{\nuc}[2]{\ensuremath{{}^{#2}\mathrm{#1}}}

\newcommand{\fm}{\ensuremath{\,\text{fm}}}
\newcommand{\fmi}{\ensuremath{\,\text{fm}^{-1}}}

\newcommand{\keV}{\ensuremath{\,\text{keV}}}
\newcommand{\MeV}{\ensuremath{\,\text{MeV}}}

\newcommand{\GeV}{\ensuremath{\,\text{GeV}}}

\newcommand{\hw}{\ensuremath{\hbar\omega}}

\newcommand{\LambdaUV}{\ensuremath{\Lambda_\text{UV}}}
\newcommand{\LambdaIR}{\ensuremath{L_\text{IR}}}

\newcommand{\LambdaNN}{\ensuremath{\Lambda_\text{NN}}}
\newcommand{\LambdaNNN}{\ensuremath{\Lambda_\text{3N}}}

\newcommand{\nicolinemediumsolid}[2][black]{\unitlength0.8ex 
  {\color{#1}\begin{picture}(6,1)
  \linethickness{0.2mm}
  \put(0,0.5){\line(1,0){6.0}}
  \put(0.01,-0.05){\parbox{6\unitlength}{\centering\color{#1}#2}}
  \end{picture}}\nolinebreak
}

\newcommand{\nicolinemediumdashed}[2][black]{\unitlength0.8ex 
  {\color{#1}\begin{picture}(6,1)
  \linethickness{0.2mm}
  \put(0,0.5){\line(1,0){1.5}}
  \put(2.2,0.5){\line(1,0){1.5}}
  \put(4.4,0.5){\line(1,0){1.5}}
  \put(0,-0.05){\parbox{6\unitlength}{\centering\color{#1}#2}}
  \end{picture}}\nolinebreak
}

\newcommand{\nicolinemediumdashdot}[2][black]{\unitlength0.8ex
  {\color{#1}\begin{picture}(6,1)
  \linethickness{0.2mm}
  \put(0,0.5){\line(1,0){0.4}}
  \put(0.9,0.5){\line(1,0){1.5}}
  \put(2.9,0.5){\line(1,0){0.4}}
  \put(3.8,0.5){\line(1,0){1.5}}
  \put(5.8,0.5){\line(1,0){0.4}}
  \put(0,-0.05){\parbox{6\unitlength}{\centering\color{#1}#2}}
  \end{picture}}\nolinebreak
}

\newcommand{\linemediumsolid}[1][black]{\unitlength1ex 
  {\color{#1}\begin{picture}(6,1)
  \linethickness{0.3mm}
  \put(0,0.5){\line(1,0){6.0}}
  \end{picture}}\nolinebreak
}

\newcommand{\linemediumdashed}[1][black]{\unitlength1ex 
  {\color{#1}\begin{picture}(6,1)
  \linethickness{0.3mm}
  \put(0,0.5){\line(1,0){1.5}}
  \put(2.2,0.5){\line(1,0){1.5}}
  \put(4.4,0.5){\line(1,0){1.5}}
  \end{picture}}\nolinebreak
}

\newcommand{\linemediumdotted}[1][black]{\unitlength1ex 
  {\color{#1}\begin{picture}(6,1)
  \linethickness{0.3mm}
  \put(0,0.5){\line(1,0){0.4}}
  \put(0.9,0.5){\line(1,0){0.4}}
  \put(1.8,0.5){\line(1,0){0.4}}
  \put(2.7,0.5){\line(1,0){0.4}}
  \put(3.6,0.5){\line(1,0){0.4}}
  \put(4.5,0.5){\line(1,0){0.4}}
  \put(5.4,0.5){\line(1,0){0.4}}
  \end{picture}}\nolinebreak
}

\newcommand{\icircle}{\ding{108}}

\newcommand{\icircleopen}{
  \setlength{\unitlength}{0.58em}
  \begin{picture}(1.0000,1.0000)
    \put(0.5,0.5){\circle{1}}
    \put(0.5,0.5){\color{white}\circle*{0.8}}
  \end{picture}
}

\newcommand{\symboldiamond}[1][black]{{\color{#1}\ding{117}}}
\newcommand{\symboltriangle}[1][black]{{\color{#1}\ding{115}}}
\newcommand{\symbolbox}[1][black]{{\color{#1}\ding{110}}}
\newcommand{\symbolcircle}[1][black]{{\color{#1}\ding{108}}}

\newcommand{\symboldiamondopen}[1][black]{{\color{#1}$\diamond$}}
\newcommand{\symboltriangleopen}[1][black]{{\color{#1}$\vartriangle$}}
\newcommand{\symbolboxopen}[1][black]{{\color{#1}$\square$}}
\newcommand{\symbolcircleopen}[1][black]{{\color{#1}\icircleopen}}

\newcommand{\vertexi}[1][black]{\symbolcircle[#1]}
\newcommand{\vertexii}[1][white]{
  \setlength{\unitlength}{1.5ex}
  \begin{picture}(1.0000,1.0000)
    \put(0.0000,0.4000){{\color{#1}\circle*{1.000}}}
    \put(0.0000,0.4000){\circle{1.000}}
  \end{picture}
  }

\definecolor{FGViolet}{rgb}{0.61,0.32,0.61}
\definecolor{FGBlue}{rgb}{0,0,0.8}
\definecolor{FGGreen}{rgb}{0.2,0.7,0.2}
\definecolor{FGOrange}{rgb}{0.95,0.5,0.1}
\definecolor{FGRed}{rgb}{0.8,0,0}
\definecolor{FGLightGray}{rgb}{0.8,0.8,0.8}

\begin{document}

\title{The In-Medium Similarity Renormalization Group: \\
A Novel Ab Initio Method for Nuclei}

\author[nscl,msu,osu]{H.\ Hergert\corref{cor}}
\ead{hergert@nscl.msu.edu}

\author[nscl,msu]{S.\ K.\ Bogner}
\ead{bogner@nscl.msu.edu}

\author[msu,nscl]{T.\ D.\ Morris}
\ead{morrist@nscl.msu.edu}

\author[ikp,emmi]{A.\ Schwenk}
\ead{schwenk@physik.tu-darmstadt.de}

\author[cns]{K.\ Tsukiyama}
\ead{tsuki.kr@gmail.com}

\address[nscl]{National Superconducting Cyclotron Laboratory, 
Michigan State University, East Lansing, MI 48824, USA}
\address[msu]{Department of Physics and Astronomy, Michigan State University,
East Lansing, MI 48824, USA}
\address[osu]{Department of Physics, The Ohio State University, 
Columbus, OH 43210, USA}
\address[ikp]{Institut f\"ur Kernphysik, Technische Universit\"at
Darmstadt, 64289 Darmstadt, Germany}
\address[emmi]{ExtreMe Matter Institute EMMI, GSI Helmholtzzentrum f\"ur
Schwerionenforschung GmbH, 64291 Darmstadt, Germany}
\address[cns]{Center for Nuclear Study, Graduate School of Science,
University of Tokyo, Hongo, Tokyo, 113-0033, Japan}

\cortext[cor]{Corresponding author.}

\begin{abstract}

\noindent
We present a comprehensive review of the In-Medium Similarity
Renormalization Group (IM-SRG), a novel \emph{ab inito} method for
nuclei. The IM-SRG employs a continuous unitary transformation of the
many-body Hamiltonian to decouple the ground state from all
excitations, thereby solving the many-body problem. Starting from a
pedagogical introduction of the underlying concepts, the IM-SRG flow 
equations are developed for systems with and without explicit spherical
symmetry. We study different IM-SRG generators that
achieve the desired decoupling, and how they affect the details of the IM-SRG
flow. Based on calculations of closed-shell nuclei, we assess possible
truncations for closing the system of flow equations in practical
applications, as well as choices of the reference state. We discuss
the issue of center-of-mass factorization and demonstrate that the
IM-SRG ground-state wave function exhibits an approximate decoupling
of intrinsic and center-of-mass degrees of freedom, similar to Coupled
Cluster (CC) wave functions. To put the IM-SRG in context with other
many-body methods, in particular many-body perturbation theory and
non-perturbative approaches like CC, a detailed perturbative analysis 
of the IM-SRG flow equations is carried out. We conclude with a
discussion of ongoing developments, including IM-SRG calculations with
three-nucleon forces, the multi-reference IM-SRG for open-shell
nuclei, first non-perturbative derivations of shell-model
interactions, and the consistent evolution of operators in the
IM-SRG. We dedicate this review to the memory of Gerry Brown, one of
the pioneers of many-body calculations of nuclei.

\end{abstract}

\maketitle

\tableofcontents

\section{In Memory of Gerry Brown}

Gerry Brown was a true giant, whose scientific contributions range
across atomic physics, condensed matter and nuclear physics, and
astrophysics. Perhaps just as much as for his research and vision,
Gerry was known as an amazing scientific mentor, supervising over 70
Ph.D.~students and a comparable number of postdocs over the years. In
actuality, the number of young people inspired by Gerry was far
greater, as the steady stream of students and postdocs who interacted
with him at NORDITA, Princeton, and Stony Brook from 1950-2009 can
attest. Gerry's warm personality and unpretentious air, together with
his breadth of knowledge, and his intuitive, physically motivated
style of doing physics, made him an ideal person for young people to
discuss with. Gerry did not suffer fools gladly; if he thought what
you were working on was wrong, uninteresting, or a dead end, you would
know it in no uncertain terms. However, if he liked what you were
doing, he would enthusiastically dole out praise and encouragement,
and offer helpful suggestions. If he {\it really} liked what you were
doing, you would earn a home-cooked dinner at his and his wife Betty's
Setauket home. It is no surprise that many of the leading figures of
the past 4+ decades broadly across nuclear physics, from nuclear
structure and nuclear astrophysics, to heavy-ion and hadronic physics,
benefited from strong interactions with Gerry at some point.

The topic of the present review is the In-Medium Similarity
Renormalization Group (IM-SRG), a powerful novel method for \emph{ab
initio} many-body calculations.  As its name implies, the IM-SRG is
strongly rooted in Renormalization Group (RG) ideas, which have made a
significant impact on nuclear structure theory since the pioneering
applications in the early 2000's.  Gerry would have been quite pleased
with the IM-SRG, as he long advocated for the increased use of RG and
Effective Field Theory (EFT) methods in nuclear physics, dating back
to when two of us (SKB and AS) were beginning Ph.D.~students at Stony
Brook in the late 1990's. It was then that Gerry provided our first
exposure to these powerful techniques, challenging us to recast in RG
language the low-momentum NN interaction $V_{{\rm low}\,k}$ and to
revisit the calculations of Fermi liquid parameters and shell model
Hamiltonians from a modern RG perspective.  This was vintage Gerry, in
that his intuitive style of doing physics told him that these problems
were intimately related to Wilsonian RG ideas, even if he didn't know
yet the details. Indeed, if pressed on any of the formalism or
technical details, he would give a wry smile and say that such things
were the responsibilities of young people to work through.

While Gerry's research interests shifted towards astrophysics,
heavy-ion and hadronic physics in his later years, the nuclear
many-body problem always held a privileged place in his heart. As
students, Gerry told us on more than one occasion that his work with
Tom Kuo in the 1960's deriving shell model Hamiltonians from the NN
interaction~\cite{Kuo:1966ij,Kuo:1967qf} was his proudest achievement.
Gerry was similarly fond of his work in the 1970's and 1980's with
Babu, B\"ackman, Niskanen, and others, where they used a combination
of microscopic many-body theory and experimental constraints to derive
the Fermi liquid parameters from the underlying NN interaction, see
Ref.~\cite{Backman:1985xy} for a review. Perhaps his sentimental
attachment to both of these problems, and to many-body problems in
general, stemmed from the fact that he fearlessly chose to work on
them at a time when many of the leading physicists, whom he deeply
admired, most notably Migdal and Wigner, told him he was crazy to work
on problems that were, in their minds, almost completely
intractable. The present-day success of \emph{ab initio} nuclear
theory owes a debt of gratitude to Gerry, both for leading the first
successful attempts to understand nuclei and nuclear matter from
nuclear forces, and for having the foresight to recognize the
important role that would be played by RG and EFT methods in the
advancement of the field.

\section{Introduction}

The quest to predict and understand the properties of nuclei starting
from the underlying nuclear forces is a long and winding road,
spanning nearly 60 years dating back to the pioneering work of
Brueckner, Bethe, and
Goldstone~\cite{Brueckner:1955rw,Bethe:1956zz,Goldstone:1957zz}. In
contrast to quantum chemistry, where predictive and accurate \emph{ab
  initio} many-body calculations were commonplace by the
1970s~\cite{Schaefer}, progress was slowed by the challenging aspects
of the nuclear problem, namely the lack of a consistent theory for the
strong inter-nucleon interactions, and the need to perform
computationally expensive (and uncontrolled) resummations to handle
the non-perturbative aspects of the problem.  Consequently, for many
years \emph{ab initio} theory languished as a predictive force, and
could only explain in semi-quantitative terms how successful
phenomenology such as the shell model and Skyrme energy-density
functionals are linked to the underlying nuclear interactions.

\begin{figure}[p]
\centering
   \begin{subfigure}[b]{1\textwidth}
  \includegraphics[width=1\linewidth]{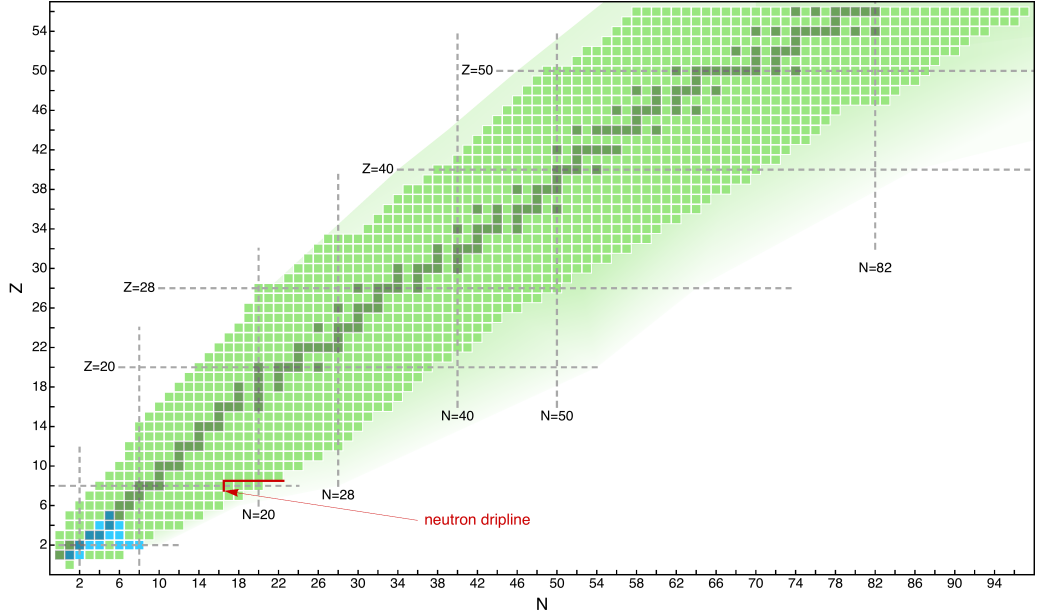}
   \caption{   \label{fig:abinitio2005} }
\end{subfigure}
\begin{subfigure}[b]{1\textwidth}
  \includegraphics[width=1\linewidth]{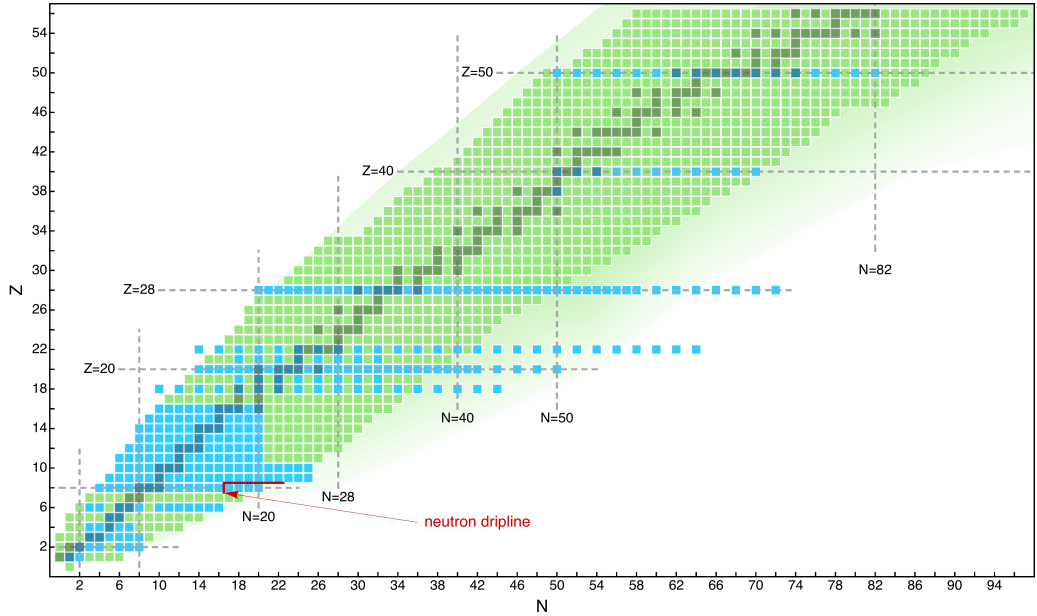}
   \caption{\label{fig:abinitio2015}}
\end{subfigure}
\caption{\label{fig:abinitio}The chart of nuclides and the reach of \emph{ab initio} calculations in (a) 2005 and (b) 2015. Nuclei for which \emph{ab initio} calculations exist are highlighted in blue. We note that the figure is for illustrative purposes only, and is based on the authors' potentially non-exhaustive survey of the literature.}
\end{figure}

As experimental efforts have shifted towards exotic nuclei, there has
been an increased urgency to develop reliable \emph{ab initio}
approaches to counter the inherent limitations of phenomenology. As
evidenced by Fig.~\ref{fig:abinitio}, tremendous progress has been
made in recent years, where the interplay of different threads, namely
rapidly increasing computational power, EFT, and RG transformations,
have enabled the development of new many-body methods and the revival
of old ones to successfully attack these
problems~\cite{Epelbaum:2009ve,Bogner:2010pq,Hammer:2013nx,Balantekin:2014gf,Hebeler:2015xq}. Remarkably,
it is now possible to perform quasi-exact calculations including
three-nucleon interactions of nuclei up to carbon or oxygen in quantum
Monte Carlo (QMC) and no-core shell model (NCSM) calculations, and
$N=Z$ nuclei up through $^{28}$Si in lattice effective field theory
with Euclidean time
projection~\cite{Carlson:2015lq,Barrett:2013oq,Roth:2011kx,Lahde:2014vn}. Moreover,
a host of approximate (but systematically improvable) methods such as
Coupled Cluster (CC), self-consistent Green's functions (SCGF),
auxiliary field diffusion Monte Carlo (AFDMC), and the IM-SRG have
pushed the frontiers of \emph{ab initio} theory well into the
medium-mass region, opening up new directions to the challenging
terrain of open-shell and exotic
nuclei~\cite{Tsukiyama:2012fk,Bogner:2014tg,Jansen:2014qf,Jansen:2015pl,Stroberg:2015qr,Soma:2013ys,Soma:2014fu,Soma:2014eu,Hergert:2013ij,Hergert:2014vn,Hergert:2015qd,Gandolfi:2014rt},
with recent highlights in the calcium
isotopes~\cite{Wienholtz:2013bh,Hagen:2015ve}.

As Gerry predicted, RG methods have played a prominent role in the
resurgence of \emph{ab initio} theory. A key to optimizing
calculations of nuclei is a proper choice of degrees of freedom.
While Quantum Chromodynamics (QCD) is the underlying theory of strong
interactions, the most \emph{efficient} low-energy degrees of freedom
for nuclear structure are the colorless hadrons of traditional nuclear
phenomenology.  But this realization is not enough.  For low-energy
calculations to be computationally efficient (or even feasible in some
cases) we need to exclude or, more generally, to \emph{decouple} the
high-energy degrees of freedom in a manner that leaves low-energy
observables invariant.

Progress on the nuclear many-body problem was hindered for decades
because nucleon-nucleon (NN) potentials that reproduce elastic
scattering phase shifts typically have strong short-range repulsion
and strong short-range tensor forces. This produces substantial
coupling to high-momentum modes, which is manifested as strongly
correlated many-body wave functions and highly nonperturbative few-
and many-body systems. For many years, the only viable option to
handle these features in a controlled manner was to use quasi-exact
methods such as QMC or NCSM, which limited the reach of \emph{ab
  initio} calculations to light $p$-shell nuclei. Powerful methods
that scale favorably to larger systems like CC and many-body
perturbation theory (MBPT) were largely abandoned in nuclear physics,
but exported to quantum chemistry, where they enjoyed immediate
success and quickly became the gold-standard for \emph{ab initio}
calculations. The success of CC and related methods in quantum
chemistry stems from the fact that Hartree-Fock is a good starting
point due to the relatively weak correlations induced by the Coulomb
interaction, in stark contrast to the nuclear case.

New approaches to nuclear forces grounded in RG ideas and techniques
have been developed in recent years that effectively make the nuclear
many-body problem look more like quantum
chemistry~\cite{Bogner:2006qf,Bogner:2007kt,Bogner:2007od,Bogner:2010pq,Jurgenson:2009bs,Roth:2011kx,Hebeler:2012ly}. The
RG allows continuous changes in ``resolution'' that decouple the
troublesome high-momentum modes and can be used to evolve interactions
to nuclear structure energy and momentum scales while preserving
low-energy observables. Such potentials, known generically as
``low-momentum interactions,'' are more perturbative and generate much
less correlated wave functions. This has played a major role in
expanding the reach of \emph{ab initio } calculations to medium-mass
nuclei, since methods that exhibit polynomial scaling can now be
converged in manageable model spaces. See
Refs.~\cite{Bogner:2010pq,Furnstahl:2012fn,Furnstahl:2013zt} for
recent reviews on the use of RG methods in nuclear physics.

As we will show in the following, the IM-SRG approach extends the RG
notion of decoupling to the many-body Hilbert space by formulating
``in-medium'' flow equations, the solution of which is equivalent to
the partial diagonalization or block-diagonalization of the many-body
Hamiltonian~\cite{Bogner:2010pq,Tsukiyama:2011uq,Tsukiyama:2012fk,Bogner:2014tg,Hergert:2013ij}. Because
of its favorable polynomial scaling with system size, and the
flexibility to target ground and excited states of both closed- and
open-shell systems, the IM-SRG provides a powerful \emph{ab initio}
framework for calculating medium-mass nuclei from first principles
that is grounded in modern RG principles. Moreover, we will show that
the IM-SRG provides a controlled, non-perturbative scheme to derive
effective valence shell model Hamiltonians and operators from the
underlying nuclear forces. We believe Gerry would have been quite
pleased!

\subsection{Organization of This Review}

This work is organized as follows. The basic IM-SRG approach is laid
out in Secs.~\ref{sec:floweq} and~\ref{sec:generators}, where we
introduce the IM-SRG flow equations for normal-ordered operators and
discuss the choice of generators. Sections~\ref{sec:numerics}
and~\ref{sec:refstate} present selected numerical results to
illustrate and elaborate on key discussion points of
Secs.~\ref{sec:floweq} and \ref{sec:generators}, in particular the
convergence behavior of IM-SRG energies for different generators, the
decoupling, and the impact of the choice of reference state. In
Sec.~\ref{sec:mbpt}, we carry out an in-depth perturbative analysis of
the IM-SRG through fourth order in MBPT. The
diagrammatic content is compared to that of CC theory, which can be
analyzed along similar lines, and we discuss the implications for
perturbative truncations of the IM-SRG flow equations. Finally, we
discuss and demonstrate the center-of-mass factorization in the IM-SRG
ground-state wave function in Sec.~\ref{sec:com}. Due to many ongoing
developments with three-nucleon forces, we restrict the results in
these sections to calculation with nucleon-nucleon interactions only.
We summarize in Sec.~\ref{sec:sum+dev} and close with a discussion of
recent advances for open-shell nuclei, including three-nucleon forces,
and for the consistent evolution of operators. Auxiliary material,
such as basic commutators required in the derivation of the IM-SRG
flow equations, are given in the appendices.

\section{\label{sec:floweq}IM-SRG Flow Equations}

\subsection{\label{sec:floweq_prelim}Preliminaries}
The Similarity Renormalization Group (SRG) was first formulated by Wegner \cite{Wegner:1994dk} and Glazek and Wilson \cite{Glazek:1993il} to study condensed matter systems and light-front quantum field theories, respectively.  From a mathematical point of view, the philosophy behind the SRG is to render the Hamiltonian $\HO(s)$ diagonal via a continuous unitary transformation 
\begin{equation}\label{eq:cut}
  \HO(s)=\UO(s)\HO(0)\UUO(s)\,,
\end{equation}
where $H(s=0)$ is the starting Hamiltonian and $s$ denotes the so-called flow
parameter, for reasons that will become apparent shortly. In practice, the demand 
for strict diagonality is usually relaxed to requiring band- or block-diagonality 
of the Hamiltonian matrix in a chosen basis, e.g., in relative momentum or harmonic
oscillator (HO) spaces. These specific cases are realized in nuclear
physics applications, where the SRG is used to decouple momentum or
energy scales, and thereby render the nuclear Hamiltonian more
suitable for \emph{ab initio} many-body calculations
\cite{Bogner:2007kt,Bogner:2010pq,Jurgenson:2009bs,Hebeler:2012ly}.

With the IM-SRG, we want to use this strategy to solve the many-body problem directly. Taking the derivative of Eq.~\eqref{eq:cut}
with respect to the flow parameter $s$, we obtain the operator flow equation
\begin{equation}\label{eq:opflow}
  \totd{}{s}\HO(s) = \comm{\etaO(s)}{\HO(s)}\,,
\end{equation} 
where the generator $\etaO(s)$ is related to the unitary transformation $\UO(s)$ by
\begin{equation}
  \eta(s)=\totd{U(s)}{s}U^{\dag}(s) = -\eta^{\dag}(s)\,.
\end{equation}
By rearranging this relation, we obtain a differential equation for $\UO(s)$ whose formal solution is given by the \emph{path-}
or \emph{S-ordered} exponential
\begin{equation}
  U(s) = \mathcal{S}\exp \int^s_0 ds' \eta(s')\,.
\end{equation}
We leave $\etaO(s)$ unspecified for now and defer the discussion of suitable choices to Sec. \ref{sec:generators}. 

Naively, one could try to solve the flow equation \eqref{eq:opflow} by choosing a suitable basis of the many-body Hilbert space and turning Eq.~\eqref{eq:opflow} into a matrix differential equation, but such an approach would ultimately amount to a
diagonalization of the many-body Hamiltonian. To make matters worse, implementing the flow means we would deal with the Hamiltonian's full spectrum rather than just some extremal eigenvalues that can be extracted efficiently in state-of-the-art, large-scale Lanczos approaches like the NCSM \cite{Navratil:2000hf,Barrett:2013oq}.

For the IM-SRG, we follow a different route, and formulate the flow equation as well as the decoupling conditions underlying the definition of $\etaO(s)$ in the language of second quantization. This approach has been very successful in producing powerful and numerically efficient many-body schemes, chief among them the CC method (see, e.g., \cite{Coester:1958dq,Coester:1960cr,Shavitt:2009,Hagen:2010uq}). 

\subsection{\label{sec:nord}Normal Ordering and Wick's Theorem}
Let us consider the usual fermionic creation and annihilation operators, $\aaO_i$ and $\aO_i$, with
\begin{equation}
  \acomm{\aaO_i}{\aaO_j} = \acomm{\aO_i}{\aO_j} = 0\,,\quad \acomm{\aaO_i}{\aO_j}=\delta_{ij}\,.
\end{equation}
The indices are collective labels for the quantum numbers of the single-particle states. Using these operators, we can construct a representation of any $A$-body operator, and a complete basis for the $A$-body Hilbert space is obtained by acting with products of $\aaO_i$ on the particle vacuum,
\begin{equation}\label{eq:vacbasis}
  \ket{\Phi\{i_1\ldots i_A\}}=\prod_{k=1}^A\aaO_{i_k}\ket{0}.
\end{equation}
The states $\ket{\Phi\{i_1\ldots i_A\}}$ are simple $A$-particle Slater determinants.

When we consider an actual $A$-body nucleus, it is inefficient to work with the basis states \eqref{eq:vacbasis} because of the existence of characteristic energy and momentum scales. The low-lying excitation spectrum is dominated by excitations of particles in the vicinity of the Fermi level, and the coupling between states, particularly between the ground state and excitations, is suppressed if their energies differ by much more than the characteristic energy $\hbar^2\lambda^2/2m$ associated with the nuclear interaction's resolution scale $\lambda$. Typical values for $\lambda$ are on the order of $3-4\,\fm^{-1}$ for interactions from chiral EFT~\cite{Epelbaum:2009ve}, or lower after softening with the free-space SRG~\cite{Bogner:2010pq}. For such interactions, we can find a single Slater determinant $\ket{\Phi}$ that is a fair approximation to the nucleus' ground state, and use $\ket{\Phi}$ rather than the particle vacuum $\ket{0}$ as a reference state to construct a complete many-body basis.

To account for the fact that $\ket{\Phi}$ is an $A$-particle state, we introduce normal-ordered operators by defining
\begin{equation}\label{eq:def_no}
  \aaO_i\aO_j \equiv \nord{\aaO_i\aO_j} +\, \contraction[1.5ex]{}{\aO}{{}^\dag_i}{\aO}\aaO_i\aO_j\,,
\end{equation}
where the contraction is the expectation value of the operator in the reference state $\ket{\Phi}$:
\begin{equation}\label{eq:def_particle_contraction}
  \contraction[1.5ex]{}{\aO}{{}^\dag_i}{\aO}\aaO_i\aO_j \equiv \matrixe{\Phi}{\aaO_i\aO_j}{\Phi} \equiv \rho_{ji} \,.
\end{equation}
By definition, the contractions are identical to the elements of the one-body density matrix of $\ket{\Phi}$ \cite{Ring:1980bb}. A normal-ordered $A$-body operator is now defined recursively by evaluating all contractions between creation and annihilation operators:
\begin{align}\label{eq:def_no_nbody}
   &\aaO_{i_1}\ldots\aaO_{i_A}\aO_{j_A}\ldots\aO_{j_1} \notag\\
    &\equiv\, \nord{\aaO_{i_1}\ldots\aaO_{i_A}\aO_{j_{A}}\ldots\aO_{j_{1}}} \notag\\
    &\hphantom{\equiv}
       + \contraction[1.5ex]{}{\aO}{{}^\dag_{i_1}}{\aO}\aaO_{i_1}\aO_{j_1} 
        \nord{\aaO_{i_2}\ldots\aaO_{i_A}\aO_{j_{A}}\ldots\aO_{j_{2}}} 
      -\; \contraction[1.5ex]{}{\aO}{{}^\dag_{i_1}}{\aO}\aaO_{i_1}\aO_{j_2} 
        \nord{\aaO_{i_2}\ldots\aaO_{i_A}\aO_{j_A}\ldots\aO_{j_{3}}\aO_{j_{1}}}
      + \text{\,singles}\notag\\
    &\hphantom{\equiv}
       + \left(
          \contraction[1.5ex]{}{\aO}{{}^\dag_{i_1}}{\aO}\aaO_{i_1}\aO_{j_1}
          \contraction[1.5ex]{}{\aO}{{}^\dag_{i_1}}{\aO}\aaO_{i_2}\aO_{j_2}
          -
          \contraction[1.5ex]{}{\aO}{{}^\dag_{i_1}}{\aO}\aaO_{i_1}\aO_{j_2}
          \contraction[1.5ex]{}{\aO}{{}^\dag_{i_1}}{\aO}\aaO_{i_2}\aO_{j_1}
        \right) 
        \nord{\aaO_{i_3}\ldots\aaO_{i_A}\aO_{j_A}\ldots\aO_{j_{3}}} + \text{\,doubles} \notag\\
    &\hphantom{\equiv}
  +\,\ldots\,+ \text{\,full contractions}\,,
\end{align}
where we have used quantum chemistry parlance (singles, doubles, etc.) for the 
number of contractions in a term.
Note that the shown double contraction corresponds to the factorization formula for the two-body density matrix of a Slater determinant,
\begin{equation}
  \rho_{j_1j_2i_1i_2}
  \equiv
  \matrixe{\Phi}{\aaO_{i_1}\aaO_{i_2}\aO_{j_2}\aO_{j_1}}{\Phi}
  = \rho_{i_1j_1}\rho_{i_2j_2}- \rho_{i_1j_2}\rho_{i_2j_1}\,.
\end{equation}

From Eq.~\eqref{eq:def_no}, it is evident that $\matrixe{\Phi}{\nord{\aaO_i\aO_j}}{\Phi}$ must vanish, and this is readily generalized for the expectation values of general normal-ordered operators in the reference state $\ket{\Phi}$,
\begin{equation}
  \matrixe{\Phi}{\nord{\aaO_{i_1}\ldots\aO_{i_1}}}{\Phi} = 0\,,
\end{equation}
which facilitates calculations of operator matrix elements in a space spanned by all possible excitations of $\ket{\Phi}$. Even more useful is Wick's theorem (see e.g.~\cite{Shavitt:2009}), which is a direct consequence of Eq.~\eqref{eq:def_no_nbody} and allows us to expand products of normal-ordered operators:
\begin{align}\label{eq:def_wick}
   & \nord{\aaO_{i_1}\ldots\aaO_{i_N}\aO_{j_{N}}\ldots\aO_{j_{1}}} 
     \nord{\aaO_{k_1}\ldots\aaO_{k_M}\aO_{l_{M}}\ldots\aO_{l_{1}}}
   \notag\\
    &=(-1)^{M\cdot N}  
     \nord{\aaO_{i_1}\ldots\aaO_{i_N}\aaO_{k_1}\ldots\aaO_{k_M}\aO_{j_{N}}\ldots\aO_{j_{1}}\aO_{l_{M}}\ldots\aO_{l_{1}}} 
    \notag\\
    &\hphantom{=}
       + (-1)^{M\cdot N}\contraction[1.5ex]{}{\aO}{{}^\dag_{i_1}}{\aO}\aaO_{i_1}\aO_{l_1} 
        \nord{\aaO_{i_2}\ldots\aaO_{k_M}\aO_{j_{N}}\ldots\aO_{l_{2}}} \notag\\
    &\hphantom{\equiv}
      + (-1)^{(M-1)(N-1)}\contraction[1.5ex]{}{\aO}{{}^\dag_{j_N}}{\aO}\aO_{j_N}\aaO_{k_1} 
        \nord{\aaO_{i_1}\ldots\aaO_{k_M}\aO_{j_{N}}\ldots\aO_{j_2}}
      \notag\\
    &\hphantom{=}
      + \text{\,singles}+ \text{\,doubles} + \ldots \,.
\end{align}
Here, the phases appear because we anti-commute the $\aaO_k$ operators past the $\aO_j$, and we encounter a new type of contraction,
\begin{equation}\label{eq:def_hole_contraction}
  \contraction[1.5ex]{}{\aO}{{}^\dag_{i}}{\aO}\aO_{i}\aaO_{j} 
  \equiv \matrixe{\Phi}{\aO_{i}\aaO_{j}}{\Phi}
  = \delta_{ij} - \rho_{ij}\,,
\end{equation}
as expected from the basic fermionic anti-commutator algebra. The important feature of Eq.~\eqref{eq:def_wick} is that only contractions between one index from each of the two strings appear in the expansion, because contractions between indices within a single normal-ordered string vanish by construction. This leads to a substantial reduction of terms in practical calculations. Another immediate consequence of Wick's theorem is that a product of normal-ordered $M$ and $N$-body operators has the general form
\begin{equation}
  \AO^{M}\BO^{N} = \sum_{k=|M-N|}^{M+N}C^{(k)}\,.
\end{equation}

\subsection{Normal-Ordered Hamiltonian}
Let us now consider an intrinsic nuclear $A$-body Hamiltonian containing both NN and 3N interactions,
\begin{equation}\label{eq:def_Hint}
  H = \left(1-\frac{1}{A}\right)T + T^{(2)} + V^{(2)} +V^{(3)}\,,
\end{equation}
where the one- and two-body kinetic energy terms are
\begin{align}
  T &\equiv \sum\frac{\pOV^2_i}{2m}\,,\\
  T^{(2)} &\equiv -\frac{1}{Am}\sum_{i<j}\pOV_i\cdot\pOV_j
\end{align}
(see Sec.~\ref{sec:com} and Ref.~\cite{Hergert:2009wh}). Choosing a single Slater determinant $\ket{\Phi}$ as the reference state, we can rewrite the Hamiltonian \emph{exactly} in terms of normal-ordered operators,
\begin{align}\label{eq:Hno}
  \HO &= E + \sum_{ij}f_{ij}\nord{\aaO_{i}\aO_{j}} + \frac{1}{4}\sum_{ijkl}\Gamma_{ijkl}\nord{\aaO_{i}\aaO_{j}\aO_{l}\aO_{k}}
    + \frac{1}{36}\sum_{ijklmn}W_{ijklmn}\nord{\aaO_{i}\aaO_{j}\aaO_{k}\aO_{n}\aO_{m}\aO_{l}}\,,
\end{align}
where the labels for the individual contributions have been chosen for historical reasons. For convenience, we will work in the eigenbasis of the one-body density matrix in the following, so that
\begin{equation}\label{eq:def_natorb}
  \rho_{ab}=n_{a}\delta_{ab}\,,\quad n_{a}\in\{0,1\}\,.
\end{equation}
The individual normal-ordered contributions in Eq.~\eqref{eq:Hno} are then given by
\begin{align}
  E &= \left(1-\frac{1}{A}\right)\sum_{a}\matrixe{a}{T}{a}n_{a}
  		+ \frac{1}{2}\sum_{ab}\matrixe{ab}{T^{(2)}\!+\!V^{(2)}}{ab}n_{a}n_{b}\notag\\
  		&\hphantom{=}+ \frac{1}{6}\sum_{abc}\matrixe{abc}{V^{(3)}}{abc}n_{a}n_{b}n_{c}\,,\label{eq:E0}\\
  f_{12} &= \left(1-\frac{1}{A}\right)\matrixe{1}{T}{2} 
  		+ \sum_{a}\matrixe{1a}{T^{(2)}\!+\!V^{(2)}}{2a}n_{a}\notag\\
  		&\hphantom{=}+ \frac{1}{2}\sum_{ab}\matrixe{1ab}{V^{(3)}}{2ab}n_{a}n_{b}\,,\label{eq:f}		\\
  \Gamma_{1234} &= \matrixe{12}{T^{(2)}\!+\!V^{(2)}}{34} + \sum_{a}\matrixe{12a}{V^{(3)}}{34a}n_{a}\,,\label{eq:Gamma}\\
  W_{123456}&=\matrixe{123}{V^{(3)}}{456}\,.
\end{align}
Due to the occupation numbers in Eqs.~\eqref{eq:E0}--\eqref{eq:Gamma}, the sums run over occupied (hole) states only. Note that the zero-, one-, and two-body parts of the Hamiltonian all contain in-medium contributions from the free-space 3N interaction. 

\subsection{\texorpdfstring{$M$}{M}-Scheme Flow Equations}
After discussing normal ordering and Wick's theorem in the previous sections, we are now ready to turn back to the operator flow equation, Eq.~\eqref{eq:opflow}. 

When carried out exactly, the IM-SRG is a continuous unitary transformation in $A$-body space, and consequently, $\eta(s)$ and $\HO(s)$ are $A$-body operators even if they initially have a lower rank at $s=0$. From the discussion of the previous sections, we see that every evaluation of the commutator on the right-hand side of Eq.~\eqref{eq:opflow} increases the particle rank of $\HO(s)$, e.g.,
\begin{equation}\label{eq:induced}
  \comm{\nord{\aaO_{a}\aaO_{b}\aO_{d}\aO_{c}}}{\nord{\aaO_{i}\aaO_{j}\aO_{l}\aO_{k}}}= \delta_{ci}\nord{\aaO_{a}\aaO_{b}\aaO_{j}\aO_{l}\aO_{k}\aO_{d}}+\ldots.
\end{equation}
All of these induced contributions will in turn contribute to the parts of $\HO(s)$ with lower particle rank in subsequent integration steps. Because an explicit treatment of all contributions up to the $A$-body level is clearly not feasible, we have to introduce a truncation to close the system of IM-SRG flow equations. We follow a simple approach, and truncate $\etaO(s)$ and $\HO(s)$ at a given particle rank $n\leq A$, which is motivated by the cluster decomposition principle for short-range interactions (see, e.g., \cite{Weinberg:1996uf}). For $n=2$, this yields the so-called IM-SRG(2) truncation, our primary truncation scheme in past works \cite{Tsukiyama:2011uq,Tsukiyama:2012fk,Hergert:2013mi}, which will serve as the basis of the discussion in the remainder of this work. On occasion, we will also consider the next truncation in the hierarchy, denoted IM-SRG(3). The corresponding flow equations are given in Appendix \ref{app:imsrg3}, but they have not been used in numerical calculations because of the computational demands associated with handling three-body operators.

Let us assume, then, that for each flow parameter $s$
\begin{align}
  \HO(s) &\approx E(s) + f(s) + \Gamma(s)\,,\\
  \etaO(s) &\approx\etaO^{(1)}(s)+\etaO^{(2)}(s)\,.
\end{align}
We introduce the permutation symbol $P_{ij}$ to interchange the attached indices in any expression, i.e.,
\begin{equation}\label{eq:def_Pij}
  P_{ij} g(\ldots,i,\ldots,j) \equiv g(\ldots,j,\ldots,i)\,,
\end{equation}
and use the fundamental commutators from Appendix \ref{app:commutators} to obtain
%
\begin{align}
  \totd{E}{s}&= \sum_{ab}(n_a-n_b)\eta_{ab} f_{ba} 
    + \frac{1}{2} \sum_{abcd}\eta_{abcd}\Gamma_{cdab} n_a n_b\bar{n}_c\bar{n}_d
    \label{eq:imsrg2_m0b}\,,\\[5pt]
%
  \totd{f_{12}}{s} &= 
  \sum_{a}(1+P_{12})\eta_{1a}f_{a2} +\sum_{ab}(n_a-n_b)(\eta_{ab}\Gamma_{b1a2}-f_{ab}\eta_{b1a2}) \notag\\ 
  &\quad +\frac{1}{2}\sum_{abc}(n_an_b\bar{n}_c+\bar{n}_a\bar{n}_bn_c) (1+P_{12})\eta_{c1ab}\Gamma_{abc2}
  \label{eq:imsrg2_m1b}\,,\\[5pt]
%
  \totd{\Gamma_{1234}}{s}&= 
  \sum_{a}\left\{ 
    (1-P_{12})(\eta_{1a}\Gamma_{a234}-f_{1a}\eta_{a234} )
    -(1-P_{34})(\eta_{a3}\Gamma_{12a4}-f_{a3}\eta_{12a4} )
    \right\}\notag \\
  &\quad+ \frac{1}{2}\sum_{ab}(1-n_a-n_b)(\eta_{12ab}\Gamma_{ab34}-\Gamma_{12ab}\eta_{ab34})
    \notag\\
  &\quad-\sum_{ab}(n_a-n_b) (1-P_{12})(1-P_{34})\eta_{b2a4}\Gamma_{a1b3}
    \label{eq:imsrg2_m2b}\,,
\end{align}
where $\nn_i=1-n_i$, and the $s$-dependence has been suppressed for brevity. To obtain ground-state energies, we integrate Eqs.~\eqref{eq:imsrg2_m0b}--\eqref{eq:imsrg2_m2b} from $s=0$ to $s\to\infty$, starting from the initial 
components of the normal-ordered Hamiltonian (Eqs.~\eqref{eq:E0}--\eqref{eq:Gamma}).

As we will discuss in more detail in Sec.~\ref{sec:mbpt}, 
Eqs.~\eqref{eq:imsrg2_m0b}--\eqref{eq:imsrg2_m2b} can easily be translated 
into Goldstone or Hugenholtz diagrams for the flowing Hamiltonian $H(s)$.
This provides us with an intuitive understanding of the mechanism through
which the IM-SRG is non-perturbatively re-summing the many-body expansion. The second
and third rows of Eq.~\eqref{eq:imsrg2_m2b}, in particular, re-sum $pp/hh$-ladder and $ph$-ring 
diagrams, respectively. Due to the use of $H(s)$, ladder-ring interference 
diagrams are generated in the limit $s\to\infty$, and therefore the IM-SRG(2) 
goes beyond the more traditional Brueckner $G$-matrix or Random-Phase 
Approximation-type summations \cite{Day:1967zl,Brandow:1967tg,Fetter:2003ve}. 
Furthermore, the commutator structure of Eq.~\eqref{eq:opflow} ensures that 
the IM-SRG is size-extensive, and only connected diagrams are generated and
re-summed by the IM-SRG flow \cite{Brandow:1967tg,Shavitt:2009}. This property
is preserved even if the commutators are truncated at a given operator rank, as in the IM-SRG(2) case presented here. 

From Eqs.~\eqref{eq:imsrg2_m0b}--\eqref{eq:imsrg2_m2b}, it is clear that the computational effort for solving the IM-SRG(2) flow equations is dominated by the two-body flow equation, which scales polynomially like $\OC(N^6)$ with the single-particle basis size $N$. This puts the IM-SRG(2) in the same category as other numerically efficient non-perturbative methods\footnote{The mentioned methods can make use of the distinction between particle and hole states in the single-particle basis to further reduce the effort. For instance, the amplitude equations of CCSD can be solved at $\OC(N_h^2N_p^4)$ cost, where typically the number of hole states $N_h$ is much smaller than the number of particle states $N_p\sim N$. Note that the construction of the CCSD effective Hamiltonian from the amplitudes requires $\OC(N^6)$ effort. In the IM-SRG(2), we are working directly with the analogous effective Hamiltonian.} like Coupled Cluster with Singles and Doubles (CCSD) \cite{Shavitt:2009,Hagen:2014ve}, the Self-Consistent Green's Function Approach (SCGF) \cite{Dickhoff:2004fk,Barbieri:2007fk,Cipollone:2013uq}, or canonical transformation theory \cite{White:2002fk,Yanai:2006uq}.

\subsection{\texorpdfstring{$J$}{J}-Scheme Flow Equations\label{sec:flow_jscheme}}
When the nuclear Hamiltonian is normal-ordered with respect to a general reference state, its manifest symmetries may become hidden and implicitly reliant on cancellations which are spoiled by the introduction of finite bases or other truncations. For systems with explicit spherical symmetry, e.g., for closed-shell nuclei, we can use a spherically symmetric $\ket{\Phi}$, and preserve the manifest rotational symmetry of $\HO$ on a term-by-term basis. In this case, the flow equations become block-diagonal in angular momentum, and independent of the angular momentum projection quantum numbers, which leads to a significant reduction in numerical effort.

In the following, the single-particle indices collectively represent the radial, angular momentum, and isospin quantum numbers $i=(k_{i}l_{i}j_{i}\tau_{i})$. Then the matrix elements of single-particle operators are diagonal in all but the radial quantum numbers, e.g.,
\begin{equation}
  f_{12}=f^{l_{1}j_{1}\tau_{1}}_{k_{1}k_{2}}\delta_{l_{1}l_{2}}\delta_{j_{1}j_{2}}\delta_{\tau_{1}\tau_{2}}\,.
\end{equation}
Likewise, two-body matrix elements are diagonal in total two-body angular momentum $J$ and independent of $M$. In this case, the IM-SRG(2) flow equations reduce to
\begin{align}
  \label{eq:imsrg2_j0b}
  \totd{E}{s}&=\sum_{ab}\hat{j}_{a}^{2}\eta_{ab}f_{ba}(n_{a}-n_{b})
          +\frac{1}{2}\sum_{abcdJ}\hJ^{2}\eta^{J}_{abcd}\Gamma^{J}_{cdab}n_{a}n_{b}\nn_{c}\nn_{d}\,,\\
  \label{eq:imsrg2_j1b}
  \totd{f_{12}}{s}&=\sum_{a}(1+P_{12})\eta_{1a}f_{a2}
            +\frac{1}{\hat{j}^{2}_{1}}\sum_{abJ}\hJ^{2}(n_{a}-n_{b})\left(\eta_{ab}\Gamma^{J}_{b1a2}-f_{ab}\eta^{J}_{b1a2}\right)\notag\\                 
           &\hphantom{=}+\frac{1}{2\hat{j}^{2}_{1}}
          \frac{1}{2}\sum_{abcJ}\hJ^{2}\big(n_{a}n_{b}\nn_{c}+\nn_{a}\nn_{b}n_{c}\big)
            \left(1+P_{12}\right)\eta^{J}_{c1ab}\Gamma^{J}_{abc2}
          \,,\\                 
  \label{eq:imsrg2_j2b}
 \totd{\Gamma^{J}_{1234}}{s}
        &=\sum_{a}\Big(\left(1-(-1)^{J-j_{1}-j_{2}}P_{12}\right)
                       \left(\eta_{1a}\Gamma^{J}_{a234}-f_{1a}\eta^{J}_{a234}\right)\notag\\
        &\hphantom{=\sum_a}- \left(1-(-1)^{J-j_{3}-j_{4}}P_{34}\right)
                    \left(\eta_{a3}\Gamma^{J}_{12a4}-f_{a3}\eta^{J}_{12a4}\right)\Big)\notag\\
        &\hphantom{=}+\frac{1}{2}\sum_{ab}\left(\eta^{J}_{12ab}\Gamma^{J}_{ab34}-\Gamma^{J}_{12ab}\eta^{J}_{ab34}\right)\left(1-n_{a}-n_{b}\right)\notag\\
        &\hphantom{=}+\sum_{abJ'}\left(n_{a}-n_{b}\right)\left(1-(-1)^{J-j_{1}-j_{2}}P_{12}\right)\notag\\
        &\hphantom{=}\qquad\times
          \widehat{J'}^2\sixj{j_{1} & j_{2} & J \\ j_{3} & j_{4} & J'}\left(\overline\eta^{J'}_{1\bar{4}a\bar{b}}\overline\Gamma^{J'}_{a\bar{b}3\bar{2}}-\overline\Gamma^{J'}_{1\bar{4}a\bar{b}}\overline\eta^{J'}_{a\bar{b}3\bar{2}}\right)\,,
\end{align}
where $\hat{j}\equiv\sqrt{2j+1}$, indices with a bar indicate time-reversed states, and the $\overline{\eta}$ and $\overline{\Gamma}$ matrix elements in the last line of Eq.~\eqref{eq:imsrg2_j2b} are obtained by a generalized Pandya transform (see, e.g., \cite{Suhonen:2007wo}),
\begin{equation}
  \overline\OO^{J}_{1\bar{2}3\bar{4}} = - \sum_{J'}\hat{J'}^{2}\sixj{j_{1} & j_{2} & J \\ j_{3} & j_{4} & J'} O^{J'}_{1432}\,.
\end{equation}
Alternatively, the particle-hole terms in the last line of Eq.~\eqref{eq:imsrg2_j2b}
can also be expressed in terms of the cross-coupled matrix elements introduced in
Ref.~\cite{Kuo:1981fk}. If the same angular-momentum coupling order is used, the
two types of matrix elements are related by a trivial phase factor $(-1)$.

\subsection{General Observables\label{sec:observables}}
In principle, the evaluation of observables other than the Hamiltonian is a straightforward task: We need to normal-order the operator $\OO(s)$ with respect to the same reference state as the Hamiltonian, and then plug it into the flow equation
\begin{equation}\label{eq:obsflow}
  \totd{}{s}\OO(s) = \comm{\etaO(s)}{\OO(s)}\,,
\end{equation} 
with $\etaO(s)$ the same as in the Hamiltonian flow equation. For consistency with the overall IM-SRG(2) scheme, $\OO(s)$ is truncated at the two-body level. We then obtain an additional set of flow equations for the normal-ordered zero-, one- and two-body parts of $\OO(s)$ which need to be solved alongside Eqs.~\eqref{eq:imsrg2_m0b}--\eqref{eq:imsrg2_m2b}. We will follow this route in later sections of this work to investigate radii (Sec.~\ref{sec:numerics_radii}) and the center-of-mass separation in the IM-SRG(2) ground-state wave function (Sec.~\ref{sec:com}). Due to the size of the system of flow equations, and the associated storage needs of numerical differential equation solvers, this procedure becomes unfeasible if we are interested in more than one or two additional operators.

Unfortunately, we cannot resort to the same strategy as in the free-space SRG case \cite{Anderson:2010br,Bogner:2010pq}, where the unitary transformation can be reconstructed from the eigenvectors of the initial and final Hamiltonians in the two-nucleon, three-nucleon,\ldots system. To do so, we would have to solve the eigenvalue problem of the Hamiltonian in the $A$-body system through exact diagonalization, and we could not even resort to large-scale NCSM machinery because it does not provide the full eigenbasis, but only the lowest eigenvalues and eigenvectors via Lanczos methods. The cost for exact diagonalization increases factorially with the single-particle basis, and it is precisely this high computational effort that motivated the development of mildly scaling methods like CC or the IM-SRG. A more efficient alternative for the evaluation of observables exists in the form of the so-called Magnus expansion \cite{Blanes:2009fk,Morris:2015ve}, which we briefly discuss in Sec.~\ref{sec:magnus}.

\section{\label{sec:generators}Choice of Generator}

\subsection{\label{sec:decoupling}Decoupling}
\begin{figure}[t]
\setlength{\unitlength}{0.8\columnwidth}
  \begin{center}
  \begin{picture}(1.0000,0.5500)
   \put(0.0350,0.0450){\includegraphics[width=0.46\unitlength]{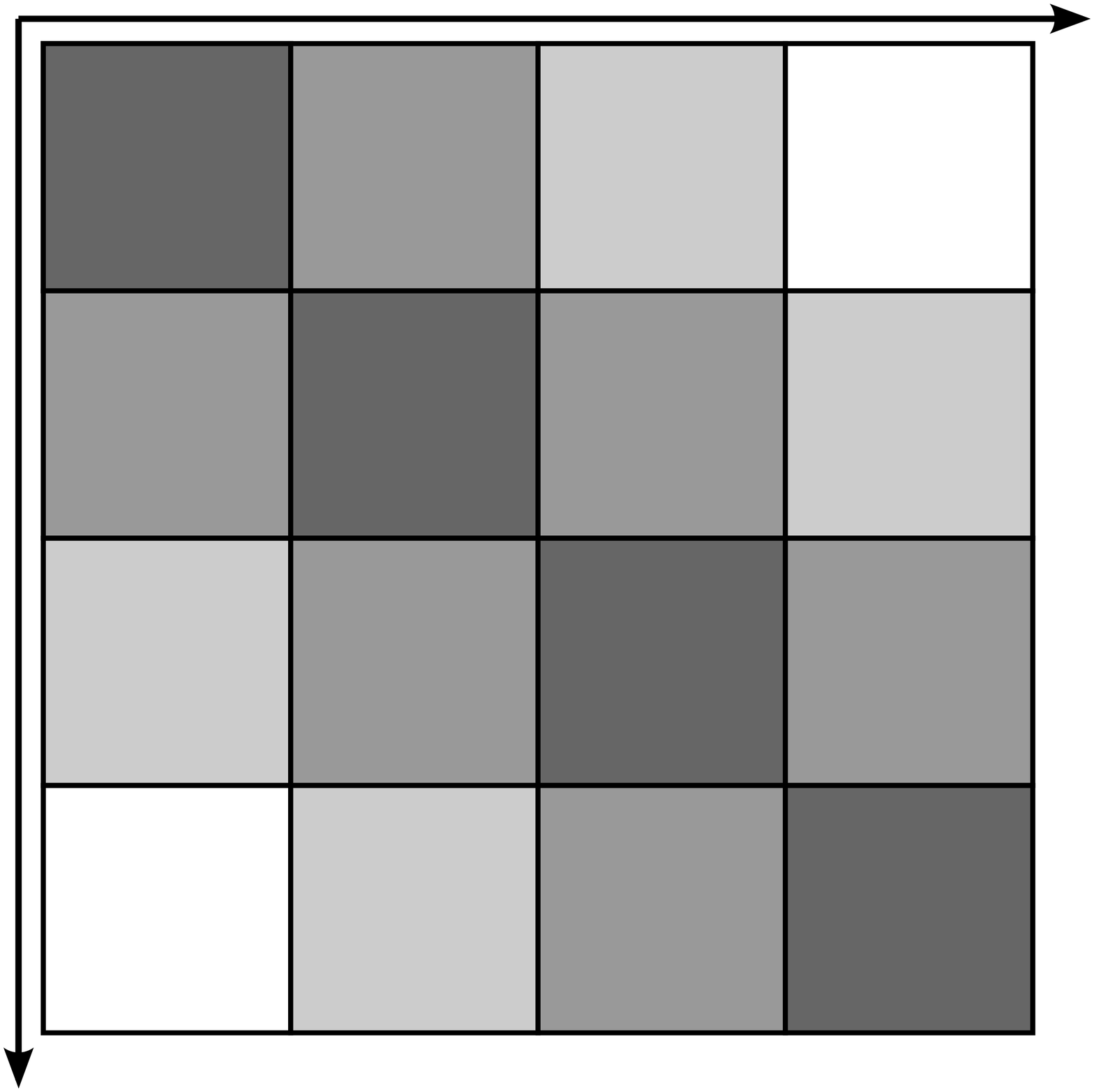}}
   \put(0.5400,0.0450){\includegraphics[width=0.46\unitlength]{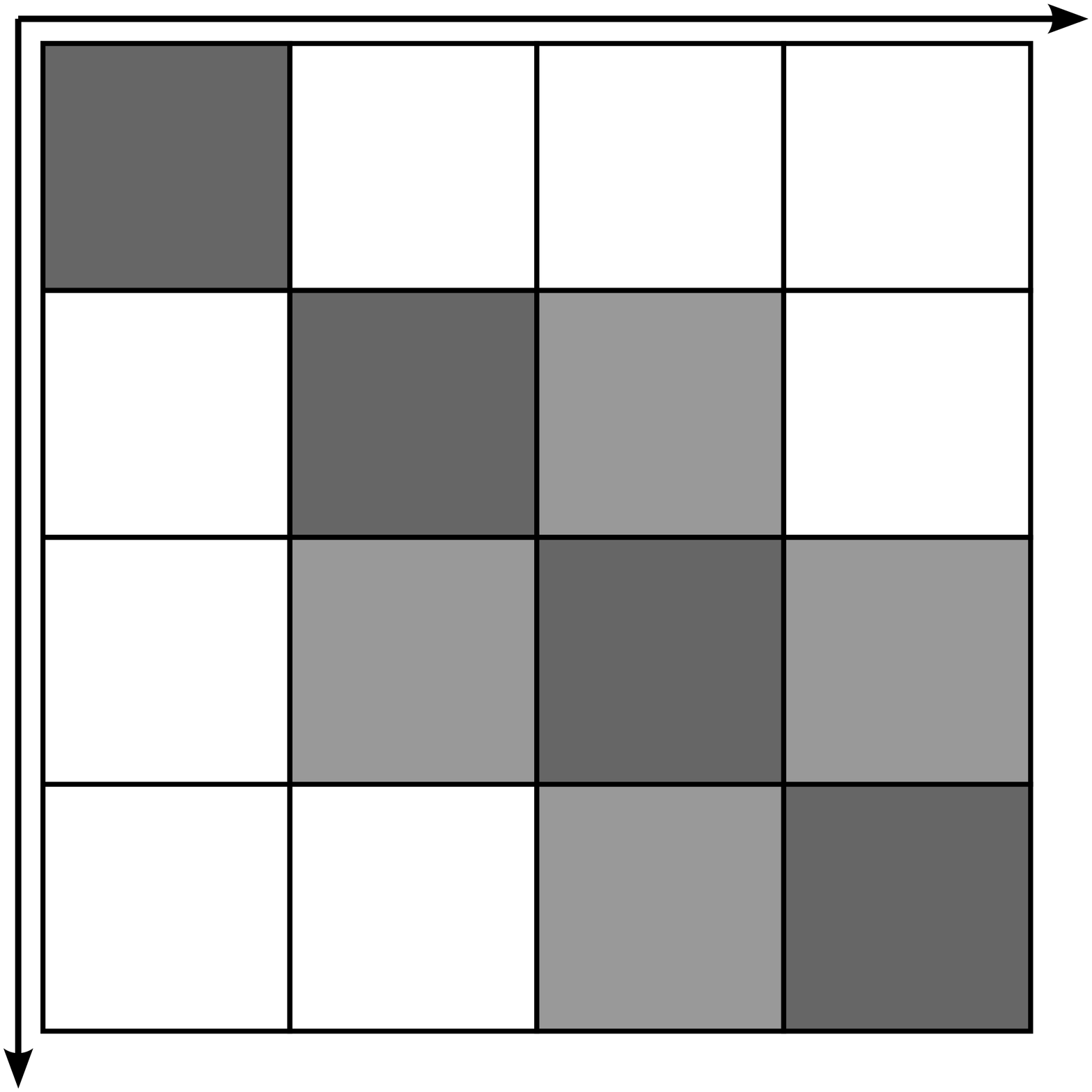}}
   \put(0.0100,0.0000){\parbox{0.5\unitlength}{\centering$\matrixe{i}{\HO(0)}{j}$}}
   \put(0.5200,0.0000){\parbox{0.5\unitlength}{\centering$\matrixe{i}{\HO(\infty)}{j}$}}
   
   \put(0.0500,0.5100){\parbox{0.11\unitlength}{\centering\footnotesize0p0h}}
   \put(0.1600,0.5100){\parbox{0.11\unitlength}{\centering\footnotesize1p1h}}
   \put(0.2630,0.5100){\parbox{0.11\unitlength}{\centering\footnotesize2p2h}}
   \put(0.3650,0.5100){\parbox{0.11\unitlength}{\centering\footnotesize3p3h}}
   \put(0.5500,0.5100){\parbox{0.11\unitlength}{\centering\footnotesize0p0h}}
   \put(0.6600,0.5100){\parbox{0.11\unitlength}{\centering\footnotesize1p1h}}
   \put(0.7630,0.5100){\parbox{0.11\unitlength}{\centering\footnotesize2p2h}}
   \put(0.8650,0.5100){\parbox{0.11\unitlength}{\centering\footnotesize3p3h}}
   \put(0.0100,0.4320){\parbox{0.11\unitlength}{\rotatebox{90}{\centering\footnotesize0p0h}}}
   \put(0.0100,0.3235){\parbox{0.11\unitlength}{\rotatebox{90}{\centering\footnotesize1p1h}}}
   \put(0.0100,0.2175){\parbox{0.11\unitlength}{\rotatebox{90}{\centering\footnotesize2p2h}}}
   \put(0.0100,0.1100){\parbox{0.11\unitlength}{\rotatebox{90}{\centering\footnotesize3p3h}}}

   \put(0.5100,0.4320){\parbox{0.11\unitlength}{\rotatebox{90}{\centering\footnotesize0p0h}}}
   \put(0.5100,0.3235){\parbox{0.11\unitlength}{\rotatebox{90}{\centering\footnotesize1p1h}}}
   \put(0.5100,0.2175){\parbox{0.11\unitlength}{\rotatebox{90}{\centering\footnotesize2p2h}}}
   \put(0.5100,0.1100){\parbox{0.11\unitlength}{\rotatebox{90}{\centering\footnotesize3p3h}}}
  \end{picture}
  \end{center}
  \caption{\label{fig:schematic}Schematic representation of the initial and final Hamiltonians, $\HO(0)$ and $\HO(\infty)$, in the many-body Hilbert space spanned by particle-hole excitations of the reference state.}
\end{figure}

After setting up the general IM-SRG flow equation framework in Sec.~\ref{sec:floweq}, we have to specify the generator $\etaO$. To this end, we first need to identify the off-diagonal parts of the Hamiltonian that the IM-SRG transformation is supposed to suppress for $s\to\infty$. The freedom to partition the Hamiltonian into suitably defined diagonal and off-diagonal pieces gives the IM-SRG flexibility to target different states, and is key to extending the method to open-shell nuclei, see Secs.~\ref{sec:mrimsrg} and \ref{sec:shell_model}. To illustrate the general idea of a targeted decoupling, let us assume our goal is to extract the ground-state energy of a closed-shell nucleus, i.e., the lowest eigenvalue of the nuclear many-body Hamiltonian. In the left panel of Fig.~\ref{fig:schematic}, we show a schematic representation of the initial Hamiltonian $\HO(0)$, in a basis consisting of $A$-particle-$A$-hole ($A$p$A$h) excitations of the reference state $\ket{\Phi}$. For the following illustration of the IM-SRG's basic concept, we assume that $\HO(0)$ has been truncated to two-body operators, that is, it can at most couple $n$p$n$h to $(n\pm2)$p$(n\pm2)$h states. The extension to three-body operators is straightforward. 

The 0p0h reference state is coupled to 1p1h and 2p2h excitations by the matrix elements
\begin{align}
  \matrixe{\Phi}{\HO(0)\nord{\aaO_{p}\aO_{h}}}{\Phi}&=f_{ph}\,,\label{eq:fod}\\
  \matrixe{\Phi}{\HO(0)\nord{\aaO_{p}\aaO_{p'}\aO_{h'}\aO_{h}}}{\Phi}&=\Gamma_{pp'hh'}\label{eq:Gammaod}\,,
\end{align}
and their Hermitian conjugates. Thus, we define the off-diagonal part of the Hamiltonian as
\begin{equation}
  \HO^{od}(s) = \sum_{ph}f_{ph}\nord{\aaO_p\aO_h} + \frac{1}{4}\sum_{pp'hh'}\Gamma_{pp'hh'}\nord{\aaO_p\aaO_{p'}\aO_{h'}\aO_h}
                +\; \text{H.c.}\label{eq:def_Hod}\,.
\end{equation}
During the flow, matrix elements between the reference state and higher excitations acquire non-zero values,
\begin{equation}\label{eq:induced_od}
  \matrixe{\Phi}{\HO(s)\nord{\aaO_{p_1}\ldots\aaO_{p_A}\aO_{h_{A}}\ldots\aO_{h_1}}}{\Phi}\neq 0\,,
\end{equation}
because $\HO(s>0)$ has induced $3-,\ldots,A-$body contributions (cf.~Eq.~\eqref{eq:induced}), just as in a 
free-space SRG evolution \cite{Bogner:2010pq,Jurgenson:2009bs,Hebeler:2012ly}. By truncating operators to 
two-body rank in the IM-SRG(2) (or any rank $n\leq A$ in a higher truncation), we force these (and other) 
matrix elements to vanish, at the cost of violating unitarity. We will have to check that this violation 
remains sufficiently small in practical calculations.

If we eliminate the matrix elements \eqref{eq:fod}, \eqref{eq:Gammaod} as $s\to\infty$, the final IM-SRG(2) 
Hamiltonian $\HO(\infty)$ has the shape shown in the right panel of Fig.~\ref{fig:schematic}: the 
one-dimensional $0$p$0$h space spanned by the reference state is completely decoupled from other 
states, and therefore an eigenspace of $\HO(\infty)$, with the eigenvalue given by the corresponding 
matrix element. In essence, this means that the IM-SRG provides a mapping between the reference 
state $\ket{\Phi}$ and an exact eigenstate $\ket{\Psi}$ of the nucleus. 

At this point, a few remarks are in order. In a finite system, i.e., in the absence of phase transitions, 
it is always possible to obtain a mapping between the reference state $\ket{\Phi}$ and an exact bound 
eigenstate $\ket{\Psi}$ of $\HO$ by performing a diagonalization, provided there are no symmetry or other 
restrictions on the $A$p$A$h basis built from $\ket{\Phi}$. Thus, the IM-SRG is guaranteed to yield an 
exact energy and wave function for the $A$-body system if the IM-SRG flow equations are not truncated. 
Induced couplings between the 0p0h reference state and excited 3p3h,\ldots,$A$p$A$h states 
(Eq.~\eqref{eq:induced_od}) are included in $\HO^{od}(s)$, and consequently suppressed for $s\to\infty$. 
By truncating the flow equations, we only obtain an approximation to an exact eigenvalue and eigenstate, 
of course. Similarly, we note that eigenvalue spectrum of $\HO(s)$ is variational only if the IM-SRG equations are solved without truncation, since the unitary equivalence with $\HO$ is otherwise only approximate. 

In general, we cannot guarantee that the IM-SRG will necessarily extract the lowest eigenstate 
of $\HO$. As in other methods which make use of a reference state, we expect to be able to reach 
the lowest eigenvalue if the reference state itself is a fair approximation to the ground state, e.g., 
a Hartree-Fock Slater determinant. We will revisit this issue in Sec.~\ref{sec:refstate}.

As indicated in Fig.~\ref{fig:schematic}, we have also eliminated the outermost side diagonals of 
$\HO(\infty)$ by suppressing the coupling between the reference state and $2$p$2$h excitations. This 
implies that any subsequent calculation which uses $\HO(\infty)$ as input, e.g., a diagonalization to 
obtain the full excitation spectrum, can be expected to converge much more rapidly (see 
Sec.~\ref{sec:numerics_decoupling}). 

Finally, we want to address why we have decided to target only a single eigenvalue and eigenstate
of the Hamiltonian by means of IM-SRG decoupling. We might wish to extend the definition 
of $H^{od}(s)$ to encompass matrix elements that mix 1p1h excitations or couple the 1p1h and 2p2h 
blocks of the Hamiltonian, in order to diagonalize $H(s)$ completely in the 0p0h and 1p1h blocks 
(cf.~Fig.~\ref{fig:decoupling}). In practice, this results in uncontrolled behavior of the generator 
and (approximate) ground and excited-state energies during the flow, because we have truncated strong induced 
three- and higher-body operators that feed back into the flow of the zero-, one-, and two-body operators that
we do track in the IM-SRG(2) scheme. To avoid these complications in our ground-state calculations, we 
restrict ourselves to decoupling the 0p0h block, similar to the CC method (see, e.g., 
\cite{Shavitt:2009}). We refer to this as the \emph{minimal decoupling} scheme.

\subsection{\label{sec:white}White Generators}
Starting from the off-diagonal Hamiltonian, Eq.~\eqref{eq:def_Hod}, we can define several classes of generators which will suppress the matrix elements of $\HO^{od}$ as we integrate the IM-SRG flow equations for $s\to\infty$. Our standard choice is motivated by the work of White on canonical transformation theory in quantum chemistry \cite{White:2002fk,Tsukiyama:2011uq}:
\begin{align}
  \etaO^\text{IA/B}(s)
  &\equiv\sum_{ph}\frac{f_{ph}(s)}{\Delta^\text{A/B}_{ph}(s)}\nord{\aaO_{p}\aO_{h}\!}
  +\sum_{pp'hh'}\frac{\Gamma_{pp'hh'}(s)}{\Delta^\text{A/B}_{pp'hh'}(s)}\nord{\aaO_{p}\aaO_{p'}\aO_{h'}\aO_{h}}-\;\text{H.c.}\,.\label{eq:eta_white}
\end{align}
The generalization to three-body or higher rank is obvious. Note that the energy denominators must cause a sign change under transposition in Eq.~\eqref{eq:eta_white} to ensure the anti-Hermiticity of $\etaO(s)$, because $f$ and $\Gamma$ are Hermitian.

The superscripts in \eqref{eq:eta_white} distinguish two different choices for the energy denominators which we will be studying in the following. They correspond to the Epstein-Nesbet and M{\o}ller-Plesset partitionings used in MBPT (see, e.g., \cite{Shavitt:2009}). A straightforward application of White's construction described in Ref.~\cite{White:2002fk} leads to the Epstein-Nesbet case, with energy denominators which are defined in terms of the diagonal matrix elements of the Hamiltonian in our chosen $n$p$n$h-representation (Fig.~\ref{fig:schematic}):
\begin{align}
  \Delta^\text{A}_{ph} &\equiv \matrixe{ph}{\HO}{ph} - \matrixe{\Phi}{\HO}{\Phi}
                  = f_p - f_h + \Gamma_{phph} = - \Delta^\text{A}_{hp} \,,\label{eq:eden1b_en}\\  
  \Delta^\text{A}_{pp'hh'} &\equiv \matrixe{pp'hh'}{\HO}{pp'hh'} - \matrixe{\Phi}{\HO}{\Phi}
                  = f_p + f_{p'} - f_h - f_{h'} - A_{pp'hh'} \equiv - \Delta^\text{A}_{hh'pp'} \,,\label{eq:eden2b_en}
\end{align}
where $f_{p}=f_{pp}, f_{h}=f_{hh}$, and
\begin{align}
  A_{pp'hh'}&\equiv\Gamma_{pp'pp'}+\Gamma_{hh'hh'}-\Gamma_{phph}
       -\Gamma_{p'h'p'h'}-\Gamma_{ph'ph'}-\Gamma_{p'hp'h}\,.\label{eq:def_A}
\end{align}
In the M{\o}ller-Plesset case, on the other hand, 
\begin{align}
  \Delta^\text{B}_{ph}     &\equiv f_p - f_h \equiv - \Delta^\text{B}_{hp} \,, \label{eq:eden1b_mp}\\  
  \Delta^\text{B}_{pp'hh'} &\equiv f_p + f_{p'} - f_h - f_{h'} \equiv - \Delta^\text{B}_{hh'pp'}\,.\label{eq:eden2b_mp}
\end{align}
By expanding the Epstein-Nesbet denominators in Eq.~\eqref{eq:eta_white} in terms of geometric series, we see that $\eta^\text{A}$ corresponds to a generator with M{\o}ller-Plesset denominators where certain MBPT diagrams have been summed to all orders.

If we want to work with the $J$-scheme flow equations \eqref{eq:imsrg2_j0b}--\eqref{eq:imsrg2_j2b}, it is not unambiguously clear how to treat the two-body matrix elements in the Epstein-Nesbet denominators \eqref{eq:eden1b_en}, \eqref{eq:eden2b_en} in the angular momentum coupling process. As a pragmatic solution to this issue, we use the monopole matrix elements
\begin{equation}
  \Gamma^{(0)}_{abcd}\equiv\frac{\sum_J(2J+1)\Gamma^J_{abcd}}{\sum_{J}(2J+1)}
\end{equation}
in Eqs.~\eqref{eq:eden1b_en}--\eqref{eq:def_A}. 

The big advantage of White-type generators in practical calculations lies in the fact that they suppress \emph{all} off-diagonal matrix elements with a decay scale identical (or close to) 1 (see Sec.~\ref{sec:scales}). Thus, the flow is \emph{not} a proper RG flow that suppresses matrix elements between states with large energy differences first. This is inconsequential if we are only interested in $\HO(\infty)$, because all unitary transformations which suppress $\HO^{od}$ must must be equivalent up to differences caused by truncating the IM-SRG flow equations.

The use of $\etaO^\text{A/B}$ has additional benefits in numerical applications: First, the effort for their construction scales as $\OC(N_h^2N_p^2)$, where $N_h\ll N$ and $N_p\sim N$ are the number of hole and particle states in a single-particle basis of dimension $N$. Second, because the generator's matrix elements are given by ratios of energies, $f$ and $\Gamma$ only contribute linearly to the magnitude of the right-hand side of the IM-SRG flow equations \eqref{eq:imsrg2_m0b}--\eqref{eq:imsrg2_m2b}. The flow equations are significantly less stiff than those for the canonical Wegner generator (Sec.~\ref{sec:generators_Wegner}), where third powers of $f$ and $\Gamma$ appear (see below). This greatly reduces the number of integration steps which are required to solve the IM-SRG flow equations. However, the White generators also have an obvious drawback: It is clear from Eq.~\eqref{eq:eta_white} that we might encounter problems with small energy denominators \cite{Tsukiyama:2012fk,Hergert:2013ij}. Fortunately, this is not the case if we deal with closed-shell nuclei, as in the present work. 

We conclude our discussion by remarking that the White-type generators $\etaO^\text{A/B}$ provide a manifest link between the IM-SRG and MBPT, which will be explored further in Section \ref{sec:mbpt}. 

\subsection{\label{sec:generators_ImTime}Imaginary-Time Generators}
A second class of generators which are used in our calculations are inspired by imaginary-time evolution techniques that are frequently used in Quantum Monte Carlo methods, for instance (see, e.g., \cite{Carlson:2015lq} and references therein). Using the off-diagonal Hamiltonian, Eq.~\eqref{eq:def_Hod}, we define
\begin{align}
  \etaO^\text{IIA/B}(s)
  &\equiv\sum_{ph} \sgn\!\left(\Delta^\text{A/B}_{ph}(s)\right)f_{ph}(s)\nord{\aaO_{p}\aO_{h}\!}\notag\\
  &\hphantom{=}
   +\sum_{pp'hh'}\sgn\!\left(\Delta^\text{A/B}_{pp'hh'}(s)\right)\Gamma_{pp'hh'}(s)\nord{\aaO_{p}\aaO_{p'}\aO_{h'}\aO_{h}}-\text{H.c.}\,,\label{eq:eta_imtime}
\end{align}
where $\Delta^\text{A/B}$ are the Epstein-Nesbet and M{\o}ller-Plesset energy denominators defined in Eqs.~\eqref{eq:eden1b_en}--\eqref{eq:eden2b_mp}. As we will see in Sec.~\ref{sec:scales}, the sign functions are necessary to ensure that off-diagonal matrix elements are suppressed instead of enhanced during the flow. The decay scale for these matrix elements is approximately given by the linear energy difference between the reference state and 1p1h or 2p2h excitations, respectively, which can be expressed via $\Delta^\text{A/B}$. Note that this implies that $\etaO^\text{IIA/B}$ generates a proper RG flow.

For the imaginary-time generator, the IM-SRG(2) flow equations are quadratic in the Hamiltonian. While this leads to a mild increase in the stiffness compared to the use of White generators, the flow is not susceptible to small or vanishing energy denominators. Combined with the low $\OC(N_h^2N_p^2)$ effort for its construction, the imaginary-time generator is a robust numerical fallback option in cases where singular White generators stall the IM-SRG flow.

\subsection{\label{sec:generators_Wegner}Wegner Generators}
In his initial work on flow equations and the SRG \cite{Wegner:1994dk}, Wegner proposed the canonical generator 
\begin{equation}\label{eq:eta_wegner}
  \etaO^\text{III}(s) = \comm{\HO^d(s)}{\HO^{od}(s)}\,.
\end{equation}
Using the definition of the off-diagonal Hamiltonian, Eq.~\eqref{eq:def_Hod}, and the commutators from Appendix \ref{app:commutators}, it is straightforward to derive the individual one-body, two-body, etc. matrix elements of $\etaO(s)$. Keeping up to two-body operators, just as in the IM-SRG(2) flow equations, we obtain
\begin{align}
  \eta_{12} &= 
  \sum_{a}(1-P_{12})f^d_{1a}f^{od}_{a2} +\sum_{ab}(n_a-n_b)(f^d_{ab}\Gamma^{od}_{b1a2}-f^{od}_{ab}\Gamma^d_{b1a2}) \notag\\ 
  &\quad +\frac{1}{2} \sum_{abc}(n_an_b\bar{n}_c+\bar{n}_a\bar{n}_bn_c) (1-P_{12})\Gamma^d_{c1ab}\Gamma^{od}_{abc2}
  \label{eq:eta_wegner_m1b}\,,\\[5pt]
  \eta_{1234}&= 
  \sum_{a}\left\{ 
    (1-P_{12})(f^d_{1a}\Gamma^{od}_{a234}-f^{od}_{1a}\Gamma^{d}_{a234} )
    -(1-P_{34})(f^{d}_{a3}\Gamma^{od}_{12a4}-f^{od}_{a3}\Gamma^d_{12a4} )
    \right\}\notag \\
  &\quad+ \frac{1}{2}\sum_{ab}(1-n_a-n_b)(\Gamma^d_{12ab}\Gamma^{od}_{ab34}-\Gamma^{od}_{12ab}\Gamma^d_{ab34})\notag\\
  &\quad-\sum_{ab}(n_a-n_b) (1-P_{12})(1-P_{34})\Gamma^d_{b2a4}\Gamma^{od}_{a1b3}
    \label{eq:eta_wegner_m2b}\,.
\end{align}
Structurally, Eqs.~\eqref{eq:eta_wegner_m1b} and \eqref{eq:eta_wegner_m2b} are identical to the flow equations except for signs stemming from the overall anti-Hermiticity of the generator. The $J$-scheme expressions for $\etaO^\text{III}(s)$ are easily obtained from Eqs.~\eqref{eq:imsrg2_j1b} and \eqref{eq:imsrg2_j2b}.

Clearly, a fixed point of the Wegner flow is reached when $\etaO(s)$ vanishes. At finite $s$, this can occur when $\HO^d(s)$ and $\HO^{od}(s)$ commute due to a degeneracy in the spectrum of $\HO(s)$, e.g., because of symmetries. A fixed point at $s\to\infty$ exists if $\HO^{od}(s)$ vanishes as required. When we work with a truncated, finite-dimensional Hilbert space, it is easy to show that \cite{Wegner:1994dk,Kehrein:2006kx}
\begin{equation} \label{eq:steepest_descent}
  \totd{}{s}\tr\left(\HO^{od}(s)\right)^2 = -2 \tr\left(\etaO^\dag(s)\etaO(s)\right)\leq 0
\end{equation}
due to $\etaO^\dag(s)\etaO(s)$ being positive semi-definite, so $\HO^{od}(s)$ is increasingly suppressed and $\HO(s)$ is indeed diagonalized. 

An interesting consequence of Eq.~\eqref{eq:steepest_descent} is that the flow generated by the generator \eqref{eq:eta_wegner} follows a steepest-descent trajectory in the manifold of unitarily transformed Hamiltonians. The flow itself has proper RG character, i.e., it eliminates off-diagonal matrix elements between states with large energy differences first, as we will demonstrate in Sec.~\ref{sec:scales}. While the proper RG behavior, the steepest-descent property Eq.~\eqref{eq:steepest_descent}, and absence of small energy denominators as in the White generators are useful formal features, Wegner generators are much less efficient in numerical applications than our other choices. The cost for constructing $\etaO^\text{III}$ is of the order $\mathcal{O}(N^6)$, with little to be gained by distinguishing particle and hole states, and therefore similar to the cost for evaluating the IM-SRG(2) flow equations themselves. In addition, the flow equations become very stiff because the RHS terms are cubic in the Hamiltonian, and the appropriate stiff ODE solvers have higher storage and computing time demands than those for non-stiff or weakly stiff cases resulting from the use of White or imaginary-time generators.

\subsection{\label{sec:scales}Decay Scales}
Further insight into the IM-SRG flows that result from our different generator choices can
be gained from a schematic analysis of the flow equations, analogous to prior work for
the free-space SRG \cite{Anderson:2008hx,Bogner:2010pq}.
Let us again assume that the Hamiltonian is split into a diagonal and an off-diagonal part,
\begin{equation}
  \HO(s)=\HO^d(s) + \HO^{od}(s)\,,
\end{equation}
where $\HO^{od}(s)$ is supposed to be suppressed as $s\to\infty$. Further, we work in the eigenbasis of $\HO^d(0)$, which is assumed to be invariant under $s$, so that at each step of the flow
\begin{equation}
  \HO^d(s)\ket{n} = E_n(s)\ket{n}\,.
\end{equation}
In this basis representation, Eq.~\eqref{eq:opflow} becomes
\begin{align}
  \totd{}{s}\matrixe{i}{\HO}{j}
  &=\sum_k\left(\matrixe{i}{\etaO}{k}\matrixe{k}{\HO}{j}-\matrixe{i}{\HO}{k}\matrixe{k}{\etaO}{j}\right)\notag\\
  &=-\left(E_i-E_j\right)\matrixe{i}{\etaO}{j}
    +\sum_k\left(\matrixe{i}{\etaO}{k}\matrixe{k}{\HO^{od}}{j}-\matrixe{i}{\HO^{od}}{k}\matrixe{k}{\etaO}{j}\right),
\end{align}
and $\matrixe{i}{\HO^{od}}{i}=0$.

Consider now a generator of White type, which can be written as
\begin{equation}
  \matrixe{i}{\etaO^I}{j}=\frac{\matrixe{i}{\HO^{od}}{j}}{E_i-E_j}\,,
\end{equation}
so that the flow equation reads
\begin{align}
  \totd{}{s}\matrixe{i}{\HO}{j}
  =-\matrixe{i}{\HO^{od}}{j}
  +\sum_k\frac{E_i+E_j-2E_k}{(E_i-E_k)(E_j-E_k)}\matrixe{i}{\HO^{od}}{k}\matrixe{k}{\HO^{od}}{j}\,.\label{eq:schema_white}
\end{align}
Let us assume that the transformation generated by $\etaO$ truly suppresses $\HO^{od}$, and consider the asymptotic behavior for large flow parameters $s>s_0\gg0$. If $||H^{od}(s_0)||\ll 1$ in some suitable norm, the second term in the flow equation can be neglected compared to the first one. In this case, Eq.~\eqref{eq:schema_white} implies
\begin{align}
  \left.\totd{E_i}{s}\right|_{s=s_0}&=2\sum_k\frac{\matrixe{i}{\HO^{od}}{k}\matrixe{k}{\HO^{od}}{i}}{E_i-E_k}
  \approx0\,,
\end{align}
and the energies stay (approximately) constant:
\begin{equation}
  E_i(s) \approx E_i(s_0)\,,\quad s>s_0\,.
\end{equation}
Consequently, Eq.~\eqref{eq:schema_white} can be integrated, and we have
\begin{equation}
  \matrixe{i}{\HO^{od}(s)}{j} \approx \matrixe{i}{\HO^{od}(s_0)}{j}\,e^{-(s-s_0)}\,,\quad s>s_0\,,
\end{equation}
as already mentioned in Sec.~\ref{sec:white}. If the initial $\HO_{od}(0)$ is perturbative, we will observe this behavior from the very onset of the flow (see the discussion in Sec.~\ref{sec:mbpt}).

For the imaginary-time generator, we have
\begin{equation}
  \matrixe{i}{\etaO^\text{II}}{j}=\sgn\left(E_i-E_j\right)\matrixe{i}{\HO^{od}}{j}\,,
\end{equation}
and the flow equation
\begin{align}
  \totd{}{s}\matrixe{i}{\HO}{j}
  &=-\left|E_i-E_j\right|\matrixe{i}{\HO^{od}}{j}\notag\\
  &\quad
    +\sum_k\left(\sgn\!\!\left(E_i-E_k\right)+\sgn\!\!\left(E_j-E_k\right)\right)\matrixe{i}{\HO^{od}}{k}\matrixe{k}{\HO^{od}}{j}\,.
    \label{eq:schema_imtime}
\end{align}
Note that the sign function in the definition of $\etaO^\text{II}$ ensures that only the absolute value of the energy difference between the states $\ket{i}$ and $\ket{k}$ appears in the first term. Integration of Eq.~\eqref{eq:schema_imtime} yields
\begin{equation}
  \matrixe{i}{\HO^{od}(s)}{j} \approx \matrixe{i}{\HO^{od}(s_0)}{j}\,e^{-|E_i-E_j|(s-s_0)}\,, \quad s>s_0\,,
\end{equation}
and off-diagonal matrix elements are suppressed, with a decay scale set by $|E_i-E_j|$.

Finally, we perform the same kind of analysis for the Wegner generator
\begin{equation}
  \matrixe{i}{\etaO^\text{III}}{j}=\matrixe{i}{\comm{\HO^d}{\HO^{od}}}{j}=(E_i-E_j)\matrixe{i}{\HO^{od}}{j}\,.
\end{equation}
The flow equation reads
\begin{align}
  \totd{}{s}\matrixe{i}{\HO}{j}
  =-\left(E_i-E_j\right)^2\matrixe{i}{\HO^{od}}{j}
  +\sum_k\left(E_i+E_j-2E_k\right)\matrixe{i}{\HO^{od}}{k}\matrixe{k}{\HO^{od}}{j}\,,\label{eq:schema_wegner}
\end{align}
and we obtain
\begin{equation}
  \matrixe{i}{\HO^{od}(s)}{j} \approx \matrixe{i}{\HO^{od}(s_0)}{j}\,e^{-(E_i-E_j)^2 (s-s_0)}\,, \quad s>s_0\,.
\end{equation}

Thus, we see that the imaginary-time and Wegner generators yield proper RG transformations, in the sense that matrix elements between states with large energy differences $\Delta E_{ij} = |E_i-E_j|$ decay at smaller flow parameters $s$ than states with small $\Delta E_{ij}$. The White generator, on the other hand, acts on all matrix elements simultaneously. If we are only interested in the limit $s\to\infty$, this should not make a difference. In Sec.\ref{sec:numerics}, we will explore to which extent this is indeed the case, and quantify differences which are caused by the terms we have omitted in our schematic analysis, and the necessary truncations in the IM-SRG flow equations.

\section{\label{sec:numerics}Numerical Explorations}
In this section, we illustrate the general properties of the IM-SRG flow equations in numerical applications, with special emphasis on a comparison of the different generators that were introduced in the previous sections. To simplify matters, we only use a two-body interaction throughout this section (see \ref{sec:numerics_implementation} for details). IM-SRG applications with three-body interactions are discussed in Sec.~\ref{sec:sum+dev} and Refs.~\cite{Hergert:2013ij,Hergert:2013mi,Hergert:2014vn,Bogner:2014tg,Caceres:2015fk,Stroberg:2015qr,Hergert:2015qd}.

\begin{figure}[p]
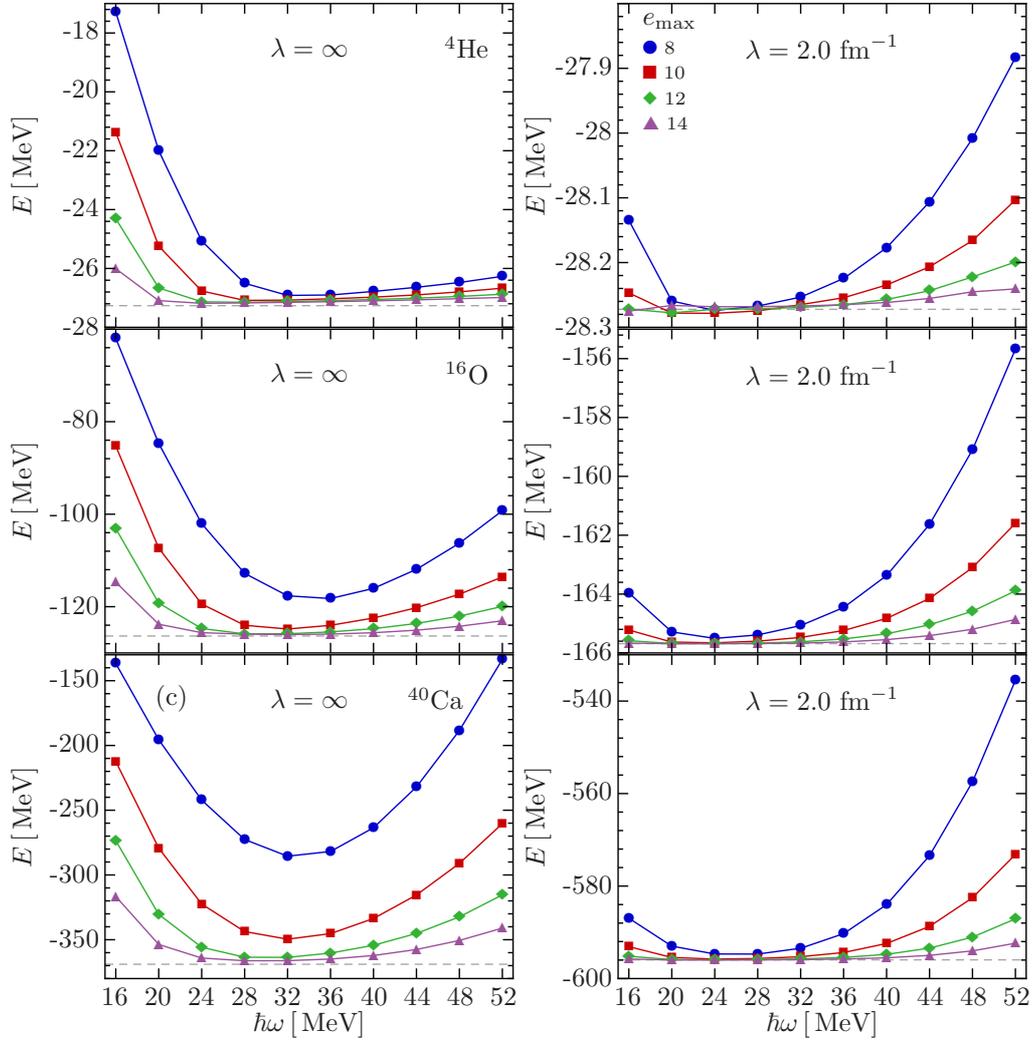

  \setlength{\unitlength}{0.500\textwidth}
  \begin{picture}(2.0000,2.0000)
    \put(0.0000,1.2800){\input{fig/chi2b_srg0000_He4_convergence.tex}}
    \put(1.0000,1.2800){\input{fig/chi2b_srg0625_He4_convergence.tex}}
    \put(0.0000,0.6400){\input{fig/chi2b_srg0000_O16_convergence.tex}}
    \put(1.0000,0.64000){\input{fig/chi2b_srg0625_O16_convergence.tex}}
    \put(0.0000,0.0000){\input{fig/chi2b_srg0000_Ca40_convergence.tex}}
    \put(1.0000,0.0000){\input{fig/chi2b_srg0625_Ca40_convergence.tex}}
  \end{picture}
  \\[-10pt]
  \caption{\label{fig:chi2b_srgXXXX_conv}
    Convergence of $\nuc{He}{4}$, $\nuc{O}{16}$, and $\nuc{Ca}{40}$ IM-SRG(2) ground-state energies w.r.t.~single-particle basis size $\eMax$, for a chiral N${}^3$LO NN interaction with $\lambda=\infty$ (left panels) and $\lambda=2.0\,\fm^{-1}$ (right panels). Notice the significant differences in the energy scales between the left and right panels. Gray dashed lines indicate energies from extrapolation the $\eMax\geq10$ data sets to infinite basis size (see text and Refs.~\cite{Furnstahl:2012ys,More:2013bh}).
  }
\end{figure}

\subsection{\label{sec:numerics_implementation}Implementation}
Like the majority of current \emph{ab initio} many-body methods, the IM-SRG is implemented in harmonic oscillator (HO) configuration spaces. The principal advantage of the HO basis is the capability to exactly separate center-of-mass and relative degrees of freedom in the evaluation of matrix elements (see, e.g., for \cite{Kamuntavicius:2001kr}). A suitable choice of model space truncation makes it possible to propagate this property to the many-body many-body wave functions, as in the NCSM for instance \cite{Barrett:2013oq}. For methods which use single-particle basis truncations, like IM-SRG, CC, SCGF, this property is lost, although an approximate factorization of center-of-mass and intrinsic wave functions is still observed empirically (see Sec.~\ref{sec:com} for a discussion).

In order to cover large mass ranges, we enforce spherical symmetry in HO bases of up to 15 major shells, and work with the $J$-scheme IM-SRG flow equations presented in Sec.~\ref{sec:flow_jscheme}. These basis sizes are sufficient to obtain satisfactory convergence even with ``bare'' interactions from chiral EFT like the \NNNLO interaction by Entem and Machleidt, with an initial cutoff $\Lambda=500\,\MeV$ \cite{Entem:2003th,Machleidt:2011bh}. For most of the following sections, we use this interaction, both at its original resolution scale, indicated by $\lambda=\infty$ in the following, and at a lower resolution scale $\lambda=2.0\,\fm^{-1}$, which is generated by a free-space SRG evolution \cite{Bogner:2007od,Bogner:2010pq}. In the latter case, we do not retain induced 3N,$\ldots$ interactions for the time being (see Sec.~\ref{sec:sum+dev} and Refs.~\cite{Hergert:2013ij,Hergert:2013mi,Hergert:2014vn,Bogner:2014tg,Caceres:2015fk,Stroberg:2015qr,Hergert:2015qd} for a more complete treatment).

To obtain reference states for the IM-SRG calculation, we solve the Hartree-Fock equations for the intrinsic Hamiltonian \eqref{eq:def_Hint}. We then transform the intrinsic Hamiltonian to the natural orbital basis defined by Eq.~\eqref{eq:def_natorb}, and carry out the normal ordering. Starting from the zero-, one-, and two-body matrix elements of the truncated normal-ordered Hamiltonian as initial values, we integrate the system of $J$-scheme flow equations \eqref{eq:imsrg2_j0b}--\eqref{eq:imsrg2_j2b} using the CVODE solver from the SUNDIALS package \cite{Hindmarsh:2005kl}. For White and imaginary-time generators, we choose the recommended Adams-Bashforth-Moulton predictor-corrector method for non-stiff systems, while the fifth-order backward-differentiation method is used for the stiff flow equations in the Wegner case. The use of simple pre-conditioners brought no appreciable speed-ups in the latter case, and more involved approximations for the Jacobian were not explored because they would spoil the $\OC(N^6)$ scaling of the IM-SRG.

To test for convergence, we calculate the energy correction from second-order MBPT for the flowing Hamiltonian $\HO(s)$, which is entirely due to the off-diagonal part of the Hamiltonian as defined in Eqs.~\eqref{eq:def_Hod}, and directly proportional to the flow of the ground-state energy, Eq.~\eqref{eq:imsrg2_j0b} (also see Sec.~\ref{sec:mbpt}). We assume that sufficient decoupling is achieved once the absolute size of the second-order correction falls below $10^{-6}\,\MeV$.

\subsection{\label{sec:numerics_convergence}Convergence}
In Fig.~\ref{fig:chi2b_srgXXXX_conv}, we show the convergence of the IM-SRG(2) ground-state energies of the closed-shell nuclei $\nuc{He}{4},\nuc{O}{16},$ and $\nuc{Ca}{40}$ with respect to the single-particle basis size $\eMax$ (see Appendix \ref{sec:numerics_implementation}). The White-Epstein-Nesbet generator, Eq.~\eqref{eq:eta_white}, was used in these calculations, and serves as our default choice in the following. For the unevolved N${}^3$LO interaction, the Hartree-Fock solutions for all three nuclei have positive energy, but the single-particle wave functions remain localized as a consequence of the expansion in a finite number of HO states. Thus, the HF states are suitable reference states, and it is evident from Fig.~\ref{fig:chi2b_srgXXXX_conv} that the IM-SRG(2) energies are converging. 

\begin{table}[t]
  \begin{tabular*}{\columnwidth}{l@{\extracolsep\fill}crr}
  \hline\hline
  Nucleus & 
  \multicolumn{1}{c}{$\lambda\,[\fm^{-1}]$} & 
  \multicolumn{1}{c}{$E_{14}\,[\MeV]$} &
  \multicolumn{1}{c}{$E_\infty\,[\MeV]$} \\
  \hline
  $\nuc{He}{4}$ & $\infty$  &  $-$27.18 & $-$27.26(3) \\
  $\nuc{O}{16}$ & $\infty$  & $-$126.01 & $-$126.3(1) \\
  $\nuc{Ca}{40}$ & $\infty$ & $-$366.23 & $-$369(1) \\
  \hline
  $\nuc{He}{4}$   & 2.0 &    $-$28.27 &  $-$28.27  \\
  $\nuc{O}{16}$   & 2.0 &   $-$165.68 & $-$165.68 \\
  $\nuc{Ca}{40}$  & 2.0 &   $-$595.98 & $-$595.95(2) \\
  $\nuc{Ni}{78}$  & 2.0 &  $-$1319.41 & $-$1319.4(1) \\
  $\nuc{Sn}{100}$ & 2.0 &  $-$1953.96 & $-$1954.3(3) \\
  $\nuc{Sn}{132}$ & 2.0 &  $-$2752.03 & $-$2753(2) \\
  \hline\hline
  \end{tabular*}
  \caption{\label{tab:chi2b_srgXXXX} IM-SRG(2) ground-state energies of selected closed-shell nuclei for the the chiral N${}^3$LO interaction by Entem and Machleidt \cite{Entem:2003th,Machleidt:2011bh}, with $\lambda=\infty$  and $\lambda=2.0\,\fm^{-1}$ (cf.~Fig.~\ref{fig:chi2b_srgXXXX_conv}). $E_{14}$ are the energies obtained for $\eMax=14$ at optimal $\hbar\omega$, and $E_\infty$ are extrapolated to infinite basis size (see text), with extrapolation uncertainties indicated in parentheses.
  }
\end{table}

We can correct for the effects of using a finite HO basis by using the methods described in Refs.~\cite{Furnstahl:2012ys,More:2013bh}. A HO basis with fixed $\eMax$ has ultraviolet and infrared cutoffs which are given by
\begin{align}
  \Lambda_\text{UV} &\equiv \sqrt{2\eMax+7}\,\hbar/\aHO\,,\\
  L_\text{IR} &\equiv \sqrt{2\eMax+7}\,\aHO\,,
\end{align}
where $\aHO=\sqrt{\hbar/m\omega}$ is the usual oscillator length, and $m$ the nucleon mass. With these definitions, we can perform a simultaneous fit of the data for (almost) all pairs $(\eMax,\hbar\omega)$ to the expression
\begin{equation}\label{eq:def_Eex}
  E(\eMax,\hbar\omega)=E_\infty + A_0e^{-\LambdaUV^2/A_1^2} + A_2 e^{-2k_\infty \LambdaIR}\,,
\end{equation}
where the energy for infinite basis size $E_\infty$, the binding momentum $k_\infty$, and the $A_i$ are treated as parameters. For the unevolved N${}^3$LO interaction, we found it necessary to exclude the $\eMax=8$ data set to obtain stable fits for $\nuc{O}{16}$ and $\nuc{Ca}{40}$, most likely because $\Lambda_\text{UV}$ is close to the cutoff of the initial interaction for $\eMax=8$ and the lower values of $\hbar\omega$ we are considering. The resulting extrapolated energies are indicated by gray dashed lines in Fig.~\ref{fig:chi2b_srgXXXX_conv}, and they fall within 1\% or less of the energies for $\eMax=14$, the largest basis size which was used in actual calculations. Both energies are reported for each nucleus in Table \ref{tab:chi2b_srgXXXX}. 

\begin{figure*}[t]
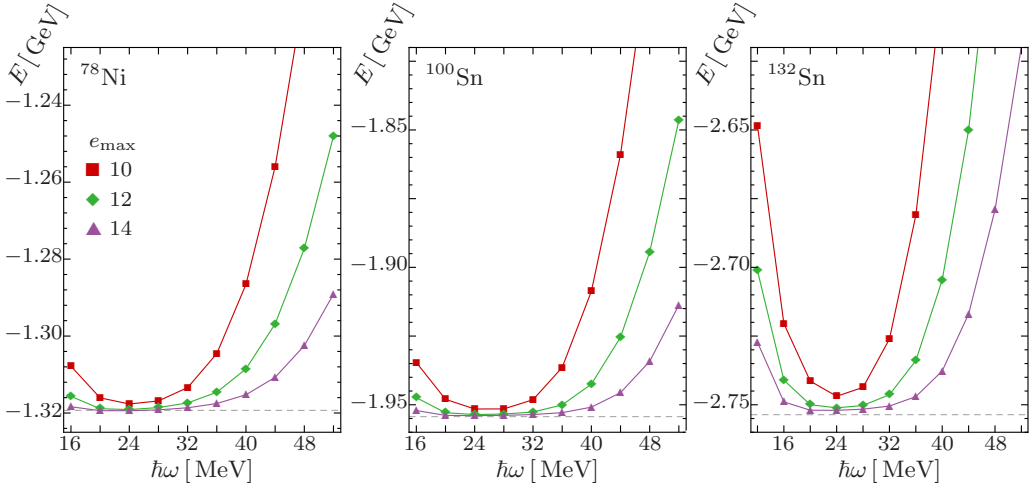

  \setlength{\unitlength}{0.3400\textwidth}
  \begin{picture}(2.9400,1.5000)
    \put(-0.0300,0.0000){\small\input{fig/chi2b_srg0625_Ni78_convergence.tex}}
    \put(0.9600,0.0000){\small\input{fig/chi2b_srg0625_Sn100_convergence.tex}}
    \put(1.9400,0.0000){\small\input{fig/chi2b_srg0625_Sn132_convergence.tex}}
  \end{picture}
  \\[-15pt]
  \caption{\label{fig:chi2b_srg0625_conv_exotics}
    Convergence of $\nuc{Ni}{78}$, $\nuc{Sn}{100}$, and $\nuc{Sn}{132}$ IM-SRG(2) ground-state energies w.r.t.~single-particle basis size $\eMax$, for the chiral \NNNLO NN interaction with $\lambda=2.0\,\fm^{-1}$. Gray dashed lines indicate energies from extrapolation the $\eMax\geq10$ data sets to infinite basis size (see text and Refs.~\cite{Furnstahl:2012ys,More:2013bh}).
  }
\end{figure*}

For $\nuc{He}{4}$, the IM-SRG(2) ground-state energy is almost $2\,\MeV$ below the exact result from a No-Core Shell Model (NCSM) calculation, which yields $-25.39\,\MeV$ for the chiral N${}^3$LO interaction we use here (see, e.g., Ref.~\cite{Jurgenson:2009bs}). We can also compare our IM-SRG(2) results with those from large-scale Coupled Cluster calculations with the same interaction \cite{Hagen:2010uq,Hagen:2014ve,Hergert:2013mi}. Superficially, IM-SRG(2) closely resembles a unitary variant of the CCSD approach (CC with singles and doubles). However, the IM-SRG(2) energies are significantly lower than the CCSD energies, which are $-23.97$, $-107.52$, and $-317.3\,\MeV$, for $\nuc{He}{4}$, $\nuc{O}{16}$, and $\nuc{Ca}{40}$, respectively. In fact, the IM-SRG(2) also provides lower ground-state energies than $\Lambda$-CCSD(T), a CC method which takes perturbative triples corrections into account, and yields $-25.51$, $-121.6$, and $-363.3\,\MeV$ for the studied nuclei. The perturbative analysis of the IM-SRG in Sec.~\ref{sec:mbpt} reveals that the bulk of the difference between IM-SRG(2) and CCSD can be explained by a systematic undercounting of certain repulsive fourth-order terms in the IM-SRG(2) truncation, which simulates the additional attraction that is otherwise gained from including triples correction. For the (comparably) hard initial interaction, the IM-SRG(2) overshoots the $\Lambda$-CCSD(T) results, while the reduced importance of higher-order MBPT corrections for soft interactions causes the IM-SRG(2) results to fall in between the CCSD and $\Lambda$-CCSD(T) results (see Secs.~\ref{sec:mbpt}, \ref{sec:sum+dev} and Refs.~\cite{Tsukiyama:2011uq,Hergert:2013mi}).

In the lower panels of Fig.~\ref{fig:chi2b_srgXXXX_conv}, we show the same kind of convergence plots for the chiral \NNNLO interaction at the reduced resolution scale $\lambda=2.0\,\fm^{-1}$. As expected, the speed of the convergence is greatly enhanced by using a softer interaction \cite{Bogner:2010pq}, which is evident from the significantly smaller energy scales in the lower panels. In Table \ref{tab:chi2b_srgXXXX}, we can see that the extrapolated energies agree with the $\eMax=14$ results within 0.01-0.1\%. For $\nuc{He}{4}$, there appear to be some deviations from the otherwise variational convergence pattern in the other cases. Of course, the IM-SRG is not strictly variational because of the truncations in the flow equations \eqref{eq:imsrg2_m0b}--\eqref{eq:imsrg2_m2b}. In the present case, however, these deviations are on the order of a 10\,$\keV$ or less, and are most likely dominated by numerical artifacts from integrating the flow equations.

For a soft interaction, the large single-particle basis sizes we have used here are sufficient to converge nuclei which are much heavier than $\nuc{Ca}{40}$. This is demonstrated in Fig.~\ref{fig:chi2b_srg0625_conv_exotics}, where we show the convergence of the IM-SRG(2) ground-state energies of the proton- or neutron-rich exotic nuclei $\nuc{Ni}{78}$, $\nuc{Sn}{100}$, and $\nuc{Sn}{132}$. The corresponding energies are included in Table \ref{tab:chi2b_srgXXXX}. Using only a softened chiral \NNNLO interaction, the binding energy of these nuclei is overestimated significantly, continuing a trend which was already noticeable for $\nuc{O}{16}$ in Fig.~\ref{fig:chi2b_srgXXXX_conv}. This is caused by the shift of repulsive strength from the off-shell two-body interaction to induced three- and higher many-body forces (which are neglected) as the resolution scale is lowered.  The severe overbinding is fixed by including the induced three-nucleon forces (see Sec.~\ref{sec:sum+dev} and \cite{Hergert:2013mi,Hergert:2013ij,Hergert:2014vn,Hergert:2015qd}). We stress that the induced 3N terms have low resolution scales as well, and do not affect the rate of convergence of the IM-SRG ground-state energies adversely. While computational issues pertaining to the storage of 3N matrix elements present a challenge, \emph{ab initio} calculations with NN+3N interactions for the $A\sim 100$ region have now become possible \cite{Binder:2013zr,Hergert:2013mi,Binder:2014fk,Hergert:2015qd}.

\subsection{\label{sec:numerics_generator}Choice of Generator}

Let us now study the effect of our choice of generator on the IM-SRG(2) ground-state energies. In Fig.~\ref{fig:chi2b_srgXXXX_generators}, we show the IM-SRG(2) ground-state energies for the five different generators discussed in Sec.~\ref{sec:generators}. Note that the panels for the White and imaginary-time generators show curves for \emph{both} the Epstein-Nesbet and M{\o}ller-Plesset choices for the energy denominators and sign functions, respectively. The resulting ground-state energies for $\nuc{Ca}{40}$ agree within 15 $\keV$. Remarkably, this agreement holds for both the softened and bare \NNNLO interactions, and irrespective of the used basis parameters $\eMax$ and $\hw$. The extrapolated energies therefore also only differ by small amounts. 

\begin{figure}[p]
  \setlength{\unitlength}{0.38\textwidth}
  \begin{picture}(2.6310,2.2500)
    \put(0.0000,0.0000){\small\input{fig/chi2b_srg0625_Ca40_genWEN.tex}}
    \put(0.8000,0.0000){\small\input{fig/chi2b_srg0625_Ca40_genIM.tex}}
    \put(1.6000,0.0000){\small\input{fig/chi2b_srg0625_Ca40_genWEG.tex}}
    \put(0.0000,0.9500){\small\input{fig/chi2b_srg0000_Ca40_genWEN.tex}}
    \put(0.8000,0.9500){\small\input{fig/chi2b_srg0000_Ca40_genIM.tex}}
    \put(1.6000,0.9500){\small\input{fig/chi2b_srg0000_Ca40_genWEG.tex}}
    \put(0.2000,2.1500){\parbox{0.8\unitlength}{\centering White -- $\eta^\text{IA/B}$}}
    \put(1.0000,2.1500){\parbox{0.8\unitlength}{\centering Im. Time -- $\eta^\text{IIA/B}$}}
    \put(1.8000,2.1500){\parbox{0.8\unitlength}{\centering Wegner -- $\eta^\text{III}$}}
  \end{picture}
  \caption{\label{fig:chi2b_srgXXXX_generators}
    IM-SRG(2) ground-state energies of $\nuc{Ca}{40}$ 
    obtained with different choices of the generator, as a function of $\hw$ and the single-particle 
    basis size $\eMax$. The interaction is the 
    chiral \NNNLO potential 
    with $\lambda=\infty$ (top panels) and 
    $\lambda=2.0\,\fm^{-1}$ (bottom panels), respectively. The dashed lines indicate extrapolated energies. For the Wegner generator, the shaded area indicates the variation from using different data sets for the extrapolation (see text).
  }
\end{figure}

It is evident from Fig.~\ref{fig:chi2b_srgXXXX_generators} that the White and imaginary-time generators give very similar results. For the bare \NNNLO interaction, the extrapolated $\nuc{Ca}{40}$ ground-state energies are $-368.9\,\MeV$ and $-367.7\,\MeV$, respectively, which is a difference of about 0.3\%. For any $\hw$ in the studied range, the energy differences between the two types of generators drop below 1\% from $\eMax=8$ onward. As expected, the differences become smaller when the resolution scale of the interaction is lowered to $\lambda=2.0\fmi$. The extrapolated energies are $-596.0\,\MeV$ and $-595.6\,\MeV$ for the White and imaginary-time generators, respectively, which amounts to a relative difference of order $10^{-4}$. The extrapolated values are affected by slightly larger differences for small and large $\hw$. Near the energy minima with respect to $\hw$, where the results are better converged, absolute differences are typically below $10\,\keV$. 

For the soft interaction, the results for the Wegner generator agree very well with those for the other generators: The extrapolated $\nuc{Ca}{40}$ ground-state energy is $-595.4\,\MeV$. The situation is quite different for the bare interaction, though. To understand what we see, we first consider the convergence pattern that is predicted for a (quasi)-variational theory by the extrapolation formula \eqref{eq:def_Eex} \cite{Furnstahl:2012ys,More:2013bh}. At fixed $\eMax$, the derivative of Eq.~\eqref{eq:def_Eex} with respect to the oscillator parameter $\hw$ indicates that the ultraviolet (UV) and infrared (IR) correction terms are minimized at large and small $\hw$, respectively. The exponents of the UV and IR terms behave like $\LambdaUV^2\sim \eMax$ and $\LambdaIR\sim ~\sqrt{\eMax}$ as $\eMax$ increases, hence we expect IR corrections to dominate eventually. Consequently, we can infer that the minimum of the energy with respect to the oscillator parameter should move to larger $\hw$ first until UV convergence is achieved, and then to smaller $\hw$ for IR convergence. 

In Fig.~\ref{fig:chi2b_srgXXXX_generators}, we only see the energy minimum move towards IR convergence at small $\hw$, which suggests that the calculation is sufficiently converged in the UV regime already for $\eMax=8$, the smallest basis shown in the figure. For the Wegner generator, the minimum is still moving to larger $\hw$ values, which suggests that UV convergence has not been achieved yet, and a slower convergence with basis size in general. If we use the data for $\eMax=8,10,12$, which behave variationally, an extrapolation to infinite basis using Eq.~\eqref{eq:def_Eex}, yields $-370.7\,\MeV$, which is compatible with the extrapolated results for the White and imaginary-time generators within uncertainties. 

Going to $\eMax=14$, we face a complication: while the energy minimum moves to larger $\hw$, the curve intersects those for smaller $\eMax$. This is not ruled out a priori, because the IM-SRG is a non-variational approach, but makes the assumptions underlying the extrapolation formula \eqref{eq:def_Eex} questionable. Setting aside the fundamental issue of applicability, we have extrapolated the energy using different subsets of our calculated data, and thereby obtain the shaded band in Fig.~\ref{fig:chi2b_srgXXXX_generators}, which represents a 10\% variation of the extrapolated energy.

To better understand the behavior of the IM-SRG flow for the Wegner generator, we have to consider how its structure differs from the other generator choices. The definition of the off-diagonal Hamiltonian $H^{od}(s)$, Eq.~\eqref{eq:def_Hod}, is the same in all cases, so we aim for the same (or at least similar) fixed points of the flow, where $\eta(\infty)=0$. However, we know that the White and imaginary-time generators are directly proportional to $H^{od}$, i.e., the only non-vanishing matrix elements are of the types $\eta_{ph/hp}$ and $\eta_{pp'hh'/hh'pp'}$. The Wegner generator, on the other hand, has many additional non-zero matrix elements coming from the evaluation of the commutator, analogous to the IM-SRG flow equation itself (cf.~Eqs.~\eqref{eq:eta_wegner_m1b}, \eqref{eq:eta_wegner_m2b})). 

\begin{figure}[t]
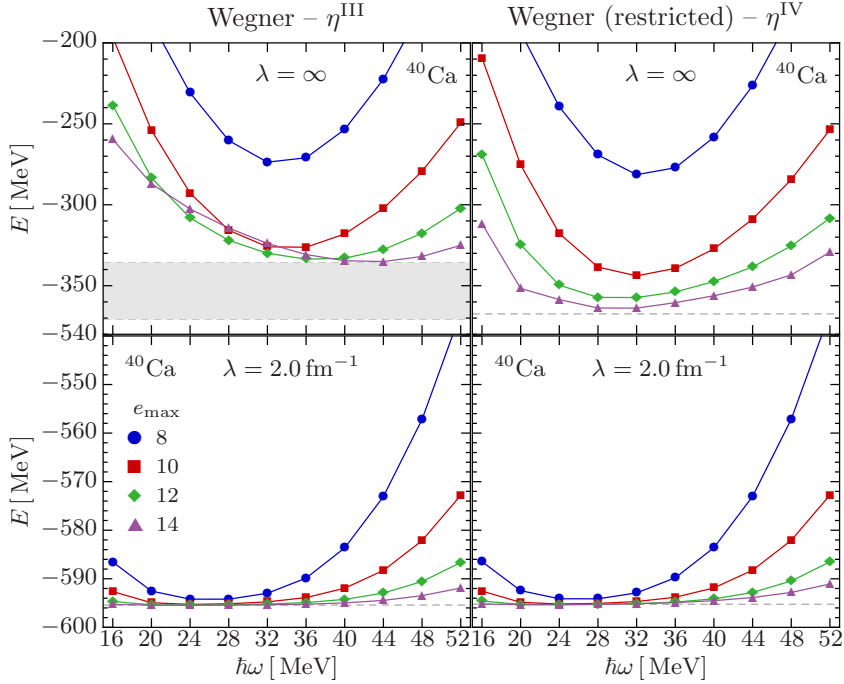

  \setlength{\unitlength}{0.45\textwidth}
  \begin{center}
    \begin{picture}(2.0000,1.6500)
      \put(0.0000,0.0000){\small\input{fig/chi2b_srg0625_Ca40_compWEG.tex}}
      \put(0.8000,0.0000){\small\input{fig/chi2b_srg0625_Ca40_compWEGPH2B.tex}}
      \put(0.0000,0.6400){\small\input{fig/chi2b_srg0000_Ca40_compWEG.tex}}
      \put(0.8000,0.6400){\small\input{fig/chi2b_srg0000_Ca40_compWEGPH2B.tex}}
      \put(0.2000,1.5200){\parbox{0.8\unitlength}{\centering Wegner -- $\eta^\text{III}$}}
      \put(1.0000,1.5200){\parbox{0.8\unitlength}{\centering Wegner (restricted) -- $\eta^\text{IV}$}}
    \end{picture}
  \end{center}
  \vspace{-20pt}
  \caption{\label{fig:chi2b_srgXXXX_wegner}
    IM-SRG(2) ground-state energies of $\nuc{Ca}{40}$ for the regular (left, as in Fig.~\ref{fig:chi2b_srgXXXX_generators}) and restricted Wegner generators (right, see text), as a function of $\hw$ and the single-particle basis size $\eMax$. The interaction is the chiral \NNNLO potential with $\lambda=\infty$ (top panels) and $\lambda=2.0\,\fm^{-1}$ (bottom panels), respectively. The dashed lines indicate extrapolated energies. 
  }
\end{figure}

It is not really a surprise, then, that the generators differ in the way they build correlation effects from the many-body perturbation series into the flowing Hamiltonian --- a difference that will be enhanced for interactions for which order-by-order convergence of that series cannot be guaranteed (cf.~Secs.~\ref{sec:numerics_decoupling} and \ref{sec:mbpt}). For illustration, Fig.~\ref{fig:chi2b_srgXXXX_wegner} compares results for the regular Wegner generator with those for a restricted version defined by
\begin{equation}
  \eta^\text{IV}_{ij} = \eta^\text{III}_{ij}\,,\quad
  \eta^\text{IV}_{ijkl} = \begin{cases} \eta^\text{III}_{ijkl} & \text{for}\quad ijkl=pp'hh',hh'pp'\,,\\ 
                                  0 & \text{else}\,, 
                    \end{cases}
\end{equation}
matching the structure of the White and imaginary-time generators. We have explored restrictions of the one-body part as well, but they cause no notable differences while the impact of the restriction in the two-body part is significant.

The convergence pattern of the restricted $\eta^\text{IV}$ is quasi-variational for both the bare and softened \NNNLO interactions, and has the energy minimum moving towards smaller $\hw$, suggesting that the calculation is converged in the UV regime, and now converging in the IR regime. The extrapolated $\nuc{Ca}{40}$ g.s. energies are $-367.4\,\MeV$ and $-595.3\,\MeV$, respectively, in very good agreement with the White and imaginary-time generators, as well as the unrestricted Wegner generator $\eta^\text{III}$ in the case of the soft interaction (also cf.~Fig.~\ref{fig:chi2b_srgXXXX_wegner}). This strongly suggests that our hypothesis was correct, and it is indeed the additional non-zero matrix elements in $\eta^\text{III}$ which introduce uncontrolled behavior. It remains to be seen whether we can reach a deeper understanding of the underlying mechanism. A likely explanation is that the truncation of the commutator \eqref{eq:eta_wegner} to one- and two-body contributions only (Eqs.~\eqref{eq:eta_wegner_m1b}, \eqref{eq:eta_wegner_m2b}) causes an imbalance in the infinite-order re-summation of the many-body perturbation series. For the time being, we have to advise against the use of the Wegner generator in IM-SRG calculations with (comparably) ``hard'' interactions that exhibit poor order-by-order convergence of the perturbation series.

\subsection{\label{sec:numerics_decoupling}Decoupling}

\begin{figure}[t]
  \setlength{\unitlength}{1.01\textwidth}
  \input{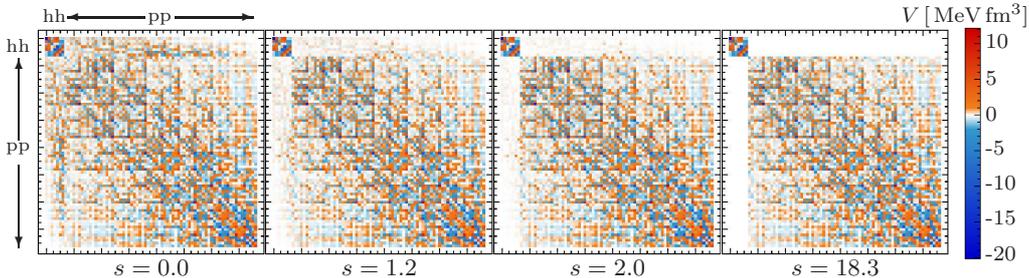}
  \\[-20pt]
  \caption{\label{fig:decoupling}Decoupling for the White generator, Eq.~\eqref{eq:eta_white}, in the $J^\pi=0^+$ neutron-neutron interaction matrix elements of $\nuc{Ca}{40}$ ($\eMax=8, \hbar\omega=20\,\MeV$, Entem-Machleidt N${}^{3}$LO(500) evolved to $\lambda=2.0\,\fm^{-1}$). Only $hhhh, hhpp, pphh,$ and $pppp$ blocks of the matrix are shown.}
\end{figure}

As discussed in Sec.~\ref{sec:decoupling}, the IM-SRG is built around the concept of decoupling the reference state from excitations, and thereby mapping it onto the fully interacting ground state of the many-body system within truncation errors. Let us now demonstrate that the decoupling occurs as intended in a sample calculation for $\nuc{Ca}{40}$ with our standard chiral \NNNLO interaction at $\lambda=2.0\,\fmi$. Figure \ref{fig:decoupling} shows the rapid suppression of the off-diagonal matrix elements in the $J^\pi=0^+$ neutron-neutron matrix elements as we integrate the IM-SRG(2) flow equations. At $s=2.0$, after only 20--30 integration steps with the White generator, the $\Gamma_{pp'hh'}(s)$ have been weakened significantly, and when we reach the stopping criterion for the flow at $s=18.3$, these matrix elements have vanished to the desired accuracy. While the details depend on the specific choice of generator, the decoupling seen in Fig. \ref{fig:decoupling} is representative for other cases.

\begin{figure}[t]
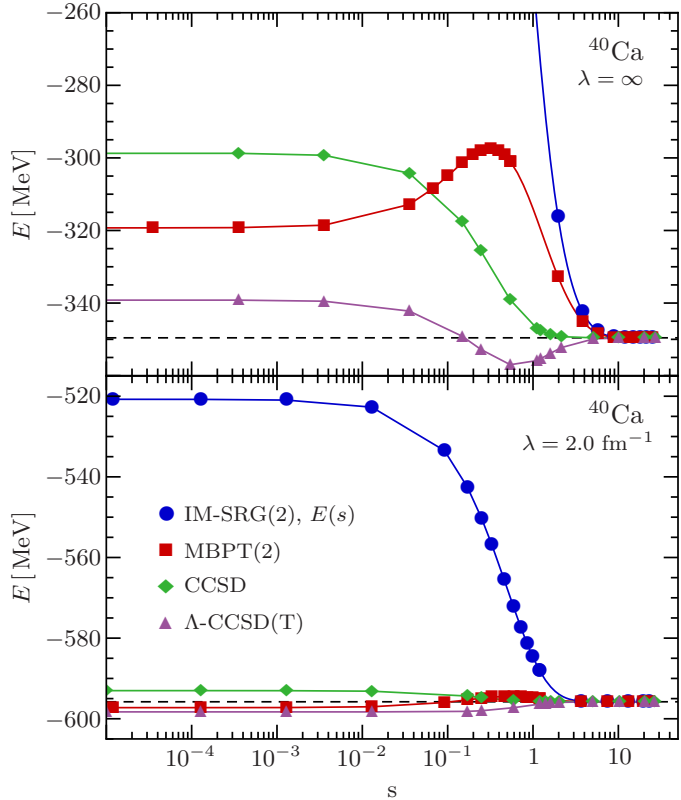

  \setlength{\unitlength}{0.7\textwidth}
  \begin{center}
    \begin{picture}(1.0000,1.2500)
      \put(-0.0500,0.5100){\small\input{fig/chi2b_srg0000_Ca40_flow_vs_s.tex}}
      \put(-0.0500,0.0000){\small\input{fig/chi2b_srg0625_Ca40_flow_vs_s.tex}}
    \end{picture}
  \end{center}
  \vspace{-40pt}
  \caption{\label{fig:flow_methods} IM-SRG(2) ground-state energy of $\nuc{Ca}{40}$ as a function of the flow parameter $s$, compared to MBPT(2), CCSD, and $\Lambda$-CCSD(T) energies with the IM-SRG-evolved Hamiltonian $\HO(s)$. We only show part of the data points to avoid clutter. Calculations were done for $\eMax=10$ and optimal $\hbar\omega=32\,\MeV$ (top) and $\hbar\omega=24\,\MeV$ (bottom), respectively, using the chiral NN interaction at different resolution scales. The dashed lines indicate the final IM-SRG(2) energies.
  }
\end{figure}

With the suppression of the off-diagonal matrix elements, the many-body Hamiltonian is driven to the simplified form first indicated in Fig.~\ref{fig:schematic}. The IM-SRG evolution not only decouples the ground state from excitations, but reduces the coupling between excitations as well. This coupling is an indicator of strong correlations in the many-body system, which usually require high- or even infinite-order treatments in approaches based on the Goldstone expansion. As we have discussed in Sec.~\ref{sec:floweq}, the IM-SRG can be understood as a non-perturbative, infinite-order re-summation of the many-body perturbation series, which builds the effects of correlations into the flowing Hamiltonian. To illustrate this, we show results from using the final IM-SRG Hamiltonian $\HO(\infty)$ in Hartree-Fock and post-HF methods in Fig.~\ref{fig:flow_methods}.

After the same 20--30 integration steps that lead to a strong suppression of the off-diagonal matrix elements (cf.~Fig.~\ref{fig:flow}), the energies of all methods collapse to the same result, which is the IM-SRG(2) ground-state energy. By construction, this is the result that would be obtained in a Hartree-Fock calculation with the IM-SRG Hamiltonian. Energy corrections due to correlations have been re-summed into the zero-body part of $\HO(\infty)$, and therefore MBPT(2) or either of the CC re-summations do not contribute additional correlation energy. The collapse of the ground-state energies occurs in the same fashion for all $(\eMax,\hw)$, although the rate and magnitude of the change in g.s. energy with the flow parameter $s$ may be quite different for each method.

Let us take a more detailed look at Fig.~\ref{fig:flow_methods}. For the bare \NNNLO interaction, the $\eMax=10$ results are not yet sufficiently converged with respect to either the single-particle basis and many-body expansions, hence the ground-state energy changes quite significantly with $s$ (cf.~Fig.~\ref{fig:chi2b_srgXXXX_conv}). For the soft \NNNLO interaction with $\lambda=2.0\fmi$, on the other hand, convergence w.r.t. basis size is already quite satisfactory at $\eMax=10$. Because this interaction is more perturbative, the small energy differences between the different many-body methods, in particular the second-order and infinite-order CC and IM-SRG re-summations, indicates good convergence of the many-body expansion\footnote{There is a caveat attached to this statement, namely that order-by-order perturbative convergence strongly depends around which reference state the perturbation expansion is constructed, see \cite{Roth:2010ys} and Sec.~\ref{sec:refstate}.} \cite{Bogner:2006qf,Bogner:2010pq}. We will return to this subject in Sec.~\ref{sec:mbpt}.

To conclude this section, we want to briefly discuss the four main scenarios that can occur when we use IM-SRG Hamiltonians as input for other many-body methods. We assume that calculations are converged w.r.t.~basis size, etc.
\begin{enumerate}
\item \emph{Full IM-SRG, exact many-body method:} For exact methods like the NCSM or No-Core Full Configuration (NCFC or FCI), the ground-state energy would be flat as a function of $s$. By performing an untruncated IM-SRG calculation, we essentially split the diagonalization of the many-body Hamiltonian into a part that is obtained by solving the IM-SRG flow equation, and a part that is obtained with traditional eigenvalue methods, with $s$ serving as an arbitrary separation point.
\item \emph{Full IM-SRG, approximate many-body method:} The ground-state energy varies with $s$, but for $s\to\infty$, the approximate many-body method yields the \emph{exact} eigenvalue due to the untruncated IM-SRG transformation. Here we see how the IM-SRG can be used to improve the input Hamiltonian for other many-body approaches.
\item \emph{Truncated IM-SRG, exact many-body method:} Again, the ground-state energy varies with $s$, and the overall variation is a measure of the extent to which the IM-SRG truncation violates exact unitarity.
\item \emph{Truncated IM-SRG, approximate many-body method:} This is the most common, and most complicated case. Because of the IM-SRG truncation, the IM-SRG will reproduce the exact ground-state energy only approximately in the limit $s\to\infty$. If the approximate many-body method contains content \emph{beyond} the truncated IM-SRG, then the result may actually degrade to some extent, whereas the IM-SRG still improves the result in the opposite scenario, but the uncertainty of $E(\infty)$ is hard to quantify unless one also uses exact many-body methods for comparison. Both of these scenarios are realized in Fig.~\ref{fig:flow_methods}: MBPT(2) is less complete than the IM-SRG(2), so the MBPT(2) energy is improved towards the exact energy. Note that this improvement can come in the form of attractive or repulsive corrections, because MBPT(2) typically underestimates the g.s.~energy for the bare interaction, but overshoots with soft interactions \cite{Roth:2006lr,Guenther:2010ge,Roth:2010ys,Bogner:2010pq,Hebeler:2011dq,Tsukiyama:2011uq,Langhammer:2012uq}. Both CCSD and $\Lambda$-CCSD(T) differ from the IM-SRG(2) at fourth order in MBPT (see Sec.~\ref{sec:mbpt}). CCSD typically underpredicts the nuclear binding energy, hence the additional correlation energy provided by the IM-SRG improvement should improve agreement with exact methods. $\Lambda$-CCSD(T) contains fourth-order 3p3h (triples) correlations, which are typically attractive, and missing in the IM-SRG(2) (cf.~Sec.~\ref{sec:mbpt}). This explains why the CCSD(T) ground-state energy actually increases (i.e., the binding energy decreases) with IM-SRG(2) input Hamiltonians as $s\to\infty$ for the soft interaction. As mentioned above, $\eMax=10$ is not yet sufficiently converged in the case of the ground-state energies shown in the top panel. For larger bases, the IM-SRG(2) Hamiltonian yields an increased $\Lambda$-CCSD(T) ground-state energy (see Sec.~\ref{sec:sum+dev} and Ref.~\cite{Hergert:2013mi}). Part of this increase is benign, because $\Lambda$-CCSD(T) is known to overestimate ground-state energies \cite{Taube:2008kx,Taube:2008vn,Shavitt:2009,Hagen:2010uq,Binder:2013fk,Hagen:2014ve}.
\end{enumerate}

\subsection{\label{sec:numerics_radii}Radii}

\begin{figure}[p]
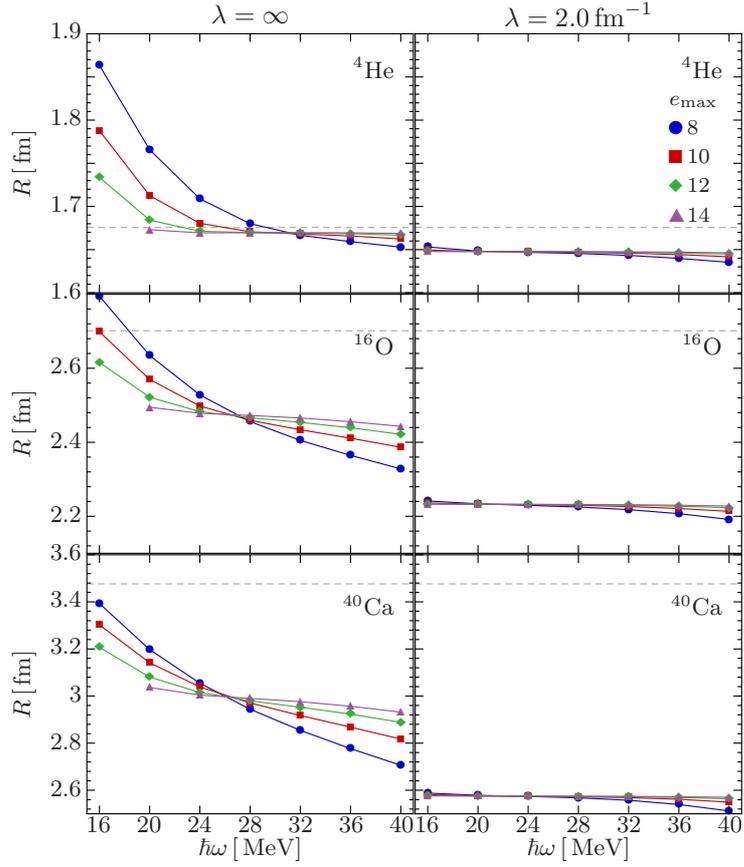

  \setlength{\unitlength}{0.400\textwidth}
  \begin{center}
    \begin{picture}(2.0000,2.2000)
      \put(0.2000,2.1500){\parbox{0.8\unitlength}{\centering$\lambda=\infty$}}
      \put(1.0000,2.1500){\parbox{0.8\unitlength}{\centering$\lambda=2.0\fmi$}}
      \put(0.0000,1.2800){\small\input{fig/chi2b_srg0000_He4_Rch.tex}}
      \put(0.8000,1.2800){\small\input{fig/chi2b_srg0625_He4_Rch.tex}}
      \put(0.0000,0.6400){\small\input{fig/chi2b_srg0000_O16_Rch.tex}}
      \put(0.8000,0.6400){\small\input{fig/chi2b_srg0625_O16_Rch.tex}}
      \put(0.0000,0.0000){\small\input{fig/chi2b_srg0000_Ca40_Rch.tex}}
      \put(0.8000,0.0000){\small\input{fig/chi2b_srg0625_Ca40_Rch.tex}}
    \end{picture}
  \end{center}
  \vspace{-10pt}
  \caption{\label{fig:chi2b_srgXXXX_rch}
    Convergence of $\nuc{He}{4}$, $\nuc{O}{16}$, and $\nuc{Ca}{40}$ IM-SRG(2) charge radii w.r.t.~single-particle basis size $\eMax$, for a chiral \NNNLO NN interaction with $\lambda=\infty$ (left panels) and $\lambda=2.0\,\fm^{-1}$ (right panels). The gray dashed lines indicate experimental charge radii from \cite{Angeli:2004ts}.
  }
\end{figure}

In Sec.~\ref{sec:observables}, we have discussed the evaluation of observables other than the ground-state energy by solving additional sets of flow equations along with those for the Hamiltonian. As an example, we show the convergence of the charge radii of $\nuc{He}{4}$, $\nuc{O}{16}$, and $\nuc{Ca}{40}$ in Fig.~\ref{fig:chi2b_srgXXXX_rch}. The results are obtained by normal-ordering and evolving the intrinsic proton mean-square radius operator,
\begin{equation}\label{eq:def_rp}
  R^2_{p} \equiv \sum_{i}\frac{1}{2}\left(1+\tau_3^{(i)}\right)\left(\rOV_i-\ROV\right)^2\,,
\end{equation}
where the isospin operator projects on protons, and $\ROV$ is the center of mass. We obtain the charge radii by applying the corrections due to the mean-square charge radii of proton and neutron (see, e.g., \cite{Kamuntavicius:1997fk}):
\begin{equation}
  R_\text{ch}\equiv\sqrt{R^2_p + r^2_p + \frac{N}{Z}r^2_n} = \sqrt{R^2_p + (0.8775\,\fm)^2 - 0.1161\,\fm^2}\,,
\end{equation}
with values of $r^2_p$ and $r^2_n$ taken from \cite{Beringer:2012fk}.

Focusing on the results for the bare \NNNLO interaction first, we find satisfactory convergence of the charge radii at a level of 1\% over a wide region of basis parameters $\hw$. For different $\eMax$, the curves intersect in the vicinity of the $\hw$ that minimizes the ground-state energies (cf.~Fig.~\ref{fig:chi2b_srgXXXX_conv}). The IM-SRG(2) result for the charge radius of $\nuc{He}{4}$ is quite close to the experimental value. It is somewhat counter-intuitive, however, that the radius is slightly underpredicted, while about $1\,$MeV binding energy is missing (see Tab.~\ref{tab:chi2b_srgXXXX}). For $\nuc{O}{16}$, the binding energy is similarly close to the experimental one, but the charge radius is already too small by almost 10\%, while overbinding and underestimation of the radius are consistent on a superficial level with $\nuc{Ca}{40}$.

Using the softened \NNNLO interaction with $\lambda=2.0\fmi$ as input, convergence of the radii improves dramatically over the bare \NNNLO case. On the scales shown in Fig.~\ref{fig:chi2b_srgXXXX_rch}, results from $\eMax=10$ onwards are all but indistinguishable. At the same time, the underestimation of the radii becomes worse, which is consistent with the increased binding energies that are reported in Sec.~\ref{sec:numerics_convergence}. Part of the problem is that the change of the resolution scale of the \NNNLO interaction induces 3N,$\ldots$ interactions which have not been taken into account. These induced interactions give repulsive contributions to the g.s.~energy, and are therefore also expected to increase the radii to some extent (see Sec.~\ref{sec:sum+dev} and Refs.~\cite{Roth:2012qf,Hagen:2012oq,Hagen:2012nx,Binder:2013zr,Hergert:2013mi,Hergert:2013ij,Cipollone:2013uq,Soma:2013ys,Binder:2014fk,Soma:2014eu,Hergert:2014vn,Hergert:2015qd}). 

Under a change of resolution scale $\lambda$, the radius operator (or any other observable) should be transformed consistently with the Hamiltonian, causing it to gain induced many-body contributions. Since RG transformations like the free-space SRG, and related methods like Lee-Suzuki, are designed to deal with high-momentum/short-distance physics, their effect on the radius and other long-ranged operators, and therefore the size of induced contributions, was expected to be small \cite{Bogner:2010pq,Stetcu:2005qh,Paar:2006zf,Anderson:2010br}. A recent free-space SRG study suggests that induced contributions may be small but not negligible in view of the discrepancies between experimental and calculated radii from state-of-the-art \emph{ab initio} many-body calculations \cite{Schuster:2014oq}.

A related issue is the use of simple one-body ans\"{a}tze like \eqref{eq:def_rp} for the mean-square proton radius and other radius or transition operators. These specific forms neglect two- and higher many-body contributions which are generated by exchange currents, for instance, and should be included in the ``bare'' operator in the first place. Chiral EFT provides a consistent framework to treat these effects on a similar footing as the interaction itself \cite{Pastore:2008uq,Pastore:2009zr,Kolling:2009yq,Song:2009ys,Pastore:2011dq,Kolling:2011bh,Rozpedzik:2011qf,Kolling:2012cr,Piarulli:2013vn,Orlandini:2011kl}, but the exploration of these structurally richer operators in nuclear many-body calculations is still in its infancy \cite{Pastore:2013nx}.

\section{\label{sec:refstate}Choice of Reference State}

\subsection{\label{sec:refstate_overview}Overview}
Reference states are a common ingredient to most many-body methods. Usually, their function
is to fix certain characteristics of the system we want to describe, e.g., the proton 
and neutron numbers of a nucleus, and to provide a starting point for the construction of a 
many-body Hilbert space that is superior to the particle vacuum. Describing many-body states 
as excitations with respect to a suitably chosen reference state allows us to account for the 
characteristic energy scales of the target nucleus, and introduce systematic truncation schemes 
based on this information. It also suggests the use of normal-ordering techniques in a natural 
fashion (cf.~Sec.~\ref{sec:nord}).

In many-body theory, we broadly distinguish two classes of reference states, namely
single- and multi-reference states. The former class exclusively consists of Slater determinants,
i.e., independent-particle states that do not describe any correlations. Conversely, multi-reference 
states are all many-body states that have non-vanishing correlations, e.g., superpositions of
Slater determinants, Hartree-Fock-Bogoliubov vacua with pairing correlations \cite{Ring:1980bb}, 
etc.

As explained in Sec.~\ref{sec:decoupling}, the IM-SRG generates a mapping between 
an \emph{arbitrary} reference state $\ket{\Phi}$ and an eigenstate $\ket{\Psi}$ 
of the Hamiltonian. In a finite system, i.e., in absence of phase transitions, and 
without symmetry constraints on the basis, such a mapping \emph{always exists}, because
we can diagonalize the Hamiltonian and construct a unitary transformation
as the dyadic product of the exact ground state and the reference state, plus suitable
additional states to complete the basis. Performing an evolution with the untruncated
IM-SRG flow equations is equivalent to such a (partial) diagonalization\footnote{Problems
can only occur if we use a pathological generator.}.

In Sections \ref{sec:floweq}--\ref{sec:numerics}, we have assumed that the reference state is
a single Slater determinant, and we will stick to this choice for most of the remainder of this 
work. A generalization of the IM-SRG framework to arbitrary correlated reference states is the
Multi-Reference IM-SRG introduced in Refs.~\cite{Hergert:2013ij,Hergert:2014vn}, which will be
the subject of Sec.~\ref{sec:mrimsrg}. An in-depth discussion of the MR-IM-SRG will be 
presented elsewhere \cite{Hergert:2015qd}.

\subsection{\label{sec:refstate_slater}Slater Determinant as Reference States}
Using a Slater determinant as the (single-)reference state is a suitable choice for systems 
with a large gap in their excitation spectrum, e.g., closed-shell nuclei. Among the Slater 
determinants, those that satisfy the Hatree-Fock conditions for a given nucleus are the most 
natural choices (cf.~Sec.~\ref{sec:numerics}), because they minimize both the mean-field 
energy and the beyond mean-field correlation energy in a variational sense. In light of
our general considerations of the IM-SRG's reference state dependence, it is worth considering 
what would happen if we used a non-HF, non-optimal Slater determinant instead.

\begin{figure}[t]
\setlength{\unitlength}{0.8\columnwidth}
  \begin{center}
  \begin{picture}(1.0000,0.4200)
   \put(-0.0500,0.0450){\includegraphics[width=0.34\unitlength]{fig/H_initial.eps}}
   \put(0.3300,0.0450){\includegraphics[width=0.34\unitlength]{fig/H_IMSRG_3ph_decoupling.eps}}
   \put(0.7100,0.0450){\includegraphics[width=0.34\unitlength]{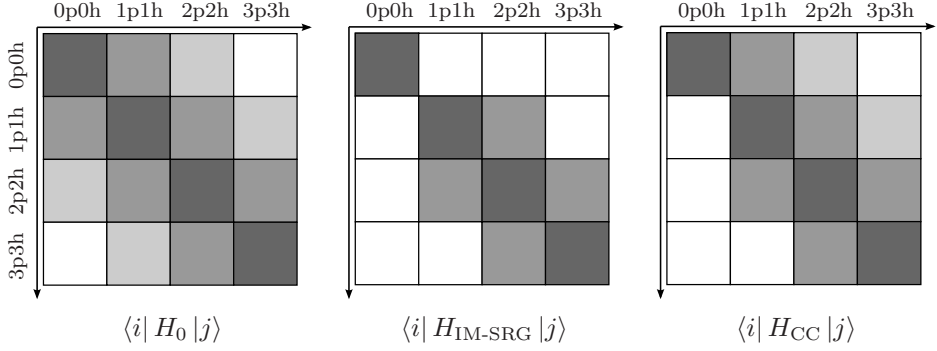}}
   \put(-0.0500,0.0000){\parbox{0.34\unitlength}{\centering$\matrixe{i}{\HO_0}{j}$}}
   \put(0.3300,0.0000){\parbox{0.34\unitlength}{\centering$\matrixe{i}{\HO_\text{IM-SRG}}{j}$}}
   \put(0.7100,0.0000){\parbox{0.34\unitlength}{\centering$\matrixe{i}{\HO_\text{CC}}{j}$}}
   
   \put(-0.0500,0.3900){\parbox{0.11\unitlength}{\centering\footnotesize0p0h}}
   \put(0.0250,0.3900){\parbox{0.11\unitlength}{\centering\footnotesize1p1h}}
   \put(0.1050,0.3900){\parbox{0.11\unitlength}{\centering\footnotesize2p2h}}
   \put(0.1800,0.3900){\parbox{0.11\unitlength}{\centering\footnotesize3p3h}}

   \put(0.3300,0.3900){\parbox{0.11\unitlength}{\centering\footnotesize0p0h}}
   \put(0.4050,0.3900){\parbox{0.11\unitlength}{\centering\footnotesize1p1h}}
   \put(0.4850,0.3900){\parbox{0.11\unitlength}{\centering\footnotesize2p2h}}
   \put(0.5600,0.3900){\parbox{0.11\unitlength}{\centering\footnotesize3p3h}}

   \put(0.7100,0.3900){\parbox{0.11\unitlength}{\centering\footnotesize0p0h}}
   \put(0.7850,0.3900){\parbox{0.11\unitlength}{\centering\footnotesize1p1h}}
   \put(0.8650,0.3900){\parbox{0.11\unitlength}{\centering\footnotesize2p2h}}
   \put(0.9400,0.3900){\parbox{0.11\unitlength}{\centering\footnotesize3p3h}}

   \put(-0.0800,0.3250){\parbox{0.11\unitlength}{\rotatebox{90}{\centering\footnotesize0p0h}}}
   \put(-0.0800,0.2450){\parbox{0.11\unitlength}{\rotatebox{90}{\centering\footnotesize1p1h}}}
   \put(-0.0800,0.1700){\parbox{0.11\unitlength}{\rotatebox{90}{\centering\footnotesize2p2h}}}
   \put(-0.0800,0.0900){\parbox{0.11\unitlength}{\rotatebox{90}{\centering\footnotesize3p3h}}}
  \end{picture}
  \end{center}
  \caption{\label{fig:cc_vs_imsrg}Schematic representation of the initial, IM-SRG, and Coupled Cluster Hamiltonians, in the many-body Hilbert space spanned by particle-hole excitations of the reference state (see text, and cf.~Sec.~\ref{sec:decoupling}).}
\end{figure}

According to a theorem by Thouless \cite{Thouless:1960fk}, any two Slater determinants 
$\ket{\Phi_A},\ket{\Phi_B}$ that are non-orthogonal and therefore have non-vanishing overlap
are related (up to a normalization constant and phase factor) by a similarity transformation:
\begin{equation}\label{eq:ref_thouless}
  \ket{\Phi_B} \sim \exp\bigg(\sum_{ph}t_{ph}\nord{\aaO_p\aO_h}\bigg)\ket{\Phi_A} \equiv e^{T^{(1)}}\ket{\Phi_B}\,.
\end{equation}
Equation \eqref{eq:ref_thouless} immediately suggests a connection with Coupled Cluster theory.
We recall that CC uses wave function ansatz
\begin{equation}
  \ket{\Psi_\text{CC}}=e^T\ket{\Phi},\qquad T=T^{(1)} + T^{(2)} + \ldots\,,
\end{equation}
with a reference Slater determinant $\ket{\Phi}$ and cluster operators
\begin{align}
  T^{(1)} &= \sum_{ph}t_{ph} \nord{\aaO_p\aO_h}\,, \\
  T^{(2)} &= \frac{1}{4}\sum_{pp'hh'}t_{pp'hh'} \nord{\aaO_p\aaO_{p'}\aO_{h'}\aO_h}\,,\\
  \ldots\notag
\end{align}
The coefficients of the cluster operators are determined by solving the algebraic system of
equations
\begin{align}
  \dmatrixe{\Phi}{e^{-T}\HO e^{T}} & = E \,,\label{eq:cc_0p0h}\\
  \matrixe{\Phi^{p}_{h}}{e^{-T}\HO e^{T}}{\Phi} &=0 \,, \label{eq:cc_1p1h}\\
  \matrixe{\Phi^{pp'}_{hh'}}{e^{-T}\HO e^{T}}{\Phi} &=0 \,, \label{eq:cc_2p2h}\\
  \ldots \notag
\end{align}
where $\ket{\Phi^{p}_{h}}=\nord{\aaO_p\aO_h}\!\ket{\Phi}, \ket{\Phi^{pp'}_{hh'}}, \ldots$, are 
particle-hole excited Slater determinants. In this way, CC generates an effective Hamiltonian 
that does not couple the reference state to excitations, which is similar in spirit to the IM-SRG
(cf.~Sec.~\ref{sec:decoupling}). However, the CC similarity transformation is non-unitary, and 
the resulting effective Hamiltonian is non-Hermitian, as is readily seen in Fig.~\ref{fig:cc_vs_imsrg}.

Since the $T^{(i)}$ are only defined in terms of particle-hole
\emph{excitation} operators, it is easy to see that cluster operators of different 
particle rank commute, 
\begin{equation}
  \comm{T^{(i)}}{T^{(j)}}=0\,,
\end{equation}
because contractions between particle creation and hole annihilation operators vanish,
and vice versa (cf.~\ref{sec:nord}). Consequently, the CC wave function can be written as
\begin{equation}
  \ket{\Psi_\text{CC}}\approx e^{T^{(1)}}e^{T^{(2)}}\ket{\Phi} = e^{T^{(2)}}e^{T^{(1)}}\ket{\Phi}\,,
\end{equation}
and we see that Thouless' theorem \eqref{eq:ref_thouless} is directly built into the CC 
formalism. Equation \eqref{eq:cc_1p1h} guarantees that $e^{T^{(1)}}\ket{\Phi}$ satisfies 
the Hartree-Fock conditions, for arbitrary Slater determinants $\ket{\Phi}$. Up to numerical
effects, CC calculations using HF or non-HF Slater determinants as input should yield the
same results.

The situation is quite different in the case of the IM-SRG. A counterpart of Thouless' 
theorem \eqref{eq:ref_thouless} for unitary transformations was proven by Rowe, Ryman,
and Rosensteel in Ref.~\cite{Rowe:1980vn}, relating normalized Slater determinants  
$\ket{\Phi_A}, \ket{\Phi_B}$ via
\begin{equation}\label{eq:rrr_theorem}
  \ket{\Phi_B} = \exp\bigg(\sum_{ph}X_{ph}\nord{\aaO_p\aO_h} - X^*_{ph}\nord{\aaO_h\aO_p}\bigg)\ket{\Phi_A} \,.
\end{equation}
Unfortunately, Eq.~\eqref{eq:rrr_theorem} does not apply to the IM-SRG in a straightforward
fashion.

As mentioned in Sec.~\ref{sec:floweq_prelim}, the unitary transformation generated by the 
IM-SRG is formally given by the $S$-ordered exponential
\begin{align}
  U(s) &= \mathcal{S}\exp \int^s_0 ds'\,\eta(s') \label{eq:def_U_pathexp}\,,
\end{align}
because the generator dynamically changes during the flow. It can be defined as a product
of infinitesimal unitary transformations,
\begin{align}       \label{eq:def_U_expds}
   U(s) &= \lim_{N\to\infty}\prod^{N}_{i=0} e^{\eta(s_i)\delta s_i}\,,\quad \sum_{i}s_i=s\,,
\end{align}
or the series expansion
\begin{align}     \label{eq:def_U_series}
   U(s) &= \sum_n \frac{1}{n!}\int^s_0 ds_1 \int^s_0 ds_2 \ldots 
          \int^s_0 ds_n \mathcal{S}\{\eta(s_1)\ldots\eta(s_n)\}\,.
\end{align}
Here, $\mathcal{S}$ ensures that 
the flow parameters in the operator products appearing in the integrands are always in 
descending order. We remind the reader that neither expression can be written 
as a proper exponential in general. Unlike the cluster operator of the CC method, the 
generator $\eta(s)$ necessarily contains particle-hole de-excitation operators, or else 
it would not be anti-Hermitian as required for a unitary transformation. Thus, it is
possible to have non-vanishing contractions between generator components of different
particle rank, and commutators of such components do not vanish in general:
\begin{equation}
  \comm{\eta^{(i)}(s)}{\eta^{(j)}(s')} \neq 0\,.
\end{equation}
As a result, $U(s)$ does not factorize automatically. In the following subsections, we
will discuss the numerical implications of this property of the IM-SRG, as well as
an approach to introduce factorization by hand.

\subsection{\label{sec:refstate_HO_vs_HF}Harmonic Oscillator vs. Hartree-Fock Slater Determinants}

\begin{figure}[t]
  \setlength{\unitlength}{0.6\textwidth}
  \begin{center}
    \begin{picture}(1.0000,1.3400)
      \put(-0.0500,0.5100){\input{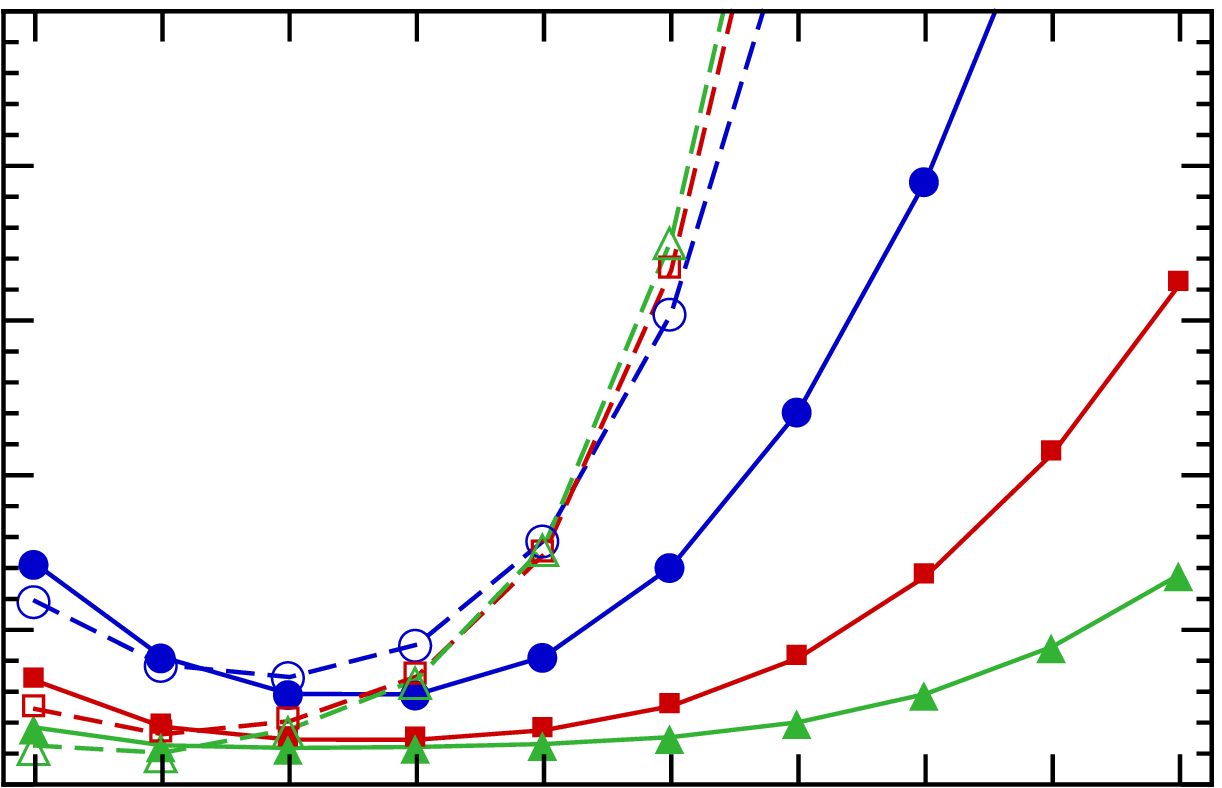}}
      \put(-0.0500,0.0000){\input{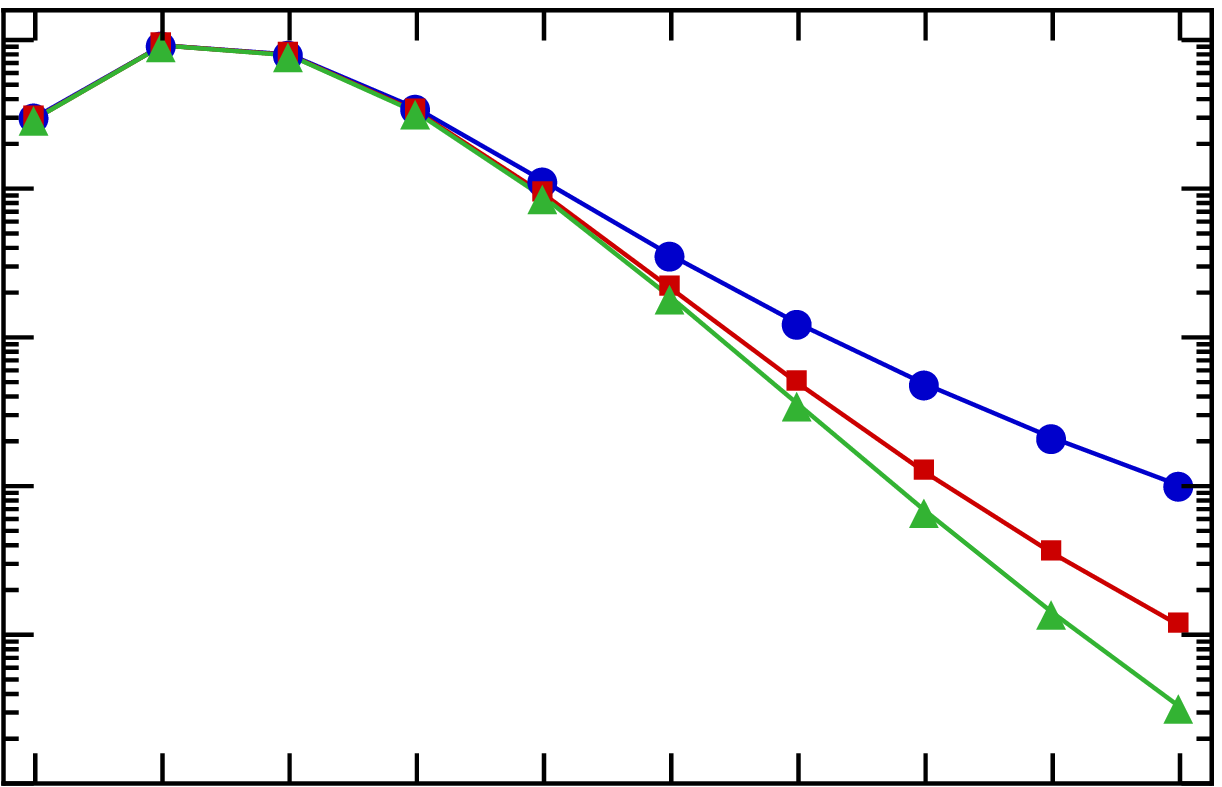}}
    \end{picture}
  \end{center}
  \vspace{-20pt}
  \caption{\label{fig:HOvsHF}Top panel: IM-SRG(2) energy of $\nuc{Ca}{40}$ with a HF (solid lines and symbols) and a HO reference state (dashed lines, open symbols), obtained with the Wegner generator. Bottom panel: 
  Overlap of the HF and HO reference states.}
\end{figure}

In previous sections, we have explained that the ground-state energies of the untruncated
IM-SRG flow equations do not depend on the choice of reference state. In practice, the
IM-SRG(2) truncation of the flow equation system (Eqs.~\eqref{eq:imsrg2_m0b}--\eqref{eq:imsrg2_m2b}) 
introduces an artificial reference-state dependence. 

In Fig.~\ref{fig:HOvsHF}, we compare ground-state energies for $\nuc{Ca}{40}$ that were 
obtained with a naive Shell Model HO Slater determinant and a HF Slater determinant, respectively. 
For oscillator parameters $16\leq\hbar\omega\leq24\,\MeV$, the two types of calculations 
essentially converge to the same ground-state energies. In this range, the HO and HF 
determinants have their largest overlap, as shown in the lower panel of Fig.~\ref{fig:HOvsHF}. 
Outside of this window, the overlap drops off steeply, which suggests that the HF single-particle 
wave functions differ appreciably from the plain HO single-particle wave functions. Of course,
we have to keep in mind that these differences are amplified exponentially when the 
many-body overlap is calculated as the product of single-particle overlaps. 

Beyond $\hw=28\,\MeV$, the IM-SRG(2) energies obtained with a HO refererence state actually grow 
with the basis size $\eMax$, which suggests that the IM-SRG is no longer targeting the 
Hamiltonian's ground state in those cases. This conclusion is supported by our inability 
to obtain converged results with White-type generators (see Eq.~\eqref{eq:eta_white}) for 
the larger $\hw$ values. The IM-SRG flow stalls because of divergences in the generator matrix 
elements. They are caused by small energy denominators that can be viewed as indicators of
level crossings in the spectrum of the evolving many-body Hamiltonian.

\begin{figure}[t]
  \setlength{\unitlength}{0.7\textwidth}
  \begin{center}
    \begin{picture}(1.3000,0.700)
      \put(0.0000,0.0000){\input{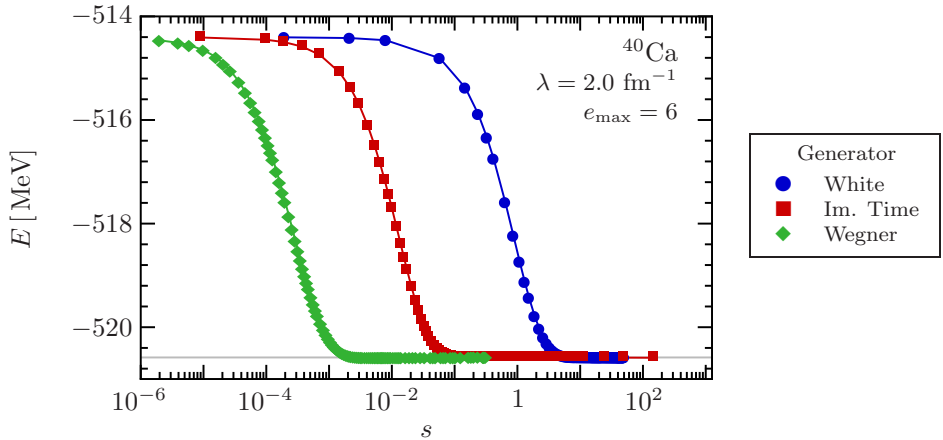}}
      \put(1.0500,0.4500){\fbox{\parbox{0.2400\unitlength}
        {\footnotesize
          \begin{tabular*}{0.1000\unitlength}{rl}
            \multicolumn{2}{c}{Generator} \\[2pt]
            \symbolcircle[FGBlue]    & White \\
            \symbolbox[FGRed]        & Im. Time \\
            \symboldiamond[FGGreen]  & Wegner 
          \end{tabular*}
        }}}
    \end{picture}
  \end{center}
  \vspace{-40pt}
  \caption{\label{fig:HFflow} IM-SRG decoupling of $1$p$1$h excitations for different generator choices, 
    starting from a HO reference state. The figure shows the $\nuc{Ca}{40}$ ground-state energy as a function 
    of the value of the flow parameter $s$. The unit of $s$ is suppressed because it differs with the choice 
    of generator. The gray line indicates the result of the Hartree-Fock calculation with the same interaction
    and basis parameters.
  }
\end{figure}

\subsection{Multi-Stage IM-SRG Evolution}
In Sec.~\ref{sec:decoupling}, we have described our philosophy of \emph{minimal decoupling}:
we choose generators that decouple only a single eigenstate from excitations instead of targeting
all eigenstates at once, in order to minimize the accumulation of errors due to the IM-SRG(2) 
truncation of the flow equations. We can take this idea a step further: in IM-SRG(2), we aim to decouple the ground state from 1p1h and 2p2h excitations \emph{simultaneously}, but it 
might be advantageous to split the transformation in two (or more) stages
if this gives us a better control over truncation errors. 

Here, we want to investigate how a sequential decoupling would render our 
results less dependent on the choice of reference state. We aim to decouple
the ground-state from the 1p1h sector first, and from 2p2h excitations in a
second stage. Recalling our analysis in Sec.~\ref{sec:decoupling}, we define 
the off-diagonal Hamiltonian for the first stage as
\begin{equation}
  f^{od}_{12} \equiv f_{12}(\nn_1 n_2 + n_1 \nn_2 )\,, 
  \qquad
  \Gamma^{od}_{1234} \equiv 0 \,.
\end{equation}
The White and imaginary-time Generators now take the simple form
\begin{alignat}{2}
  \etaO^\text{IA/B}_{12}
  &=\frac{f^{od}_{12}}{\Delta^\text{A/B}_{12}}\,, \qquad&\etaO^\text{IA/B}_{1234}&=0\,,\\[5pt]
  \etaO^\text{IIA/B}_{12}
  &=\;\sgn\!\left(\Delta^\text{A/B}_{12}\right)f^{od}_{12}\,,\qquad& \etaO^\text{II}_{1234}&=0\,.
\end{alignat}
Note that these generators are one-body operators, hence there are
\emph{no induced three-body terms} in the IM-SRG flow equations (unless
we include an initial three-body force), and we can solve them without
truncation errors. 

According to Eq.~\eqref{eq:eta_wegner}, the Wegner generator would have a 
non-vanishing two-body contribution that induces three-body terms during 
the flow. Taking a cue from the discussion in Sec.~\ref{sec:numerics_generator},
we avoid these induced terms by explicitly setting the two-body part to zero:
\begin{alignat}{2}
  \eta^\text{III}_{12} &= 
  \sum_{a}(1-P_{12})f^d_{1a}f^{od}_{a2} - \sum_{ab}(n_a-n_b)f^{od}_{ab}\Gamma_{b1a2}\,,
  &\qquad
  \eta^{III}_{1234}&\equiv0\,.
\end{alignat}

For the three types of one-body generators, the flow equations 
\eqref{eq:imsrg2_m0b}--\eqref{eq:imsrg2_m2b} reduce to\footnote{Note that this is \emph{not}
 an IM-SRG(1) truncation, because the two-body part of the flowing 
Hamiltonian, $\Gamma$, explicitly appears in the flow equation. This is necessary to
generate the HF resummation as we evolve $s\to\infty$.}
\begin{align}
  \totd{E}{s}&= \sum_{ab}(n_a-n_b)\eta_{ab} f_{ba}\,,\label{eq:imsrghf_m0b}\\[5pt]
  \totd{f_{12}}{s} &= 
  \sum_{a}(1+P_{12})\eta_{1a}f_{a2} +\sum_{ab}(n_a-n_b)\eta_{ab}\Gamma_{b1a2}\,,\label{eq:imsrghf_m1b}\\[5pt]
  \totd{\Gamma_{1234}}{s}&= 
  \sum_{a}\left\{ 
    (1-P_{12})\eta_{1a}\Gamma_{a234} - (1-P_{34})\eta_{a3}\Gamma_{12a4}\label{eq:imsrghf_m2b}
    \right\}\,.
\end{align}
When we normal-order the initial Hamiltonian with respect to a HO Slater determinant 
and integrate this system of flow equations until $\eta$ vanishes, we reproduce the 
HF ground-state energies, as illustrated for $\nuc{Ca}{40}$ in Fig.~\ref{fig:HFflow}. 
This does not come as a surprise, because the HF procedure optimizes the single-particle
wave function until the variational ground state is decoupled from 1p1h 
excitations\footnote{The Rowe-Ryman-Rosensteel theorem, Eq.~\eqref{eq:rrr_theorem},
connects the HO and HF determinants through a unitary transformation that is a proper 
exponential, which implies that the equivalent (up to small numerical differences) IM-SRG 
transformation generated by Eqs.~\eqref{eq:imsrghf_m0b}--\eqref{eq:imsrghf_m2b}
can also be expressed as a proper rather than a path-ordered exponential. For 
further discussion, we refer the reader to Refs.~\cite{Magnus:1954xy,Blanes:2009fk,Morris:2015ve}.}.

The final IM-SRG Hamiltonian can then be used as input to a second evolution, 
with the aim of decoupling the ground-state from the 2p2h sector as well. We
could try to achieve this by using pure two-body generators, but the commutator
$\comm{\eta^{(2)}}{H}$ re-induces contributions to the one-body Hamiltonian matrix,
which in turn cause some small coupling between the ground state and 1p1h states to
re-appear. This is avoided if we use the full IM-SRG(2) flow equations. 

\begin{figure}[t]
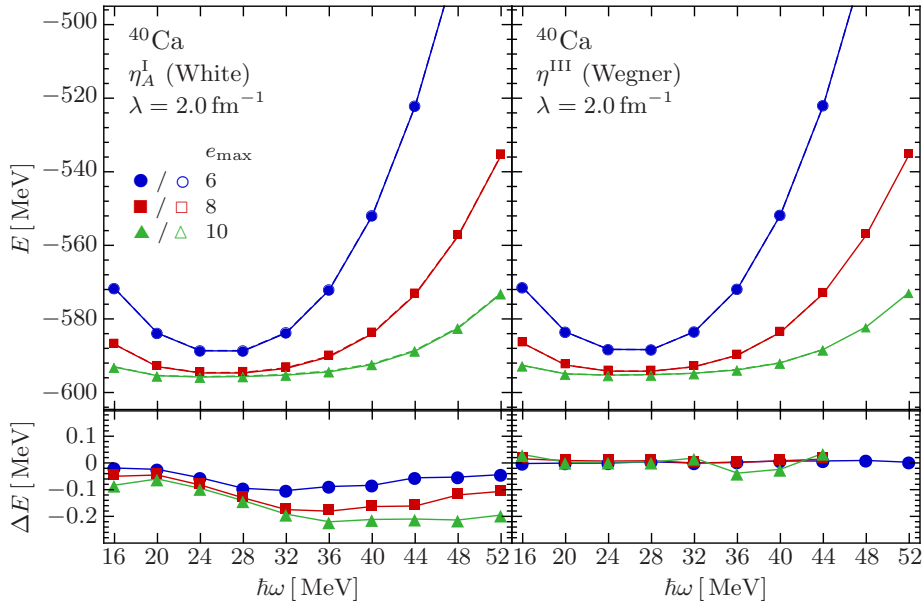

  \setlength{\unitlength}{0.5\textwidth}
  \begin{picture}(2.0000,1.2000)
    \put(0.1050,0.0000){\small\input{fig/chi2b_srg0625_Ca40_white_hf_vs_ho2s_dE.tex}}
    \put(0.1050,0.2600){\small\input{fig/chi2b_srg0625_Ca40_white_hf_vs_ho2s.tex}}
    \put(0.9000,0.0000){\small\input{fig/chi2b_srg0625_Ca40_wegner_hf_vs_ho2s_dE.tex}}
    \put(0.9000,0.2600){\small\input{fig/chi2b_srg0625_Ca40_wegner_hf_vs_ho2s.tex}}
  \end{picture}
  \\[-20pt]
  \caption{\label{fig:chi2b_srg0625_2stage}
    Top panel: Comparison of $\nuc{Ca}{40}$ ground-state energies from one-stage IM-SRG(2) evolution 
    with a HF reference state (solid symbols and lines) and two-stage evolution with a HO 
    reference state (open symbols, dashed lines), for White and Wegner generators (see text).
    Two-stage evolution for $hw=48,52\,\MeV$ did not converge in the $\eMax=8,10$ bases.
    Bottom panel: Difference of the ground-state energies from the two evolution 
    procedures.
  }
\end{figure}

In Fig.~\ref{fig:chi2b_srg0625_2stage}, we show $\nuc{Ca}{40}$ ground-state energies 
from the full two-stage evolution based on a HO reference state. We recall from the 
previous section that single-stage evolutions using the White generator failed 
for the majority of studied $\hw$ values, and energies obtained with the Wegner 
generator exhibited a growing trend with increasing basis size. The two-stage 
evolution fixes both pathologies, and unsurprisingly, the results compare extremely 
well to those form a single-stage IM-SRG(2) evolution based on a HF reference state.
For the White generator, the energy difference between the two types of calculation
is $200\,\keV$ or less (i.e., less than 0.04\% of the calculated g.s. energy), and
results for the Wegner generator are practically identical, except for two cases
at large $\hw$ where the calculation did not converge properly.

Our findings for the two-stage evolution show that the reference-state dependence
is essentially due to uncontrolled effects from omitted induced contributions when
we use a HO reference state. While the present case is somewhat artificial --- 
we essentially use the IM-SRG to perform a Hartree-Fock calculation in a more
complicated way than usual --- it nevertheless suggests that multi-stage evolution
may be a useful strategy for controlling truncation errors, especially in more
complex decoupling scenarios and generator splittings, e.g., for the decoupling
of valence spaces for the nuclear Shell Model, as discussed in Sec.~\ref{sec:shell_model}.


\section{\label{sec:mbpt}Perturbative Analysis of the Flow Equations}

\begin{figure}[t]
  \setlength{\unitlength}{1.1\textwidth}
  \begin{center}
    \input{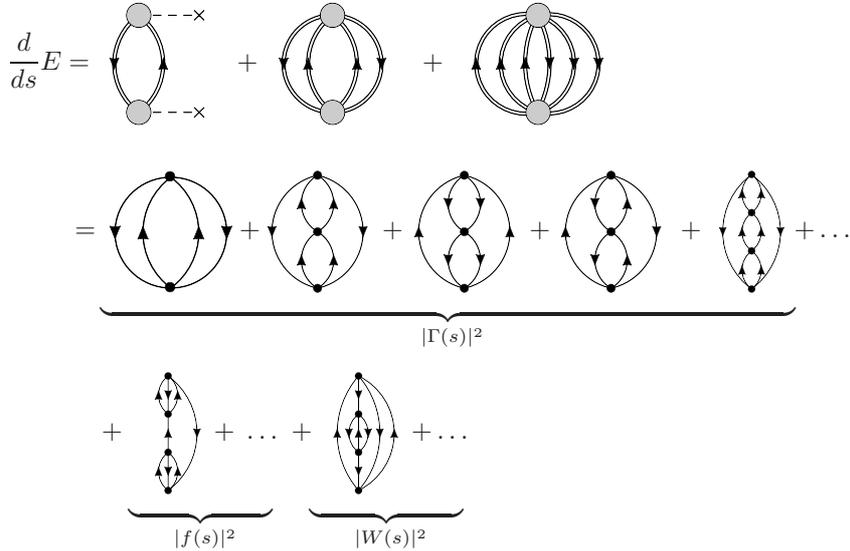}
  \end{center}
  \caption{\label{fig:flow}Schematic illustration of the energy flow equation \eqref{eq:flowpt} for the White generator with M{\o}ller-Plesset energy denominators (Eq.~\eqref{eq:eta_white}) in terms of Hugenholtz diagrams (see text). The grey vertices represent $\HO(s)$, and the double lines indicate energy denominators calculated with $f(s)$. On the second line, the flow equation is expanded in terms of $\HO(s-\delta s)$ (simple black vertices) and the corresponding energy denominators from $f(s-\delta s)$ (single lines). The braces indicate which term of $H(s)$ is expanded, and dots represent higher order diagrams generated by the integration step $s-\delta s \to s$.}
\end{figure}

\subsection{\label{sec:mbpt_intro}Overview}
The expressions for the White-type generators discussed in Sec. \ref{sec:white} are a manifest link between the IM-SRG and Many-Body Perturbation Theory (MBPT). For the sake of discussion, we focus on the White generator with M{\o}ller-Plesset energy denominators, keeping the short-hands $\Delta_{ph}$, $\Delta_{pp'hh'}$, etc., but dropping the superscript $B$. The generator with Epstein-Nesbet energy denominators can always be connected to this case by series expansion, e.g.,
\begin{equation}
  \frac{1}{f_p-f_h+\Gamma_{phph}} = \frac{1}{f_p-f_h}\sum_k \left(\frac{\Gamma_{phph}}{f_p-f_h}\right)^k\,.
\end{equation}

Let us now consider the flow equation for the ground-state energy \eqref{eq:imsrg2_m0b}, but broaden our perspective beyond the IM-SRG(2) truncation to keep track of the induced three-body contribution (cf.~Eq.~\eqref{eq:imsrg3_0b} and the discussion in Sec.~\ref{sec:decoupling}). Plugging in the White-M{\o}ller-Plesset generator with explicit three-body contribution, 
we obtain
\begin{align}
  \totd{E}{s}=&2\sum_{ph}\frac{|f_{ph}|^{2}}{\Delta_{ph}}
        +\frac{1}{2}\sum_{pp'hh'}\frac{|\Gamma_{pp'hh'}|^{2}}{\Delta_{pp'hh'}}
        +\frac{1}{18}\sum_{pp'hh'}\frac{|W_{pp'p''hh'h''}|^{2}}{\Delta_{pp'p''hh'h''}} 
        \label{eq:flowpt}\,.
\end{align}
The right-hand side of Eq.~\eqref{eq:flowpt} has the structure of the second-order MBPT correction to the ground-state energy, but the matrix elements and energy denominators depend on the flow parameter $s$. Thus, Eq.~\eqref{eq:flowpt} implies that the ground-state energy $E(s)$ is RG improved with contributions from higher orders of MBPT during the flow. 

In the following discussion, we characterize all operators in terms of the same dimensionless book-keeping parameter $g$. We also assume that the initial Hamiltonian satisfies the hierarchy $f^{d} > \Gamma > W$ throughout the flow. The hierarchy of $\Gamma$ and $W$, in particular, is compatible with the natural hierarchy of chiral two- and three-nucleon forces \cite{Epelbaum:2009ve,Bogner:2010pq}. Initially,
\begin{align}
  E(0) = \OC(g^0),\quad f^d(0) = \OC(g^0)\,, \quad\Gamma(0) &= \OC(g)\,.\label{eq:mbpt_count1}
\end{align}
If we do not include an initial three-body term, and choose a HF Slater determinant $(f^{od}=\{f_{ph},f_{hp}\})$ as the IM-SRG reference state, we also have
\begin{equation}
  f^{od}(0) = 0,\quad W(0) = 0\,.
\end{equation}
From the flow equations \eqref{eq:imsrg2_m0b}--\eqref{eq:imsrg2_m2b} (or \eqref{eq:imsrg3_0b}--\eqref{eq:imsrg3_3b}), we can conclude that corrections to $\Gamma(s)$ are of order $\OC(g)$. Corrections to $f(s)$ are $\OC(g^2)$ because they are generated by terms which are quadratic in $\Gamma(s)$, and the same reasoning holds for the induced off-diagonal and three-body matrix elements,  
\begin{equation}
  f^{od}(s)=\OC(g^2)\,,\quad W(s)=\OC(g^2)\,,\quad\text{for}\;s>0\, \label{eq:mbpt_count2}
\end{equation}
 (also cf.~Sec.~\ref{sec:scales}). This establishes that the three terms in the flow equation \eqref{eq:flowpt} are of order $\OC(g^4), \OC(g^2)$, and $\OC(g^4)$, respectively.

In Fig.~\ref{fig:flow}, the effect of integrating Eq.~\eqref{eq:flowpt} by a single step $s-\delta s \to s$ is illustrated schematically in terms of Hugenholtz diagrams (see, e.g., \cite{Negele:1998ve,Shavitt:2009}). Expanding the $\HO(s)$ vertices in terms of $\HO(s-\delta s)$ vertices, we see that that the $\Gamma(s)$ term has contributions from $\OC(g^2)$ through $\OC(g^4)$. Expanding in $\HO(s-2\delta s)$ instead, we would get additional higher order diagrams, and so forth. Thus, we perform a (partial)re-summation of the many-body perturbation series by integrating the IM-SRG flow equations from $s=0$ to $\infty$. 

Figure \ref{fig:flow} shows that all topologies for second- and third-order energy diagrams are generated, and we will demonstrate below that we build up the complete energy through $\OC(g^3)$ when we integrate Eq.~\eqref{eq:flowpt}. The $\Gamma(s)$ term also generates fourth-order diagrams with up to $4$p$4$h/quadruples excitations, but $f(s)$ and $W(s)$ terms clearly contribute at fourth order as well. The former are included in the IM-SRG(2), which is therefore third-order correct, similar to Coupled Cluster with singles and doubles (CCSD). To obtain a formally correct fourth-order energy, we need to keep the induced three-body terms, e.g., use the IM-SRG(3) truncation or some appropriate approximation, as in CC with singles, doubles, and perturbative triples (CCSD(T)), for instance.

We stress, however, that the perturbative analysis will \emph{not} provide us with a means to judge the IM-SRG truncation error in nuclear physics applications, aside from a guaranteed linear scaling of the error with the particle number $A$ due to size extensivity \cite{Bartlett:1981zr,Shavitt:2009}. Even for nuclear Hamiltonians with low resolution scales, the MBPT series does not converge order by order for general reference states, as shown in Refs.~\cite{Roth:2010ys,Langhammer:2012uq}. The choice of reference state plays
an important role in this context (cf.~Sec.~\ref{sec:refstate}).

In the remainder of this section, we will analyze the IM-SRG in greater detail. The main goal of this analysis is to provide an understanding of how the IM-SRG relates to other diagrammatic methods like finite-order MBPT, the Self-Consistent Green's Function approach \cite{Dickhoff:2004fk,Soma:2011vn,Cipollone:2013uq}, or CC, which can be analyzed diagrammtically along the same lines as the IM-SRG (see, e.g., \cite{Shavitt:2009}). 

As mentioned above, we choose a HF Slater determinant as the reference state $\ket{\Phi}$ for the IM-SRG and the MBPT expansion. Then $f_{ph}(s)$ vanishes for $s=0$ (because of the HF equations) and $s\to\infty$ (because of the IM-SRG decoupling condition), and we will only have to discuss canonical HF MBPT diagrams in the language of \cite{Shavitt:2009}. The inclusion of non-HF (where $f_{ph}\neq0$) and non-canonical HF diagrams (where $f_{pp'}, f_{hh'}$ are non-diagonal) is straightforward but tedious because their number grows much more rapidly than the number of canonical HF diagrams \cite{Shavitt:2009}.

\subsection{\label{sec:mbpt_pc}Power Counting}
In the following discussion, we will use superscripts to indicate the order of individual terms in the IM-SRG flow equations. 
Let us first address the subtleties in the power counting that was defined in Eqs.~\eqref{eq:mbpt_count1} and \eqref{eq:mbpt_count2}. The natural orbitals for a HF Slater determinat $\ket{\Phi}$ are the HF orbitals, which means that $f(0)$ is diagonal in the particle and hole blocks of the s.p. basis, and $f_{ph}(0)=f_{hp}(0)=0$. Since these are the off-diagonal matrix elements defining the one-body part of the generator \eqref{eq:eta_white}, $\eta_{ab}$ vanishes as well, and the one-body flow equation at $s=0$ becomes
\begin{align}
  \left.\totd{f_{12}}{s}\right|_{s=0} &= \sum_{abc}(n_an_b\bar{n}_c+\bar{n}_a\bar{n}_bn_c)(1+P_{12})\eta^{[1]}_{c1ab}\Gamma^{[1]}_{abc2}+\ldots\,.\label{eq:pt1_0b_init}
\end{align}
Thus, corrections to $f$ start at $\OC(g^2)$ (cf.~Sec.~\ref{sec:scales}), and we have
\begin{align}
  f_{pp'}(s) &= \overline{f}^{[0]}_{p}\delta_{pp'} + f^{[2]}_{pp'}(s) + \ldots\,, \\
  f_{hh'}(s) &= \overline{f}^{[0]}_{h}\delta_{hh'} + f^{[2]}_{hh'}(s) + \ldots\,, \\
  f_{ph}(s)  &= f^{[2]}_{ph}(s) + \ldots\,,
\end{align} 
where the notation $\overline{f}^{[0]}$ indicates that the term does not depend on $s$. It immediately follows that
corrections and $s$-dependence of the M{\o}ller-Plesset energy denominators also appear at $\OC(g^2)$,
\begin{align}
  \Delta_{ab}(s) &= \overline{\Delta}_{ab}^{[0]} + \Delta_{ab}^{[2]}(s) + \ldots\,,\\
  \Delta_{abcd}(s) &= \overline{\Delta}_{abcd}^{[0]} + \Delta_{abcd}^{[2]}(s) + \ldots\,.
\end{align}
Consequently,the generator matrix elements are given by
\begin{align}
  \eta_{ph} &= \frac{f^{[2]}_{ph}}{\overline{\Delta}^{[0]}_{ph}} + \frac{f^{[3]}_{ph}}{\overline{\Delta}^{[0]}_{ph}}
                 +\frac{f^{[4]}_{ph}}{\overline{\Delta}^{[0]}_{ph}} + 
                 \frac{f^{[2]}_{ph}\Delta^{[2]}_{ph}}{\big(\overline{\Delta}^{[0]}_{ph}\big)^2}
                 + \OC(g^5)
                 \label{eq:eta_wmp1b_pt}\,,\\
  \eta_{pp'hh'} &= \frac{\Gamma^{[1]}_{pp'hh'}}{\overline{\Delta}^{[0]}_{pp'hh'}} 
                      + \frac{\Gamma^{[2]}_{pp'hh'}}{\overline{\Delta}^{[0]}_{pp'hh'}}
                      + \frac{\Gamma^{[3]}_{pp'hh'}}{\overline{\Delta}^{[0]}_{pp'hh'}}
                      + \frac{\Gamma^{[1]}_{pp'hh'}\Delta^{[2]}_{pp'hh'}}{\big(\overline{\Delta}^{[0]}_{pp'hh'}\big)^2}
                 + \OC(g^4)
                 \label{eq:eta_wmp2b_pt}\,,
\end{align}
and their Hermitian conjugates. Based on these considerations, we will proceed to discuss the one- and two-body flow equations at increasing orders $\OC(g^n)$. Since the energy flow equation does not feed back into the flow for $f$ and $\Gamma$, we will discuss it separately afterwards.

\subsection{\texorpdfstring{$\OC(g)$}{O(g)} Flow}
As shown in the previous section, corrections to the one-body Hamiltonian $f$ only begin to contribute at $\OC(g^2)$, hence
\begin{equation}
  \dot{f}^{[1]}_{12} = 0 \quad\Rightarrow\quad f^{[1]}_{12}(s) = 0\,,
\end{equation}
where the dot indicates the derivative with respect to $s$. The first-order contribution to the two-body flow comes from the first line of Eq.~\eqref{eq:imsrg2_m2b}:
\begin{align}
    &\dot{\Gamma}^{[1]}_{1234}= -\sum_{a}\left\{ 
    (1-P_{12})(\fz_{1a}\eta^{[1]}_{a234} )
    -(1-P_{34})(\fz_{a3}\eta^{[1]}_{12a4} )
    \right\}\,,
\end{align}
where we have used Eqs.~\eqref{eq:eta_wmp1b_pt} and \eqref{eq:eta_wmp2b_pt}, and $f^{[1]}=0$. Since $\etaO$ only has pphh and hhpp matrix elements and $f^{[0]}$ is diagonal, we have
\begin{align}
  \dot{\Gamma}^{[1]}_{pp'hh'}&=
    -\left(\fz_p + \fz_{p'} - \fz_h -\fz_{h'}\right)
    \eta^{[1]}_{pp'hh'}\notag\\
    &=-\Deltaz_{pphh'}\eta^{[1]}_{pp'hh'}\,,
\end{align}
and an analogous equation for the Hermitian conjugate, while $\dot{\Gamma}^{[1]}_{1234}=0$ otherwise. Thus, the flow equations can be integrated easily, and we obtain
\begin{align}
  \Gamma^{[1]}_{abcd}(s)=\Gammao_{abcd}\times
    \begin{cases} 
       e^{-s} & \text{for}\,abcd=pp'hh',hh'pp'\,,\\
       1      & \text{otherwise}\,,
    \end{cases}\label{eq:Gamma_1st}
\end{align}
with
\begin{equation}\label{eq:Gammao}
  \Gammao_{abcd}\equiv\Gamma_{abcd}(0)\,.
\end{equation}

\subsection{\label{sec:mbpt2_flow}\texorpdfstring{$\OC(g^2)$}{O(g2)} Flow}

\begin{figure}[t]
  \setlength{\unitlength}{\textwidth}
  \begin{center}
  \input{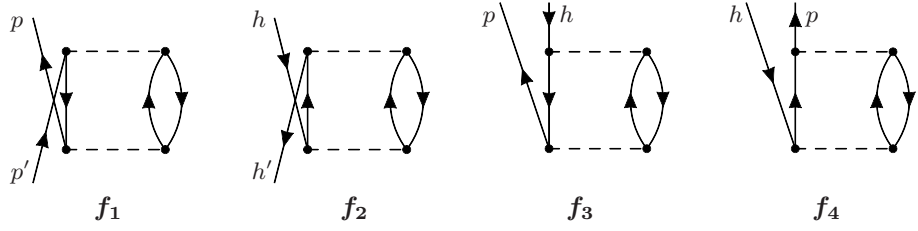}
  \end{center}
  \vspace{-5pt}
  \caption{\label{fig:mbpt2_f}
    Antisymmetrized Goldstone diagrams for the $\OC(g^2)$ effective one-body Hamiltonian (see text). Interpretation rules are summarized in Appendix \ref{app:diagram}.
  }
\end{figure}

We begin our discussion with the second-order contribution to $f$. Using Eq.~\eqref{eq:Gamma_1st}, the IM-SRG flow equation \eqref{eq:imsrg2_m1b} yields
\begin{align}
  \dot{f}^{[2]}_{pp'}&=\frac{1}{2}\sum_{p''hh'}\left(\eta^{[1]}_{p''phh'}\Gamma^{[1]}_{hh'p''p'}
                      +\eta^{[1]}_{p''p'hh'}\Gamma^{[1]}_{hh'p''p}\right)\notag\\
    &=\frac{1}{2}\sum_{p''hh'}\Gammao_{p''phh'}\Gammao_{hh'p''p'}\!\left(\frac{e^{-2s}}{\Deltaz_{p''ph'hh'}} + 
        \frac{e^{-2s}}{\Deltaz_{p''p'h'hh'}}\right)\notag\\
    &\equiv 2 \ft_{pp'}e^{-2s}\,.
\end{align}
The flow equations for the other matrix elements of $f^{[2]}(s)$ have the same structure, consisting of an $s$-independent amplitude and a function containing a decaying exponential in $s$. With the initial value condition $f^{[2]}(0)=0$, we obtain
\begin{align}
  f^{[2]}_{ab}(s)=\ft_{ab}\times
    \begin{cases}
      (1-e^{-2s}) & \text{for}\,ab=pp',hh'\,, \\
      se^{-s}     & \text{for}\,ab=ph,hp\,.
    \end{cases}
\end{align}
For $s\to\infty$, the IM-SRG builds up and adds the amplitudes $\ft_{pp'}$ and $\ft_{hh'}$ to the effective one-body Hamiltonian, which precisely correspond to the second-order contributions from MBPT. We can express them succinctly in terms of the antisymmetrized Goldstone diagrams shown in Fig. \ref{fig:mbpt2_f}:
\begin{align}
  \ft_{pp'}&=\frac{1}{2}\left(\left(f_1\right)_{pp'}+\left(p\leftrightarrow p'\right)\right)\,,\label{eq:ft_pp}\\
  \ft_{hh'}&=\frac{1}{2}\left(\left(f_2\right)_{hh'}+\left(h\leftrightarrow h'\right)\right)\,,\label{eq:ft_hh}\\
  \ft_{ph} &=\left(f_3\right)_{ph}+\left(f_4\right)_{hp}\,.\label{eq:ft_ph}
\end{align}  
The rules for interpreting such diagrams are derived in most many-body texts, so we only summarize them in Appendix \ref{app:diagram} for convenience. For the particle-hole matrix elements, we have
\begin{equation}
  f^{[2]}_{ph}(0) = f^{[2]}_{ph}(\infty) = 0\,,
\end{equation}
because we start with a HF Slater determinant and demand that the reference state is again decoupled from $1$p$1$h excitations for $s\to\infty$. At intermediate stages of the flow, the amplitudes $\ft_{ph}$ and $\ft_{hp}$ contribute to the build-up of higher-order MBPT diagrams.

\begin{figure*}[t]
  \setlength{\unitlength}{\textwidth}
  {\small\input{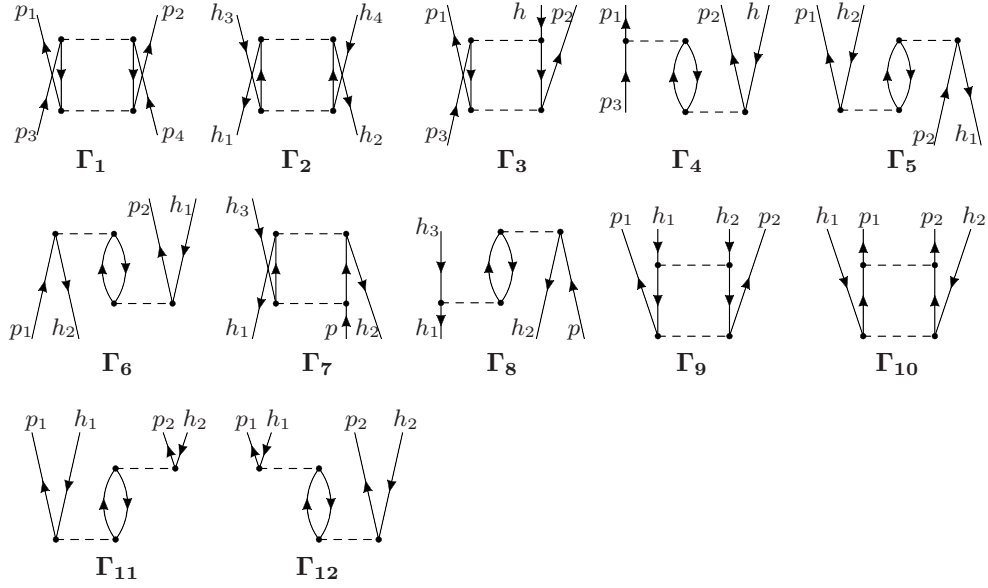}}
  \\[-30pt]
  \caption{\label{fig:mbpt2_Gamma}
    Antisymmetrized Goldstone diagrams for the $\OC(g^2)$ effective two-body vertex $\Gamma$ (see text). Interpretation rules are summarized in Appendix \ref{app:diagram}.
  }
\end{figure*}

For the second-order two-body vertex $\Gamma^{[2]}$, the same kind of analysis yields
\begin{align}
  \Gamma^{[2]}_{abcd}(s)=
    \Gammat_{abcd}\times
    \begin{cases}
      (1-e^{-2s}) & \text{for}\;abcd=p_{1}p_{2}p_{3}p_{4},\\
                                & \quad h_{1}h_{2}h_{3}h_{4},\\
                                & \quad p_{1}h_{1}p_{2}h_{2}, \ldots\,,\\
      (1-e^{-s})  & \text{for}\;abcd=p_{1}p_{2}p_{3}h,\\
                                & \quad h_{1}h_{2}h_{3}p,\ldots\,,\\
      se^{-s}     & \text{for}\;abcd=p_{1}p_{2}h_{1}h_{2},\\
                                & \quad \ldots\,,
    \end{cases}
\end{align}
where the dots indicate all allowed permuations and Hermitian conjugates of the explicitly given indices. The corresponding amplitudes are
\begin{align}
 \Gammat_{p_1p_2p_3p_4} 
      &= \frac{1}{2}\left((\Gamma_1)_{p_1p_2p_3p_4} + (\Gamma_1)_{p_3p_4p_1p_2}\right)\,, \label{eq:Gammat_pppp}\\
 \Gammat_{h_1h_2h_3h_4} 
      &= \frac{1}{2}\left((\Gamma_2)_{h_1h_2h_3h_4} + (\Gamma_2)_{h_3h_4h_1h_2}\right)\,, \label{eq:Gammat_hhhh}\\
 \Gammat_{p_1p_2p_3h}   &= (\Gamma_3)_{p_1p_2p_3h} + (1-P_{p_1p_2})(\Gamma_4)_{p_1p_2p_3h}\,, \label{eq:Gammat_ppph}\\
 \Gammat_{p_1h_1p_2h_2} 
      &= \frac{1}{2}\left((\Gamma_5)_{p_1 h_1 h_2 p_2} + (\Gamma_6)_{h_2 p_2 p_1 h_1 }\right)\,, \label{eq:Gammat_phph}\\
 \Gammat_{h_1h_2h_3p}   &= (\Gamma_7)_{h_1h_2h_3p} + (1-P_{h_1h_2})(\Gamma_8)_{h_1h_2h_3p}\,, \label{eq:Gammat_hhhp}\\
 \Gammat_{p_1p_2h_1h_2} &= (\Gamma_9)_{p_1p_2h_1h_2} + (\Gamma_{10})_{h_1h_2p_1p_2} 
                        +   
                           (1-P_{p_1p_2})(\Gamma_{11}+\Gamma_{12})_{p_1p_2h_1h_2}\,, 
                           \label{eq:Gammat_pphh}
\end{align}
where we refer to the diagrams in Fig.~\ref{fig:mbpt2_Gamma}. Expressions for the remaining combinations of indices can be obtained by using the antisymmetry and Hermiticity of $\Gammat_{abcd}$. Equations \eqref{eq:Gammat_pppp}--\eqref{eq:Gammat_pphh} 
are given in a hybrid form, i.e., they contain explicit Hermitian conjugates and line permutations of the diagrams. This allows us to express our analytic expressions for the amplitudes in terms of the minimal set of diagrams in Fig.~\ref{fig:mbpt2_Gamma}. If one envisions the inverse problem of constructing the IM-SRG flow equations from diagrams, one would of course include all possible diagram topologies, and express the amplitudes purely as sums of diagrams before deriving analytic expressions.

As in the schematic discussion of the energy flow equation in Sec. \ref{sec:mbpt_intro}, we also want to keep track of induced three-body terms. The IM-SRG(3) flow equation for the three-body vertex, Eq.~\eqref{eq:imsrg3_3b}, reveals that there are $\OC(g^2)$ contributions from products of $\eta^{[1]}_{abcd}(s)$ and $\Gamma^{[1]}_{abcd}(s)$, hence we have to analyze $W^{[2]}$. However, we will limit the discussion to the matrix elements of $W^{[2]}$ which can actually contribute to the fourth-order corrections to the ground-state energy (see Fig.~\ref{fig:schematic} and the discussion Sec.~\ref{sec:mbpt_intro}). Integrating the $\OC(g^2)$ three-body flow equation, we obtain
\begin{align}
  W^{[2]}_{abcdef}(s)=
    \Wt_{abcdef}\times
    \begin{cases}
      (1-e^{-2s}) & \text{for}\;abcdef=\\
                                & \quad p_{1}p_{2}h_{1}h_{2}p_{3}p_{4}\,,\\
                                & \quad h_{1}h_{2}p_{1}p_{2}h_{3}h_{4}\,,\\
                                & \quad \ldots\,,\\
      se^{-s}     & \text{for}\;abcdef=\\
                                & \quad p_{1}p_{2}p_{3}h_{1}h_{2}h_{3},\\
                                & \quad \ldots\,,
    \end{cases}
\end{align}
where the dots again indicate allowed Hermitian conjugates and permutations of indices. In terms of the diagrams shown in Fig.~\ref{fig:mbpt2_W}, the amplitudes are
\begin{align}
  \Wt_{p_1p_2h_1h_2p_3p_4}
      &=\frac{1}{2}\left((W_1)_{p_1p_2h_1h_2p_3p_4} + (W_1)_{h_2p_3p_4p_1p_2h_1}\right)\,,\label{eq:mbpt2_Wpphhpp}\\
  \Wt_{h_1h_2p_1p_2h_3h_4}
      &=\frac{1}{2}\left((W_2)_{h_1h_2p_1p_2h_3h_4} + (W_2)_{p_2h_3p_4h_1h_2p_1}\right)\,,\label{eq:mbpt2_Whhpphh}\\
  \Wt_{p_1p_2p_3h_1h_2h_3}
  &=P(p_1p_2/p_3)P(h_1h_2/h_3)\left(W_3+ W_4\right)_{p_1p_2p_3h_1h_2h_3}\,,
      \label{eq:mbpt2_Wppphhh}
\end{align}
where we have defined the three-body permutation symbols
\begin{align}
  P(ij/k) &\equiv 1-P_{ik}-P_{jk}\,, \label{eq:def_Pijk1}\\
  P(i/jk) &\equiv 1-P_{ij}-P_{ik}\,. \label{eq:def_Pijk2}
\end{align}

\begin{figure}[t]
  \setlength{\unitlength}{0.6\textwidth}
  \begin{center}
    \small\input{fig/mbpt2_W}
  \end{center}
  \vspace{-5pt}
  \caption{\label{fig:mbpt2_W}
    Antisymmetrized Goldstone diagrams for the $\OC(g^2)$ effective three-body vertex $W$ (see text). Interpretation rules are summarized in Appendix \ref{app:diagram}.
  }
\end{figure}

\subsection{\label{sec:mbpt3_flow}\texorpdfstring{$\OC(g^3)$}{O(g3)} Flow}
\newcommand{\diag}{
  \setlength{\unitlength}{0.175\columnwidth}
  \begin{picture}(1.0000,0.9500)
    \put(0.0000,0.0000){\includegraphics[width=\unitlength]{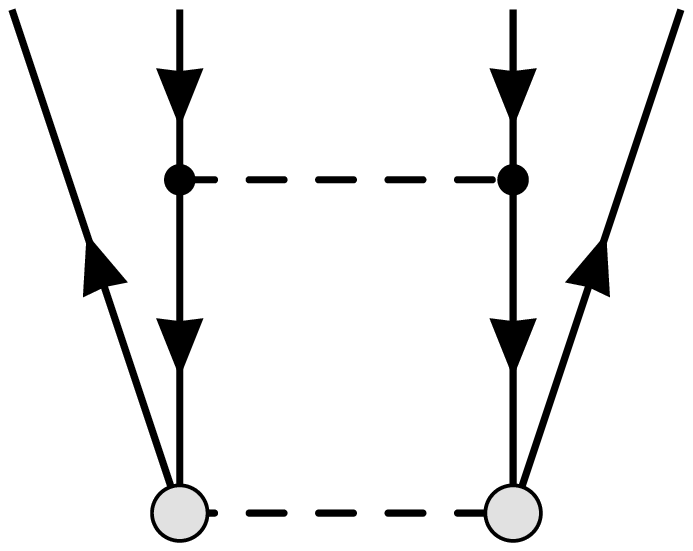}}
  \end{picture}
}
\newcommand{\diaga}{
  \setlength{\unitlength}{0.175\columnwidth}
  \begin{picture}(1.0000,0.9500)
    \put(0.0000,0.0000){\includegraphics[width=\unitlength]{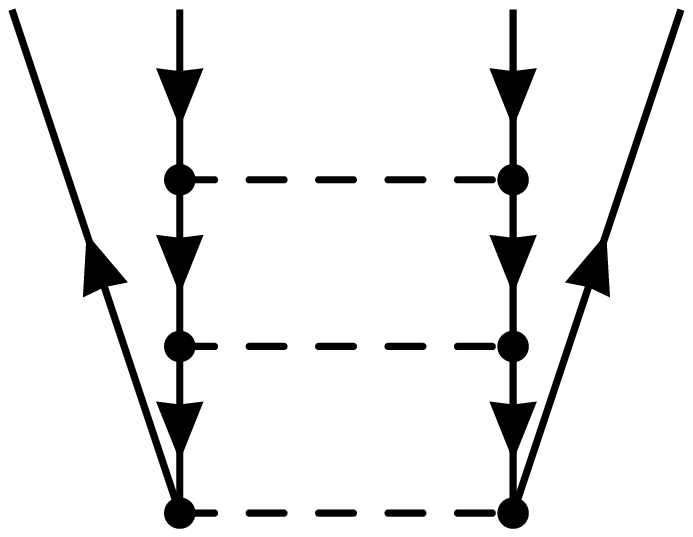}}
  \end{picture}
}
\newcommand{\diagb}{
  \setlength{\unitlength}{0.175\columnwidth}
  \begin{picture}(1.0000,0.9500)
    \put(0.0000,0.0000){\includegraphics[width=\unitlength]{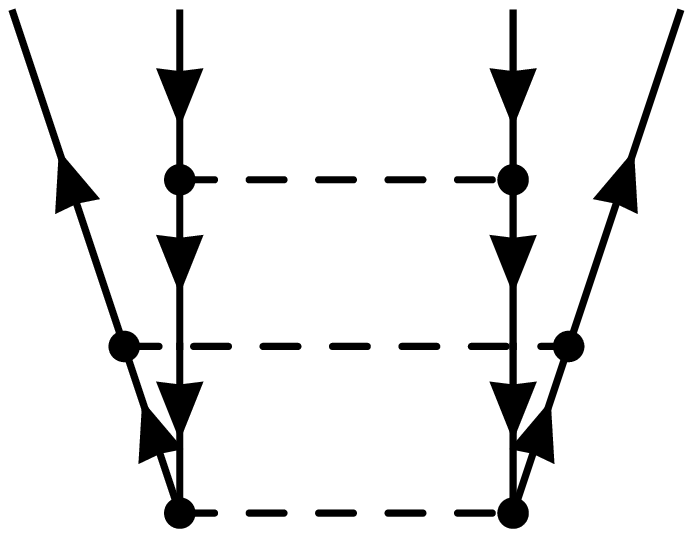}}
  \end{picture}
}
\newcommand{\diagc}{
  \setlength{\unitlength}{0.175\columnwidth}
  \begin{picture}(1.0000,0.9500)
    \put(0.0000,0.0000){\includegraphics[width=\unitlength]{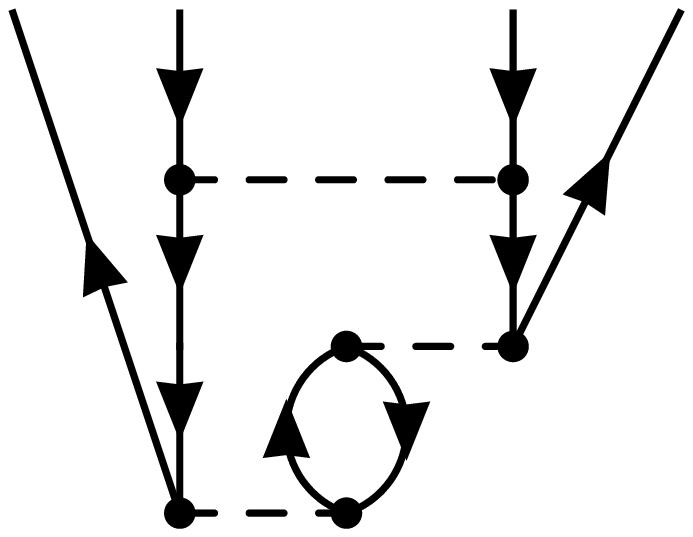}}
  \end{picture}
}
\begin{figure}[t]
  \setlength{\unitlength}{\textwidth}
  \begin{center}
    {\small\input{fig/mbpt3_Gamma}}
  \end{center}
  \vspace{-10pt}
  \caption{\label{fig:mbpt3_Gamma}
    Antisymmetrized Goldstone diagrams for the $\OC(g^3)$ effective two-body vertex $\Gamma$ (see text). 
    Black (\vertexi) and gray vertices (\!\vertexii[FGLightGray]\!\!\!) correspond to $\Gammao$ (Eq.~\eqref{eq:Gammao}),$\ft$ (Eqs.~\eqref{eq:ft_pp}--\eqref{eq:ft_ph}), $\Gammat$ (Eqs.~\eqref{eq:Gammat_pppp}--\eqref{eq:Gammat_pphh}), and $\Wt$ (Eqs.~\eqref{eq:mbpt2_Wpphhpp}--\eqref{eq:mbpt2_Wppphhh}), respectively. Interpretation rules are summarized in Appendix \ref{app:diagram}.
  }
\end{figure}

The analysis of the third-order one- and two-body flow equations is straightforward, but the number of terms (or diagrams) we have to consider increases significantly. Here, we content ourselves with analyzing $\Gamma^{[3]}_{pp'hh'}(s)$, the only missing ingredient for the discussion of the energy flow equation through $\OC(g^4)$, as in the overview presented in Sec.~\ref{sec:mbpt_intro}. Using our results from the previous sections, the two-body flow equation can be written as
\begin{align}
 \dot{\Gamma}^{[3]}_{p_1p_2h_1h_2}
    &\equiv-\Gamma^{[3]}_{p_1p_2h_1h_2} + \left(\overline{A} + \overline{D}\right)_{p_1p_2h_1h_2}se^{-s}
    + \left(\overline{B}+\overline{C}\right)_{p_1p_2h_1h_2}\left(e^{-3s}-e^{-s}\right)\,,\label{eq:mbpt3_dGamma}
\end{align}
which is solved by
\begin{align}
 \Gamma^{[3]}_{p_1p_2h_1h_2}(s)
    &=\left(\overline{A} + \overline{D}\right)_{p_1p_2h_1h_2}\frac{s^2}{2}e^{-s}
    - \left(\overline{B}+\overline{C}\right)_{p_1p_2h_1h_2}\left(\frac{e^{-3s}-e^{-s}}{2}+se^{-s}\right)\,.
    \label{eq:mbpt3_Gpphh}
\end{align}

The amplitudes $\overline{A}$ to $\overline{D}$ are given by the diagrams shown in Fig.~\ref{fig:mbpt3_Gamma}, where black and grey indices indicate the first- and second-order vertices, respectively:
\begin{align}
  \overline{A}_{p_1p_2h_1h_2}&=
      \left(1-P_{p_1p_2}\right)(A_1)_{p_1p_2h_1h_2}
    + \left(1-P_{h_1h_2}\right)(A_2)_{p_1p_2h_1h_2} \notag\\
    &\hphantom{=}+\left(A_3+A_4\right)_{p_1p_2h_1h_2}
    + \left(1-P_{p_1p_2}\right)\left(1-P_{h_1h_2}\right)\left(A_5\right)_{p_1p_2h_1h_2}\,,\label{eq:mbpt3_Gpphh_A}\\
  \overline{B}_{p_1p_2h_1h_2}&=
      -\etat_{p_1p_2h_1h_2}\Deltat_{p_1p_2h_1h_2}\notag\\
    &\hphantom{=} +\left(1-P_{p_1p_2}\right)(B_1)_{p_1p_2h_1h_2}
    + \left(1-P_{h_1h_2}\right)(B_2)_{p_1p_2h_1h_2} \notag\\
    &\hphantom{=}+\left(B_3+B_4\right)_{p_1p_2h_1h_2}
    + \left(1-P_{p_1p_2}\right)\left(1-P_{h_1h_2}\right)\left(B_5\right)_{p_1p_2h_1h_2}\,,\label{eq:mbpt3_Gpphh_B}\\
  \overline{C}_{p_1p_2h_1h_2}&=
     (1-P_{h_1h_2})\left(C_1\right)_{p_1p_2h_1h_2}
    + (1-P_{p_1p_2})\left(C_2\right)_{p_1p_2h_1h_2}\,,\label{eq:mbpt3_Gpphh_C}\\
  \overline{D}_{p_1p_2h_1h_2}&=
     (1-P_{h_1h_2})\left(D_1\right)_{p_1p_2h_1h_2}
    + (1-P_{p_1p_2})\left(D_2\right)_{p_1p_2h_1h_2}\,.\label{eq:mbpt3_Gpphh_D}
\end{align}
$\overline{A}$ and $\overline{B}$ are contained in the standard IM-SRG(2) truncation, whereas $\overline{C}$ and $\overline{D}$ are leading-order induced three-body terms. In particular, the former is a product of $\Wt$ and the two-body generator,
\begin{align}
  \overline{C}_{p_1p_2h_1h_2}&=\frac{1}{2}(1-P_{h_1h_2})\!\!\sum_{p'p''h'}\!\Wt_{p_1h'p_2p'h_1p''}\etao_{p'p''h'h_2}\notag\\
                            &\hphantom{=}+\frac{1}{2}(1-P_{p_1p_2})\!\!\sum_{h'h''p'}\!\Wt_{h_1p'h_2h'p_1h''}\etao_{h'h''p'p_2}\,,
  \label{eq:mbpt3_def_C}
\end{align}
while the latter is a product of $\Gammao$ and the three-body generator instead:
\begin{align}
  \overline{D}_{p_1p_2h_1h_2}&=\frac{1}{2}(1-P_{h_1h_2})\!\!\sum_{h'h''p'}\!\etat_{h'h_2h''p_1p_2p'}\Gammao_{h_1p''h'h''}\notag\\
                            &\hphantom{=}+\frac{1}{2}(1-P_{p_1p_2})\!\!\sum_{p'p''h'}\!\etat_{p'p''p_2h_1h'h_2}\Gammao_{p_1h'p'p''}\,.
  \label{eq:mbpt3_def_D}                          
\end{align}
This distinction is of little consequence in the present analysis, but may become important if the Hamiltonian and the generator are not truncated to the same particle rank. Note, however, that the diagrams for $\overline{C}$ and $\overline{D}$ have different topologies: The former couples the reference state to an excited $2$p$2$h state via intermediate $2$p$2$h excitations, whereas the latter has  intermediate $3$p$3$h states.

By expanding the grey $\Gammat$ vertices in Fig.~\ref{fig:mbpt3_Gamma} in terms of $\Gammao$, we can also see how the IM-SRG flow performs a non-perturbative resummation of the MBPT series, as indicated in Sec.~\ref{sec:mbpt_intro}. The diagram $A_3$, for instance, is expanded as
\begin{align}
  \vcenter{\hbox{\diag}} & = \vcenter{\hbox{\diaga}} + \vcenter{\hbox{\diagb}}
        + \vcenter{\hbox{\diagc}} + \ldots\,,
\end{align}
and contains ladder diagrams, as well as diagrams where ladder and polarization configurations interfere. Such interference diagrams set the IM-SRG apart from the traditional $G$-matrix and RPA approaches, which only resum ladders and polarization diagrams, respectively \cite{Fetter:2003ve}.

\subsection{Energy through \texorpdfstring{$\OC(g^4)$}{O(g4)}}
\newcommand{\diaghh}{
  \setlength{\unitlength}{0.135\columnwidth}
  \begin{picture}(1.0000,1.4000)
    \put(0.0000,0.0000){\includegraphics[width=\unitlength]{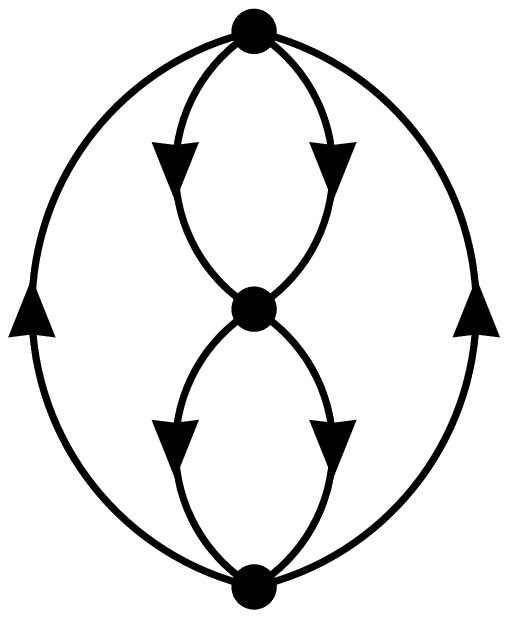}}
  \end{picture}
}
\newcommand{\diagpp}{
  \setlength{\unitlength}{0.135\columnwidth}
  \begin{picture}(1.0000,1.4000)
    \put(0.0000,0.0000){\includegraphics[width=\unitlength]{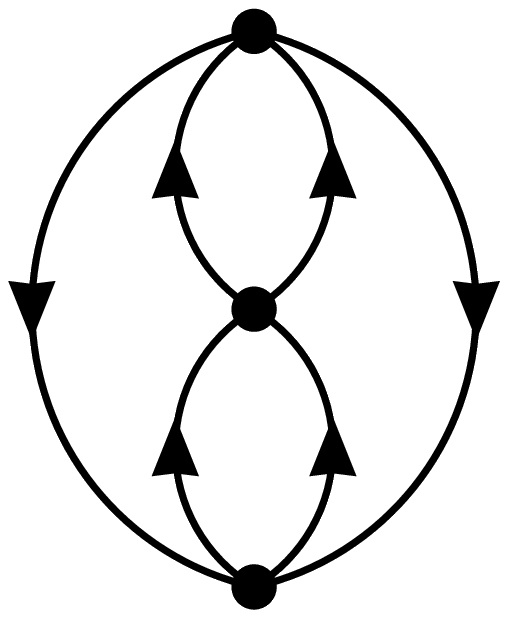}}
  \end{picture}
}
\newcommand{\diagph}{
  \setlength{\unitlength}{0.135\columnwidth}
  \begin{picture}(1.0000,1.4000)
    \put(0.0000,0.0000){\includegraphics[width=\unitlength]{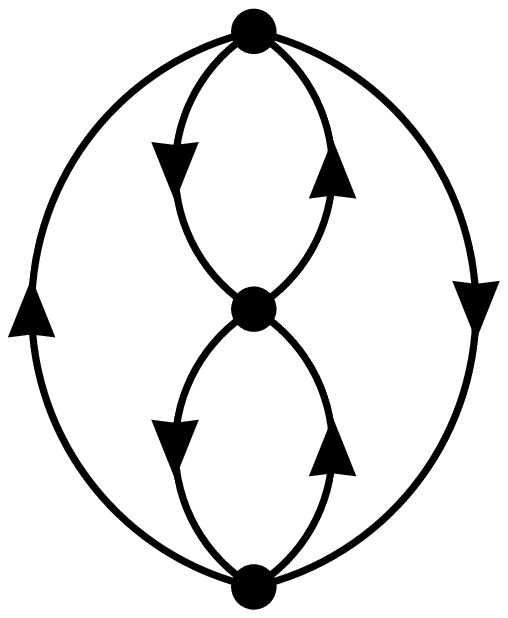}}
  \end{picture}
}
Let us now consider the energy flow equation. At $\OC(g^2)$, we have
\begin{align}
  \dot{E}^{[2]}&=\frac{1}{2}\sum_{h_1h_2p_1p_2}\eta^{[1]}_{h_1h_2p_1p_2}\Gamma^{[1]}_{p_1p_2h_1h_2}\notag\\
               &=\frac{1}{2}\sum_{h_1h_2p_1p_2}\etao_{h_1h_2p_1p_2}\Gammao_{p_1p_2h_1h_2}e^{-2s}\,.
\end{align}
Integrating this equation with $E^{[2]}(0)=0$\,, we obtain
\begin{equation}
  E^{[2]}(s)=\frac{1}{4}\left(1-e^{-2s}\right)\sum_{h_1h_2p_1p_2}\frac{\Gammao_{h_1h_2p_1p_2}\Gammao_{p_1p_2h_1h_2}}{\Deltaz_{h_1h_2p_1p_2}}\,,
\end{equation}
i.e., $E^{[2]}(\infty)$ is just the standard second-order MBPT correction to the energy of the reference state (cf.~Fig.~\ref{fig:flow}).

Likewise, the flow equation for the $\OC(g^3)$ energy reads
\begin{align}
  \dot{E}^{[3]}&=\frac{1}{2}\sum_{h_1h_2p_1p_2}\left(\eta^{[1]}_{h_1h_2p_1p_2}\Gamma^{[2]}_{p_1p_2h_1h_2} + 
               \eta^{[2]}_{h_1h_2p_1p_2}\Gamma^{[1]}_{p_1p_2h_1h_2} \right)\notag\\
               &=\sum_{h_1h_2p_1p_2}\frac{\Gammao_{h_1h_2p_1p_2}\Gammat_{p_1p_2h_1h_2}}{\Deltaz_{h_1h_2p_1p_2}}e^{-2s}
\end{align}
and integration yields
\begin{align}
  E^{[3]}(s) = \frac{1}{4}\left(1 - (2s+1)e^{-2s}\right)\!\!\!\sum_{h_1h_2p_1p_2}\!\!\!\frac{\Gammao_{h_1h_2p_1p_2}\Gammat_{p_1p_2h_1h_2}}{\Deltaz_{h_1h_2p_1p_2}}\,.                
\end{align}
For $s\to\infty$, 
\begin{equation}
  E^{[3]}(\infty)=\frac{1}{4}\sum_{h_1h_2p_1p_2}\frac{\Gammao_{h_1h_2p_1p_2}\Gammat_{p_1p_2h_1h_2}}{\Deltaz_{h_1h_2p_1p_2}}\,,
\end{equation}
and plugging in $\Gammat$ from Eq.~\eqref{eq:Gammat_pphh}, this immediately becomes
\begin{equation}
  E^{[3]}(\infty)=\vcenter{\hbox{\diaghh}}+\vcenter{\hbox{\diagpp}}+\vcenter{\hbox{\diagph}}\,,
\end{equation}
the standard third order energy correction.

\begin{figure*}[p]
  \setlength{\unitlength}{\textwidth}
  \input{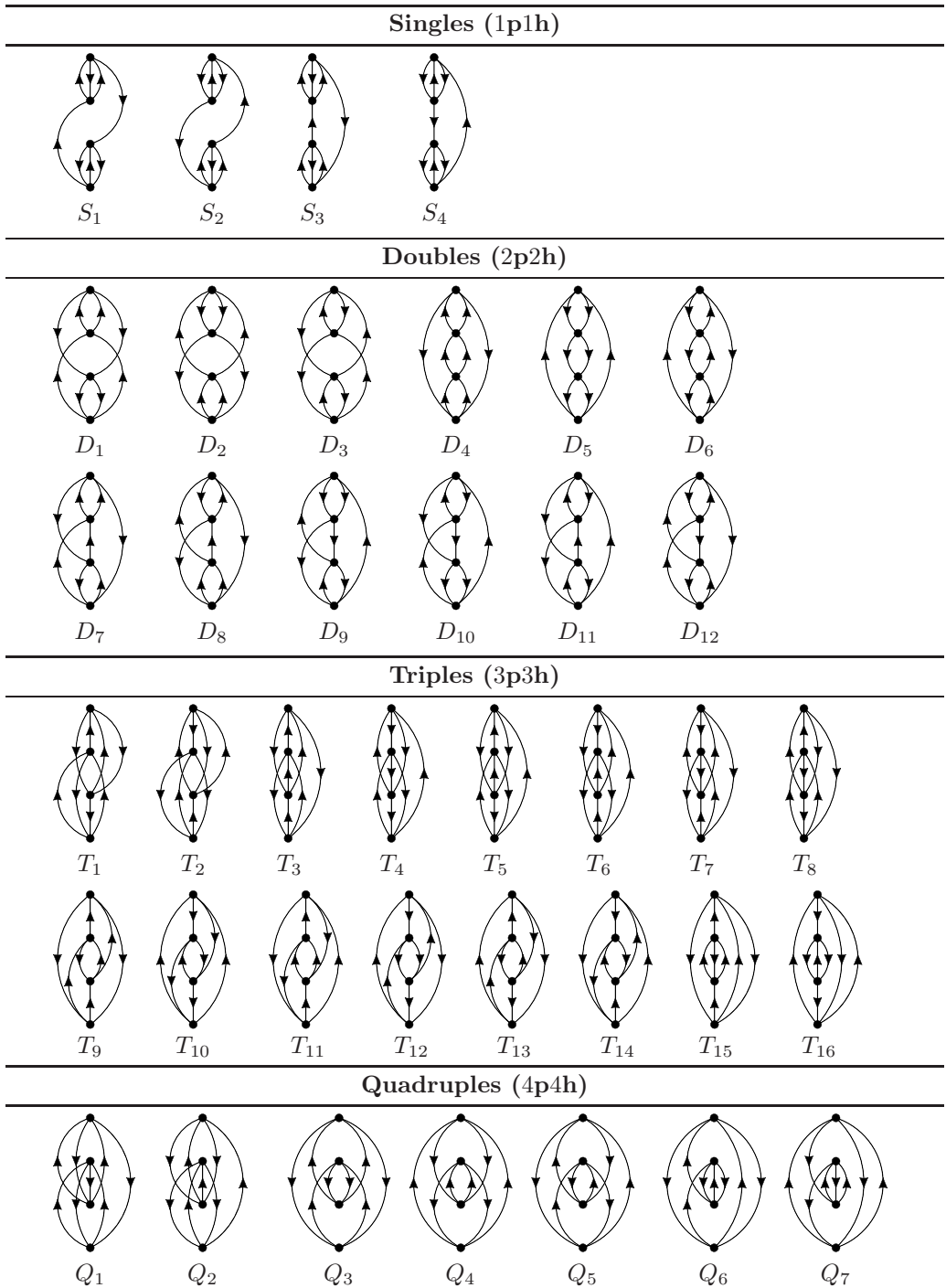}
  \caption{\label{fig:mbpt4}Connected Hugenholtz diagrams for the fourth-order energy correction $E^{(4)}$ (cf. \cite{Shavitt:2009}).
  }
\end{figure*}

At $\OC(g^4)$, we have to consider products of $\eta^{[2]}$ and the second-order Hamiltonian contributions $f^{[2]}, \Gamma^{[2]},$ and $W^{[2]}$ (cf.~Fig.~\ref{fig:flow}), as well as the cross terms
\begin{align}
  E^{[4]}_{3-1}
  &=\frac{1}{2}
  \sum_{p_1p_2h_1h_2}\left(\eta^{[3]}_{h_1h_2p_1p_2}\Gamma^{[1]}_{p_1p_2h_1h_2} + [\eta\leftrightarrow\Gamma]\right)\notag\\
  &=\frac{1}{2}
  \sum_{p_1p_2h_1h_2}\left(-\Gamma^{[1]}_{h_1h_2p_1p_2}\frac{\Delta^{[2]}_{h_1h_2p_1p_2}}{\Deltaz_{h_1h_2p_1p_2}} + 2\Gamma^{[3]}_{h_1h_2p_1p_2}\right)
  \times\frac{\Gamma^{[1]}_{p_1p_2h_1h_2}}{\Deltaz_{h_1h_2p_1p_2}}\,.
\end{align}
The first term is due to the expansion of the energy denominator in $\eta^{[3]}$ to second order (cf.~Sec.~\ref{sec:mbpt_pc}). However, it is easy to see that contributions from this term cancel in the sum, because $\Delta^{[0/2]}_{pp'hh'}$ is antisymmetric under transposition while $\Gamma^{[1]}_{pp'hh'}$ is symmetric. Thus, the energy flow equation becomes
\begin{align}
  \dot{E}^{[4]} &=
  2s^2e^{-2s}\sum_{ph}\etat_{hp}\ft_{ph}
  +\frac{s^2}{2}e^{-2s}\sum_{p_1p_2h_1h_2}\etat_{h_1h_2p_1p_2}\Gammat_{p_1p_2h_1h_2}\notag\\
  &\hphantom{=}+\frac{s^2}{18}e^{-2s}\sum_{p_1p_2p_3h_1h_2h_3}\etat_{h_1h_2h_3p_1p_2p_3}\Wt_{p_1p_2p_3h_1h_2h_3}\notag\\
  &\hphantom{=}+\sum_{p_1p_2h_1h_2}\frac{\Gammao_{p_1p_2h_1h_2}}{\Delta_{h_1h_2p_1p_2}}\left[ \frac{s^2}{2}e^{-2s}\left(\overline{A}+\overline{D}\right)_{h_1h_2p_1p_2}\right.\notag\\
  &\hphantom{=}\hspace{10em}\left.
               -\left(\frac{e^{-4s}-e^{-2s}}{2}+se^{-s}\right)\left(\overline{B}+\overline{C}\right)_{h_1h_2p_1p_2}
               \right]\,.
\end{align}
Integrating and taking the limit $s\to\infty$, we obtain the fourth-order energy correction
\begin{align}
  E^{[4]}(\infty) &=
  \frac{1}{2}\sum_{ph}\etat_{hp}\ft_{ph}
  +\frac{1}{8}\sum_{p_1p_2h_1h_2}\etat_{h_1h_2p_1p_2}\Gammat_{p_1p_2h_1h_2}\notag\\
  &\hphantom{=}+\frac{1}{72}\sum_{p_1p_2p_3h_1h_2h_3}\etat_{h_1h_2h_3p_1p_2p_3}\Wt_{p_1p_2p_3h_1h_2h_3}\notag\\
  &\hphantom{=}+\frac{1}{8}\sum_{p_1p_2h_1h_2}\left[ \left(\overline{A}-\overline{B}\right)_{h_1h_2p_1p_2}
               +\left(\overline{D}-\overline{C}\right)_{h_1h_2p_1p_2}
               \right]\frac{\Gammao_{p_1p_2h_1h_2}}{\Delta_{h_1h_2p_1p_2}}\,\notag\\
  &\equiv E^{[4]}_f + E^{[4]}_\Gamma + E^{[4]}_W + E^{[4]}_A + E^{[4]}_B + E^{[4]}_C + E^{[4]}_D\,. \label{eq:mbpt4_energy}
\end{align}

In Fig.~\ref{fig:mbpt4}, we show all fourth-order Hugenholtz energy diagrams for the canonical HF case (see Sec.~\ref{sec:mbpt_intro} and Ref.~\cite{Shavitt:2009}). It is a straightforward but arduous task to identify the diagrammtic content of the individual contributions to $E^{[4]}$ by plugging the expressions for the amplitudes from the previous sections into Eq.~\eqref{eq:mbpt4_energy}. We find
\begin{align}
  E^{[4]}_f      &= \frac{1}{2}\sum_{i=1}^4 S_i\,,\label{eq:def_E4f}\\
  E^{[4]}_\Gamma &= \frac{1}{2}\sum_{i=1}^{12} D_i\,,\\
  E^{[4]}_W      &= \frac{1}{2}\sum_{i=1}^{16} T_i\,,\label{eq:def_E4W}\\
  E^{[4]}_A      &= \frac{1}{2}\left(\sum_{i=1}^4S_i + \sum_{i=1}^{12}D_i\right)\,,\label{eq:def_E4A}\\
  E^{[4]}_B      &= Q_3 +  Q_4 +  Q_5 + \frac{1}{2}\left(Q_1 + Q_2 + Q_6 + Q_7\right)\,,\\
  E^{[4]}_C      &= \frac{1}{2}\left(Q_1 + Q_2 + Q_6 + Q_7\right)\,,\label{eq:def_E4C}\\
  E^{[4]}_D      &= \frac{1}{2}\sum_{i=1}^{16}T_i\,,\label{eq:def_E4D}
\end{align}
so $E^{[4]}(\infty)$ contains all required diagrams, and is indeed the complete fourth-order energy. 

\subsection{\label{sec:mbpt_discussion}Discussion}
As concluded on general grounds in Sec.~\ref{sec:mbpt_intro}, the IM-SRG(2) energy 
is complete to third order in MBPT, but misses certain contributions in fourth order. 
Our detailed analysis shows that
\begin{align}
  E^{[4]}_{\text{IM-SRG(2)}} &= E^{[4]}_f + E^{[4]}_\Gamma + E^{[4]}_A + E^{[4]}_B \notag\\
  &= \sum_{i=1}^4S_i + \sum_{i=1}^{12}D_i + Q_3+Q_4+Q_5 + \frac{1}{2}\left(Q_1+Q_2+Q_6+Q_7\right)
  \,,\label{eq:mbpt4_energy_imsrg2}
\end{align}
i.e., IM-SRG(2) contains the complete fourth-order singles and doubles contributions, 
as well as the symmetric and \emph{half} of the asymmetric quadruples diagrams shown 
in Fig.~\ref{fig:mbpt4}. 

In the discussion of Fig.~\ref{fig:flow_methods} in Sec.~\ref{sec:numerics_decoupling}, 
we have observed that the IM-SRG(2) ground-state energy of $\nuc{Ca}{40}$ for the chiral
NN Hamiltonian with $\lambda=2.0\fmi$ lies between Coupled Cluster results at the CCSD 
and $\Lambda-$CCSD(T) level \cite{Shavitt:2009,Taube:2008kx,Taube:2008vn}. Overall, the
three methods agree within a few percent of the total ground-state energy. This pattern 
has consistently emerged in all our IM-SRG calculations for finite nuclei with softened 
chiral interactions (resolution scales $\lambda\sim 2\fmi$), both with and without 3N 
forces \cite{Tsukiyama:2011uq,Hergert:2013mi, Hergert:2013ij,Hergert:2014vn}. The 
diagrammatic content of these methods through fourth order explains this behavior, at 
least qualitatively. In terms of the quantities \eqref{eq:def_E4f}--\eqref{eq:def_E4D} 
defined in the previous subsection, the fourth-order energy contributions to CCSD and 
$\Lambda-$CCSD(T) are 
\begin{align}
  E^{[4]}_{\text{CCSD}} &= E^{[4]}_f + E^{[4]}_\Gamma + E^{[4]}_A + E^{[4]}_B + E^{[4]}_C \notag\\
                        &= \sum_{i=1}^4S_i + \sum_{i=1}^{12}D_i + \sum_{i=1}^{7}Q_i \,,\label{eq:def_E4CCSD}
\end{align}
and 
\begin{align}
  E^{[4]}_{\Lambda-\text{CCSD(T)}} &= E^{[4]}_f + E^{[4]}_\Gamma + E^{[4]}_W + E^{[4]}_A 
  + E^{[4]}_B + E^{[4]}_C + E^{[4]}_D \notag\\
  &= \sum_{i=1}^4S_i + \sum_{i=1}^{12}D_i + \sum_{i=1}^{16}T_i + \sum_{i=1}^{7}Q_i 
      \,,\label{eq:def_E4CCSDT}
\end{align}
respectively. In a typical calculation, CCSD ground-state energies are too high due to
missing correlation energy from attractive fourth-order 3p3h (triples) configurations
that are included in $\Lambda-$CCSD(T) through $E^{[4]}_{W,D}$. In all our calculations, 
the asymmetric quadruples diagrams $Q_{1,2,6,7}$ (cf.~Fig.~\ref{fig:mbpt4}) are 
repulsive. The IM-SRG(2) misses half of this repulsion, namely the $E^{[4]}_C$ term,
and mocks up missing attraction from the triples terms $E^{[4]}_{W,D}$ in this way.

\begin{figure}[t]
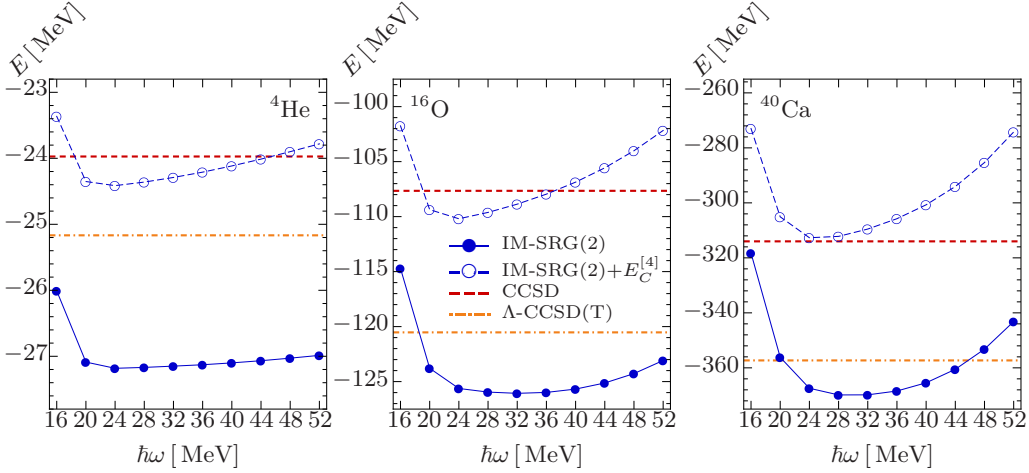

  \setlength{\unitlength}{0.34\textwidth}
  \begin{picture}(2.9400,1.450)
    \put(-0.0700,0.0000){\small\input{fig/chi2b_srg0000_He4_quad.tex}}
    \put(0.9150,0.0000){\small\input{fig/chi2b_srg0000_O16_quad.tex}}
    \put(1.9200,0.0000){\small\input{fig/chi2b_srg0000_Ca40_quad.tex}}
  \end{picture}
  \caption{\label{fig:chi2b_srg0000_quad}
    Effect of fourth-order quadruples (4p4h) contribution $E_C^{[4]}$, Eq.~\eqref{eq:def_E4C} on the ground-state 
    energies of $\nuc{He}{4}$, $\nuc{O}{16}$, and $\nuc{Ca}{40}$ (see text): Comparison of IM-SRG(2) with and
    without $E_C^{[4]}$, calculated with the initial Hamiltonian $\HO(0)$, to CCSD and $\Lambda-$CCSD(T). 
    All calculations used the chiral \NNNLO Hamiltonian with $\lambda=\infty$ in an $\eMax=14$ single-particle
    basis. The shown CC values were taken at optimal $\hw$.
  }
\end{figure}

Let us now consider the implications of our analysis for calculations with the unevolved 
chiral \NNNLO Hamiltonian. Referring back to Fig.~\ref{fig:flow_methods} again, there is 
a larger variation between the $\nuc{Ca}{40}$ ground-state energies from IM-SRG(2), CCSD, 
and $\Lambda-$CCSD(T). This is expected because of the Hamiltonian's higher resolution scale, 
which adversely affects the many-body convergence. We find an IM-SRG(2) ground-state 
energy that is lower than that of $\Lambda-$CCSD(T), which contains the complete fourth-order 
energy and is therefore expected to be a better approximation to the exact result from the
MBPT point of view. A similar observation was made for $\nuc{He}{4}$ in the first published 
IM-SRG study \cite{Tsukiyama:2011uq}, where the IM-SRG(2) ground-state energy of $-27.6\,\MeV$ 
was found to be about $2\,\MeV$ lower than the $\Lambda-$CCSD(T) and exact NCSM results. This 
motivated the development of a perturbative truncation scheme that is discussed in 
Sec.~\ref{sec:mbpt_truncation}, but no longer used in practice.

In Fig.~\ref{fig:chi2b_srg0000_quad}, we show the effect of adding the fourth-order quadruples
term $E^{[4]}_C$ to the IM-SRG(2) ground-state energies of $\nuc{He}{4}$, $\nuc{O}{16}$, and
$\nuc{Ca}{40}$. In light of our perturbative analysis, especially Eqs.\eqref{eq:mbpt4_energy_imsrg2}
and \eqref{eq:def_E4CCSD}, it is not surprising that the repulsive contributions from this term 
shift the ground-state energies in close proximity to the the CCSD results, which are shown for 
reference. The agreement is not exact due to fifth- and higher-order differences in the perturbative 
content of IM-SRG(2) and CCSD.

Finally, we want to remark on the different origins of the induced three-body vertices which 
contribute to $E^{[4]}_C$ and $E^{[4]}_D$, as pointed out in the discussion of Eqs.~\eqref{eq:mbpt3_def_C} 
and \eqref{eq:mbpt3_def_D} in Sec.~\ref{sec:mbpt3_flow}. This is relevant for asymmetric 
truncations of $\HO$ and $\eta$ at different particle rank, and the development of approximations 
to the full IM-SRG(3) scheme by the selective addition of terms to the IM-SRG(2) flow equations.
$E^{[4]}_C$ is a product of $\Wt$ and the two-body generator, while $E^{[4]}_D$ is a product of 
$\Gammat$ and the three-body generator. Thus, it is sufficient to consider only the induced 
three-body interaction $W$ to fully include the fourth-order quadruples\footnote{In 
Ref.~\cite{Evangelista:2012uq}, Evangelista and Gauss have demonstrated that $E^{[4]}_C$ 
is not included in a modified CCSD scheme if intermediate terms in the nested commutators 
are only expanded up to two-body operators. These intermediates correspond to the pieces of 
$W$ that are induced by the commutator of two-body operators, hence the mechanism for generating 
$E^{[4]}_C$ is very similar in CC and IM-SRG.}. A full inclusion of fourth-order 
triples requires the induced three-body interaction as well as the use of a three-body generator.

\subsection{\label{sec:mbpt_truncation}Perturbative Truncations}
As discussed repeatedly throughout this work (see, e.g., Secs.~\ref{sec:numerics_decoupling},
\ref{sec:mbpt_intro}), order-by-order convergence of a many-body perturbation expansion 
strongly depends on the resolution scale of the Hamiltonian, and the choice of reference 
state on which the perturbation series is constructed. This is particularly true for the 
case of nuclear Hamiltonians \cite{Bogner:2006qf,Ramanan:2007vn,Bogner:2010pq,Roth:2010ys,
Langhammer:2012uq}. Nevertheless, it is worthwhile to attempt and organize the right-hand side 
of the IM-SRG flow equation --- essentially, the $\beta$ function of the IM-SRG flow (see, 
e.g., \cite{Peskin:1997az,Zinn-Justin:2007kx}) --- in terms of a perturbative expansion, 
which is a common feature of RG approaches throughout all fields of physics. 

\begin{figure}[t]
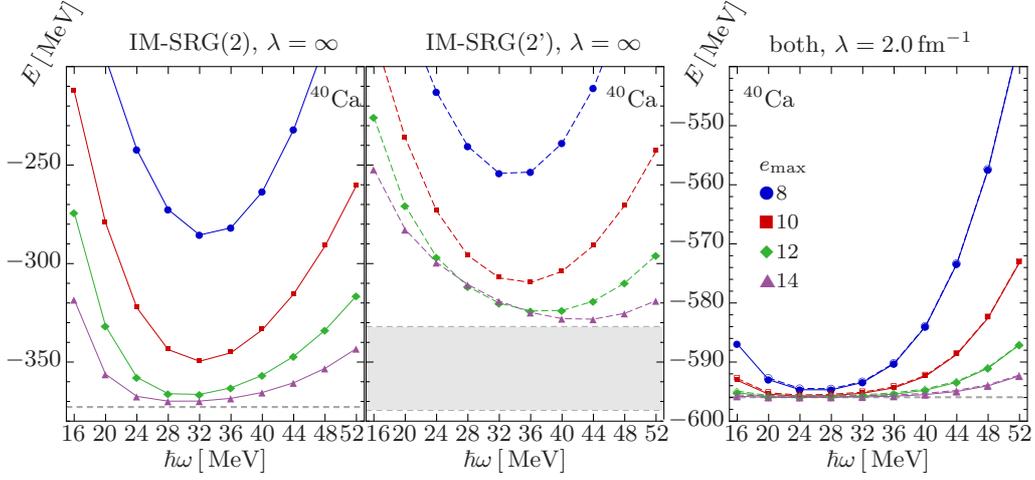

  \setlength{\unitlength}{0.365\textwidth}
  \begin{picture}(2.7390,1.3000)
    \put(-0.0500,0.0000){\small\input{fig/chi2b_srg0000_Ca40_truncIMSRG2.tex}}
    \put(0.7500,0.0000){\small\input{fig/chi2b_srg0000_Ca40_truncIMSRG2p.tex}}
    \put(1.7200,0.0000){\small\input{fig/chi2b_srg0625_Ca40_trunc.tex}}
    \put(0.2000,1.2000){\parbox{0.8\unitlength}{\small\centering IM-SRG(2), $\lambda=\infty$}}
    \put(1.0000,1.2000){\parbox{0.8\unitlength}{\small\centering IM-SRG(2'), $\lambda=\infty$}}
    \put(1.900,1.2000){\parbox{0.8\unitlength}{\small\centering both, $\lambda=2.0\fmi$}}
  \end{picture}
  \caption{\label{fig:chi2b_srgXXXX_2vs2p}
    Comparison of $\nuc{Ca}{40}$ ground-state energies of the regular IM-SRG(2) (solid lines) and perturbative IM-SRG(2')  truncations (dashed lines). The default White generator $\eta^\text{IA}$, Eq.~\eqref{eq:eta_white}, was used in both cases.
    The interaction is the chiral \NNNLO potential with $\lambda=\infty$ (left and center panels) and $\lambda=2.0\,\fm^{-1}$ (right panel), respectively. The dashed lines indicate extrapolated energies. For the IM-SRG(2') truncation, the shaded area indicates the variation from using different data sets for the extrapolation (see text).
  }
\end{figure}

Based on the power counting from Eqs.~\eqref{eq:mbpt_count1} and \eqref{eq:mbpt_count2}, 
an earlier work \cite{Tsukiyama:2011uq} introduced a perturbative truncation which eliminates 
terms of $\OC(g^3)$ from the flow equations \eqref{eq:imsrg2_m0b}--\eqref{eq:imsrg2_m2b}:
\begin{align}
  \totd{E}{s}&= 
    \frac{1}{2} \sum_{abcd}\eta_{abcd}\Gamma_{cdab} n_a n_b\bar{n}_c\bar{n}_d
    \,, \label{eq:imsrg2p_m0b}\\[5pt]
%
  \totd{f_{12}}{s} &= 
  \sum_{a}(1+P_{12})\eta_{1a}f_{a2} \notag\\
  &\hphantom{=}+\sum_{abc}(n_an_b\bar{n}_c+\bar{n}_a\bar{n}_bn_c) (1+P_{12})\eta_{c1ab}\Gamma_{abc2}
  \,,\label{eq:imsrg2p_m1b}\\[5pt]
%
  \totd{\Gamma_{1234}}{s} &= 
  -\sum_{a}\left\{ 
    (1-P_{12})f_{1a}\eta_{a234}
    -(1-P_{34})f_{a3}\eta_{12a4}
    \right\}\notag \\
  &\quad+ \frac{1}{2}\sum_{ab}(1-n_a-n_b)(\eta_{12ab}\Gamma_{ab34}-\Gamma_{12ab}\eta_{ab34})\notag\\
  &\quad-\sum_{ab}(n_a-n_b) (1-P_{12})(1-P_{34})\eta_{b2a4}\Gamma_{a1b3}
    \,.\label{eq:imsrg2p_m2b}
\end{align}
We will refer to this truncation scheme as IM-SRG(2') in the following\footnote{
Note that the labeling was reversed in Ref.~\cite{Tsukiyama:2011uq}, which primarily 
used this perturbative truncation scheme for numerical calculations.}. 

The integration of the IM-SRG(2') flow equation yields a third-order complete energy, 
while certain contributions from fourth order onward are missing. Using the same 
definitions as in Eq.~\eqref{eq:mbpt4_energy}, we find that
\begin{align}
  E^{[4]}_{\text{IM-SRG(2')}} &= E^{[4]}_\Gamma + (E^{[4]}_A - E^{[4]}_f) + E^{[4]}_B \notag\\
   &= \sum_{i=1}^{12}D_i + Q_3+Q_4+Q_5 + \frac{1}{2}\left(Q_1+Q_2+Q_6+Q_7\right) \,,\label{eq:mbpt4_energy_imsrg2p}
\end{align}
i.e., the IM-SRG(2') does not contain the fourth-order singles contribution. This is 
caused by the absence of the single-particle term in the energy flow equation 
\eqref{eq:imsrg2p_m0b}, and the diagrams $A_1$ and $A_2$ from the amplitude 
$\overline{A}$ (see Fig.~\ref{fig:mbpt3_Gamma} and Eq.~\eqref{eq:mbpt3_Gpphh_A}).

In Fig.~\ref{fig:chi2b_srgXXXX_2vs2p}, we compare $\nuc{Ca}{40}$ ground-state 
energies obtained with the regular and perturbative truncations. For the soft 
\NNNLO interaction with $\lambda=2.0\fmi$, shown in the right panel, the two 
truncations give almost identical results. The agreement between ground-state 
energies is on the level of $10^{-4}$ or better, with extrapolated energies for 
$\nuc{Ca}{40}$ differing by only $2\,\keV$. 

For the bare interaction, on the other hand, the truncation schemes behave quite 
differently. The IM-SRG(2) ground-state energy has a quasi-variational convergence 
pattern, which allows us a stable extrapolation to infinite HO basis size. The 
IM-SRG(2') truncation's ground-state energy minimum is still moving to larger $\hw$ 
for the considered bases, indicating a lack of UV convergence, and the variational
pattern breaks down as we increase $\eMax$ from 12 to 14. Extrapolation from different 
subsets of the calculated energies using Eq.~\eqref{eq:def_Eex} produces large 
uncertainties which are indicated by the shaded band in Fig.~\ref{fig:chi2b_srgXXXX_2vs2p}.

\begin{figure}[t]
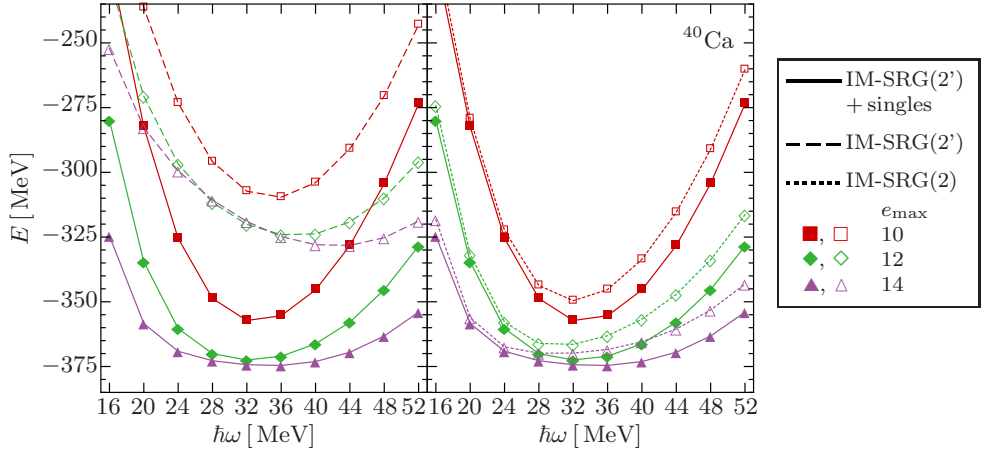

  \setlength{\unitlength}{0.4\textwidth}
  \begin{picture}(2.30000,1.2000)
    \put(0.0500,0.0000){\small\input{fig/chi2b_srg0000_Ca40_IMSRG2p_singles_A.tex}}
    \put(0.8450,0.0000){\small\input{fig/chi2b_srg0000_Ca40_IMSRG2p_singles_B.tex}}
    \put(1.9000,0.7000){\fbox{\parbox{0.4500\unitlength}
      {\footnotesize
        \begin{tabular*}{0.2000\unitlength}{@{}r@{\extracolsep{0.2em}}l}
          \linemediumsolid[black]   & IM-SRG(2')\\
                                    & +\,singles\\[5pt]
          \linemediumdashed[black]  & IM-SRG(2')\\[5pt]
          \linemediumdotted[black]  & IM-SRG(2) 
        \end{tabular*}
        \\
        \begin{tabular*}{0.2000\unitlength}{rl}
          & $\eMax$\\
          {\symbolbox[FGRed]}, {\scriptsize\symbolboxopen[FGRed]} & 10\\
          {\!\!\symboldiamond[FGGreen]}, {\large\symboldiamondopen[FGGreen]} & 12\\
          {\scriptsize\symboltriangle[FGViolet]}, \symboltriangleopen[FGViolet] & 14
        \end{tabular*}
      }}}
  \end{picture}
  \\[-20pt]
  \caption{\label{fig:chi2b_srg0000_Ca40_singles}
    Effect of adding the fourth-order singles (1p1h) contribution (cf.~Eqs.~\eqref{eq:def_E4f}, 
    \eqref{eq:def_E4A} and \eqref{eq:mbpt4_energy_imsrg2p}) to the IM-SRG(2') ground-state 
    energy of $\nuc{Ca}{40}$ (see text). The singles contributions for different $\hw$ were 
    calculated with the initial Hamiltonian $\HO(0)$. All shown results were obtained for 
    the chiral \NNNLO Hamiltonian with $\lambda=\infty$.
  }
\end{figure}

As discussed above, the IM-SRG(2') ground-state energy, Eq.~\eqref{eq:mbpt4_energy_imsrg2p}, 
does not contain the fourth-order singles. In Fig.~\ref{fig:chi2b_srg0000_Ca40_singles}, we
demonstrate that the omission of this contribution accounts for the bulk of the energy 
difference between IM-SRG(2) and IM-SRG(2'), using $\nuc{Ca}{40}$ as an example. Moreover, 
the addition of the fourth-order singles restores the variational behavior of the ground-state 
energy as a function of the single-particle basis size $\eMax$.

\begin{figure}[t]
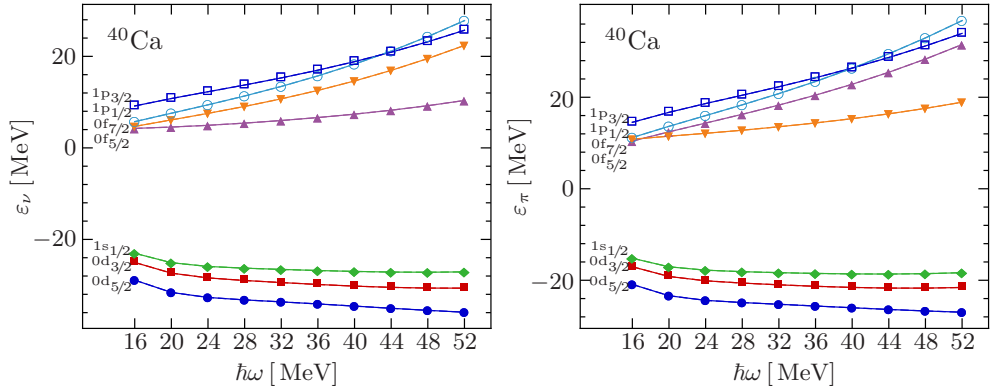

  \setlength{\unitlength}{0.5\textwidth}
  \begin{picture}(2.0000,0.8800)
    \put(-0.0200,0.0000){\small\input{fig/chi2b_srg0000_Ca40_trunc_neut_espe.tex}}
    \put(0.9500,0.0000){\small\input{fig/chi2b_srg0000_Ca40_trunc_prot_espe.tex}}
  \end{picture}
  \vspace{-30pt}
  \caption{\label{fig:chi2b_srg0000_Ca40_espe}
    Effective neutron (left panel) and proton (right panel) single-particle energies of 
    $\nuc{Ca}{40}$ from IM-SRG(2) (solid lines) and IM-SRG(2') (dashed lines) calculations using
    the chiral \NNNLO interaction with $\lambda=\infty$ in an $\eMax=14$ single-particle basis.
  }
\end{figure}

Compared to the regular IM-SRG(2), the IM-SRG(2') flow equations lack $\OC(g^3)$ contractions
of $f$ and $\Gamma$ with the two- and one-body parts of $\eta$, respectively. The effect of
this omission on the two-body matrix element is hard to analyze in greater detail, in part 
due to their sheer number. To test the impact of the missing terms on the flowing one-body 
Hamiltonian, we calculate the Baranger effective single-particle energies (ESPEs) by diagonalizing 
the final $f(\infty)$ in both truncations (see \cite{Baranger:1970bh,Duguet:2012ys,Duguet:2015lq}). 
The neutron and proton $sd-$ and $pf-$shell ESPEs in $\nuc{Ca}{40}$ are shown in 
Fig.~\ref{fig:chi2b_srg0000_Ca40_espe}, and we find that the results obtained with IM-SRG(2) 
and IM-SRG(2') are practically indistinguishable.

We conclude by following up on the perturbative analysis of the difference between 
IM-SRG(2) and CC results with the unevolved chiral \NNNLO Hamiltonian that was begun 
in Sec.~\ref{sec:mbpt_discussion}. In Ref.~\cite{Tsukiyama:2011uq}, the overestimation 
of the $\nuc{He}{4}$ ground-state energy in IM-SRG(2) calculations when compared to 
$\Lambda-$CCSD(T) and exact NCSM results was the main motivation for the 
investigation of the IM-SRG(2') truncation. The IM-SRG(2') result closely
matches the CCSD result for $\nuc{He}{4}$, $-23.98\,\MeV$, but the present discussion
reveals this agreement as accidental, an artifact of the omission of attractive 
fourth-order singles producing a similar change in the ground-state energy as the 
addition of the repulsive quadruples term $E^{[4]}_C$ (see the discussion in 
Sec.~\ref{sec:mbpt_discussion}). While both truncations work equally well for
sufficiently soft, perturbative nuclear Hamiltonians, the IM-SRG(2) truncation remains
well-behaved at higher resolution scales, at the same computational cost, which 
is why we favor this truncation scheme in practical applications.

\section{\label{sec:com}Center-of-Mass Factorization}

\subsection{\label{sec:com_problem}Center-of-Mass Problem in Finite Nuclei}
Nuclear interactions are invariant under translations and Galilean boosts 
(in a non-relativistic formalism), hence the nuclear $A$-body Hamiltonian 
can be split into a sum of center-of-mass and intrinsic parts,
\begin{equation}
  \HO = \Hcm + \Hint\,.
\end{equation}
We can write the free center-of-mass Hamiltonian as
\begin{equation}
  \Hcm = \frac{\POV^2}{2mA}\,,\label{eq:def_Hcm_free}
\end{equation}
and express the intrinsic Hamiltonian in a manifestly symmetry-invariant way as
\begin{align}
  \Hint&= \sum_{i<j} \frac{\qOV_{ij}^2}{2\mu} + \sum_{i<j} V^{(2)}(\vec{\xi}_{ij}) + \sum_{i<j<k}V^{(3)}(\vec{\xi}_{ij},\vec{\xi}_{ijk}) \label{eq:Hint_jacobi}\,.
\end{align}
Here, $\mu$ is the reduced nucleon mass, and we have introduced the relative momenta 
\begin{equation}
  \qOV_{ij} = \frac{1}{2}(\pOV_i - \pOV_j),
\end{equation}
and Jacobi coordinates
\begin{align}
  \vec{\xi}_{ij} &\equiv \rOV_i - \rOV_j\,, \\
  \vec{\xi}_{ijk} &\equiv \frac{1}{2} \left(\rOV_i + \rOV_j \right) - \rOV_k \,.
\end{align}
For the sake of simplicity, we neglect the proton-neutron mass difference as well as 
a possible dependence of the interaction on momentum transfers, which is a common feature
of EFT-based approaches. 

In (very) light nuclei, it is possible to use wave functions that are defined in terms of 
the Jacobi coordinates and solve the nuclear many-body problem while preserving translational
invariance explicitly (see, e.g., \cite{Navratil:2000hf}). However, the proper 
antisymmetrization of such wave functions becomes extremely difficult as the number of 
particles increases \cite{Navratil:2000hf}. Moreover, the very nature of the Jacobi
coordinates prevents us from referring to an independent-particle description of the 
nucleus when we set up the basis for our many-body Hilbert space.

Alternatively, we can work with single-particle coordinates and wave functions, or
the corresponding creation and annihilation operators in second quantization. In this
case, the intrinsic Hamiltonian can be written as
\begin{align}
  \Hint &= \left(1-\frac{1}{A}\right)\sum_{i} \frac{\pOV_i^2}{2m} + 
            \sum_{i<j} \left( V^{(2)}(\rOV_i,\rOV_j) - \frac{2}{Am} \pOV_i\cdot\pOV_j \right)
            + \sum_{i<j<k}V^{(3)}(\rOV_i,\rOV_j,\rOV_k) \label{eq:Hint}\,.
\end{align}
Antisymmetrized many-body states are readily constructed in such an approach, even for a 
large number of particles. Translational invariance could be achieved by using single-particle 
plane wave states, but such a basis is ill-suited to describe finite objects like 
nuclei. The natural and convenient choice are localized single-particle bases, which provide 
good convergence for nuclear bound states. The drawback is that even an infinitely large 
localized basis cannot produce a many-body wave function whose center-of-mass 
part is a plane wave, because plane waves states are not contained in the Hilbert space of 
square-integrable wave functions\footnote{To properly work with such states, 
a rigged Hilbert space must be introduced (see, e.g., \cite{Madrid:2005rt} and references 
therein).}. 

In practical calculations, we can only work with finite localized single-particle bases 
in any case. This automatically means that the center-of-mass is artifically 
localized, as if it were 
moving in an effective potential whose range is controlled by the basis 
parameters\footnote{Similar observations are used to construct extrapolations of bound-state 
energies to infinite harmonic oscillator bases in 
Refs.~\cite{Furnstahl:2012ys,More:2013bh,Furnstahl:2014vn,Wendt:2015zl}.}. For nuclear
bound states, the individual nucleons are clustered into a nucleus in the vicinity of the 
origin of the chosen coordinate system, and the center-of-mass will be localized in the 
nucleus' interior (see Sec.~\ref{sec:com_localization}).

Unfortunately, the energy scales of center-of-mass motion and intrinsic nuclear excitations are of
comparable size. When we solve the eigenvalue problem for the intrinsic Hamiltonian,
we obtain spurious copies of the intrinsic spectrum that are built on different 
center-of-mass states, and this has to be taken into account when we analyze
excitations or transitions.

\subsection{Factorization of Center-of-Mass and Intrinsic Wave Functions}

Since the center-of-mass and relative degrees of freedom of the nuclear Hamiltonian
are independent (see Eqs.~\eqref{eq:def_Hcm_free},~\eqref{eq:Hint_jacobi}), we have
\begin{equation}
  \comm{\Hcm}{\Hint}=0\,,\label{eq:comm_Hcm_Hint}
\end{equation}
and many-body states factorize into center-of-mass and intrinsic parts:
\begin{equation}
  \ket{\Psi} = \ket{\Psi_\text{cm}}\otimes\ket{\Psi_\text{int}}\,.
\end{equation}
Note that we can replace $\Hcm$ by any operator which is exclusively defined in terms of 
$\POV$ and the center-of-mass coordinate $\ROV$ without spoiling the 
factorization of the many-body state.

While Eq.~\eqref{eq:comm_Hcm_Hint} holds on the operator level, there are aspects
of practical many-body calculations that can violate this condition, and spoil the
factorization of the center-of-mass and intrinsic wave functions. First, and maybe
foremost, is the use of a finite Hilbert space. Let us introduce an operator $\PC$ 
that projects the Hamiltonian on a truncated model space ($\PC$-space). Factorization 
of the center-of-mass and intrinsic wave functions in the model space is achieved if 
\begin{align}
  \comm{\PC\Hcm\PC}{\PC\Hint\PC}=
    \PC\Hcm\PC\Hint\PC - \PC\Hint\PC\Hcm\PC=
    0\,,\label{eq:commP_Hcm_Hint}.
\end{align}
This condition is satisfied if either 
\begin{equation}
  \comm{\Hcm}{\PC}=0 \label{eq:comm_Hcm_P}
\end{equation}
and/or 
\begin{equation}
  \comm{\Hint}{\PC}=0\,. \label{eq:comm_Hint_P}
\end{equation}
Equations \eqref{eq:comm_Hcm_P} and \eqref{eq:comm_Hint_P} imply that the $\PC$-space
is constructed from complete ei\-gen\-spaces of either (or both) of the operators. Because
the eigenspaces of $\Hint$ are not known a priori, it is only practical to focus on
satisfying the condition \eqref{eq:comm_Hcm_P}, and the aforementioned freedom of
choice for $\Hcm$ becomes a valuable tool in this case. In the next subsection, we
will discuss how this leads to the concept of the $\Nmax$-complete model space. Such
spaces are used by the No-Core Shell Model.

Second, effective Hamiltonians and operators are introduced in either an explicit or 
an implicit fashion in most many-body approaches. In the traditional Shell Model with a 
core, there is a projection $\PC$ onto the finite space as discussed above, followed by
a subsequent splitting in a $\PC'$ valence and $\QC$ excluded space, and usually,
neither of these projections is compatible with the condition \eqref{eq:commP_Hcm_Hint}.
In IM-SRG or the Coupled Cluster method, the effective operators are obtained through 
unitary or similarity transformations, e.g.,
\begin{alignat}{2}
  \HO(s) &= \UO(s)H\UUO(s)\,,\qquad& \UO(s)\UUO(s) = 1\,,\\
  \HO_\text{CC} &= \UO H\UO^{-1}\,,\qquad& \UO\UO^{-1} = 1\,.
\end{alignat}
Equation \eqref{eq:comm_Hcm_Hint} will be invariant under these tranformations as
long as unitarity is preserved, or the transformation is invertible, respectively.
Formally, it would even be possible to achieve this in the presence of rank or 
basis truncations in the IM-SRG generator or CC cluster operators, if we could
work with the representations of these objects in the full $A$-body Hilbert space
basis. However, the whole point of methods like IM-SRG and CC is to not construct
such representations. Practical evaluations of effective operators or wave function
properties violate unitarity due to the truncation of induced operators. In IM-SRG(2),
for instance,
\begin{equation}
  \comm{\Hcm(s)}{\Hint(s)} = \comm{\Hcm^{(1)}(s)+\Hcm^{(2)}(s)}{\Hint^{(1)}(s)+\Hint^{(2)}(s)} 
  \neq 0
\end{equation}
and there are non-vanishing terms which couple the center-of-mass and intrinsic 
coordinates. In Sec.~\ref{sec:com_approxfactor}, we will present numerical evidence
that this coupling is weak, and approximate factorization holds.

\subsection{\texorpdfstring{$\Nmax$}{Nmax}-Complete Spaces}
The diagnostics for center-of-mass factorization are inspired by considerations for
so-called $\Nmax$-complete model spaces, which are used, e.g., in the No-Core Shell
Model (NCSM). The basis of such spaces consists of Slater determinants that are
constructed by distributing $A$ nucleons over HO single-particle states in all
possible ways, subject to the (energy) truncation 
\begin{equation}\label{eq:def_Nmax}
  \sum_{i=1}^Ae_i\leq \Nmax\,.
\end{equation}
Here, $e_i$ are the energy quantum numbers characterizing the eigenstates of the HO
single-particle Hamiltonian
\begin{equation}
  \hO_i \equiv \frac{\pOV^2_i}{2m} + \frac{1}{2}m\omega^2\rOV_i^2 \,.
\end{equation}
The Slater determinants themselves are eigenstates of a system of $A$ oscillators that
is described by the Hamiltonian
\begin{align}
  \HO^\text{HO} &\equiv \sum_{i=1}^A \frac{\pOV_i^2}{2m} + \frac{1}{2}\sum_{i=1}^Am\omega^2\rOV^2_i\notag\\
                &=\sum_{i=1}^A \frac{\pOV_i'^2}{2m} + \frac{1}{2}\sum_{i=1}^Am\omega^2\rOV'^2_i + \frac{\POV^2}{2Am} + \frac{1}{2}Am\omega^2\ROV^2\,, \label{eq:HHO}
\end{align}
where we have switched to the center-of-mass and intrinsic coordinates:
\begin{equation}
  \rOV_i' \equiv \rOV_i - \ROV\,,\quad \pOV'_i \equiv \pOV_i - \frac{1}{A}\POV\,.
\end{equation}
It is a unique feature of the harmonic oscillator that the intrinsic and center-of-mass 
Hamiltonians for systems of many oscillators are oscillators themselves. This property
makes it possible to define the well-known Talmi-Moshinsky transformation 
\cite{Moshinsky:1959cp,Tobocman:1981uq,Kamuntavicius:2001kr}
between single-particle based and center-of-mass/intrinsic HO many-body states, which is 
heavily employed in the calculation of interaction matrix elements in nuclear 
physics. 

Noting that
\begin{equation}
  \Hcm^\text{HO} \equiv \frac{\POV^2}{2Am} + \frac{1}{2}Am\omega^2\ROV^2 = 
  \frac{1}{A}\sum_{i}\hO_i + \sum_{i<j}\left(\frac{1}{Am}\pOV_i\cdot\pOV_j 
  + \frac{m\omega^2}{A}\rOV_i\cdot\rOV_j\right)\,,
\end{equation}
and using the relation
\begin{equation}
  \comm{\rOV^2_i}{\sum_{j<k}\pOV_j\cdot\pOV_k} 
    = i \sum_{j}^{j\neq i}\left(\rOV_i\cdot\pOV_j + \pOV_j\cdot\rOV_i\right)
     = - \comm{\pOV_i^2}{\sum_{j<k}\rOV_j\cdot\rOV_k}\,, 
\end{equation}
it is easy to show that
\begin{equation}
  \comm{\HO^\text{HO}}{\Hcm^\text{HO}} = \comm{\Hint^\text{HO}}{\Hcm^\text{HO}} = 0\,,
\end{equation}
i.e., the $A$-nucleon Slater determinants are shared eigenstates of $\HO^\text{HO},
\Hcm^\text{HO}$, and $\Hint^\text{HO}$. We can now define a projection operator
\begin{equation}\label{eq:def_P_NCSM}
  \PC_\text{NCSM} \equiv 
    \sum_{\alpha}^{N_\alpha\leq \Nmax}\ketbra{\alpha}{\alpha}\,,
    \quad N_\alpha \equiv \sum_{i=1}^A e^\alpha_i\,,
\end{equation}
where $e^\alpha_i$ denotes the HO energy quantum number of particle $i$ in the
Slater determinant labeled by $\ket{\alpha}$. $\PC_\text{NCSM}$ is constructed
from dyadic products of eigenstates of $\Hcm^\text{HO}$, hence it is clear that
\begin{equation}\label{eq:comm_Hcm_P_HO}
  \comm{\Hcm^\text{HO}}{\PC_\text{NCSM}} = 0\,.
\end{equation}

Recall from the previous section that $\Hcm$ can be chosen freely if our goal is
to solve the eigenvalue problem of the intrinsic Hamiltonian and work in an
untruncated many-body Hilbert space. Gloeckner and Lawson originally suggested the
use of
\begin{equation} \label{eq:lawson}
  \beta \Hcm \equiv \beta\left(\Hcm^\text{HO} - \frac{3}{2}\hbar\omega\right)
\end{equation}
for the traditional nuclear Shell Model with a core \cite{Gloeckner:1974gb}, and
this is also the natural choice for the NCSM. Together with the use of an $\Nmax$-complete
model space defined through Eq.~\eqref{eq:def_P_NCSM}, this modification ensures
that Eq.~\eqref{eq:comm_Hcm_P_HO} is satisfied. Consequently, the factorization
of the center-of-mass and intrinsic states is preserved in the truncated space,
as discussed in Sec.~\ref{sec:com_problem}. Spurious center-of-mass excitations
that appear in the spectrum of $\HO_\text{NCSM}$ can be shifted to arbitrary high
energies (``projected'' out) through a suitable choice of the parameter $\beta$.

\subsection{\label{sec:com_approxfactor}Approximate Factorization for IM-SRG}
\addtocontents{toc}{\protect\setcounter{tocdepth}{2}}

The energy truncation used to define $\Nmax$-complete model spaces in the previous
section is only appropriate for Shell Model-like diagonalization approaches. A much
larger class of many-body methods can only be defined with a pure single-particle
basis truncation: This class encompasses Hartree-Fock and beyond mean-field methods
like Many-Body Perturbation Theory (MBPT), truncated Configuration Interaction (CI), 
Self-Consistent Green's Functions (SCGF), Coupled Cluster (CC), and the IM-SRG. As discussed
repeatedly throughout this work, a major appeal of these methods is their polynomial
scaling with the single-particle basis, which greatly extends the range of nuclei
they can be applied to. In general, the basis and many-body truncations employed by
these methods do not satisfy the conditions \eqref{eq:comm_Hcm_P} and \eqref{eq:comm_Hint_P},
and the center-of-mass and intrinsic degrees of freedom of the many-body wave function
do not factorize exactly. Thus, we need to investigate the impact of this imperfect
factorization on our IM-SRG results.

\subsubsection{Characteristic Energy Scale of the Center-of-Mass Hamiltonian}

Most studies of center-of-mass contaminations in the literature make use of Lawson's
method as sketched in the previous section, i.e., the Hamiltonian \eqref{eq:lawson} is
added to $\Hint$ and the sensitivity of ground- and excited-state energies to variations
of the parameter $\beta$ is probed \cite{Gloeckner:1974gb,Roth:2009oh,Hagen:2009fk,Hagen:2010uq}.
The authors of Ref.~\cite{Hagen:2009fk} pointed out that the choice of the oscillator
frequency $\omega$ in Eq.~\eqref{eq:lawson} is not obvious, and that one may overestimate
the center-of-mass contamination of states by using the frequency of the underlying
single-particle basis. Instead, they propose that the true scale for center-of-mass
excitations is given by $\hbar\widetilde{\omega}$, where $\widetilde{\omega}$ is a
function of the basis frequency $\omega$. 

\begin{figure}[t]
  \setlength{\unitlength}{0.6\textwidth}
  \begin{center}
  \begin{picture}(1.0000,1.0500)
    \put(-0.0500,0.4000){\input{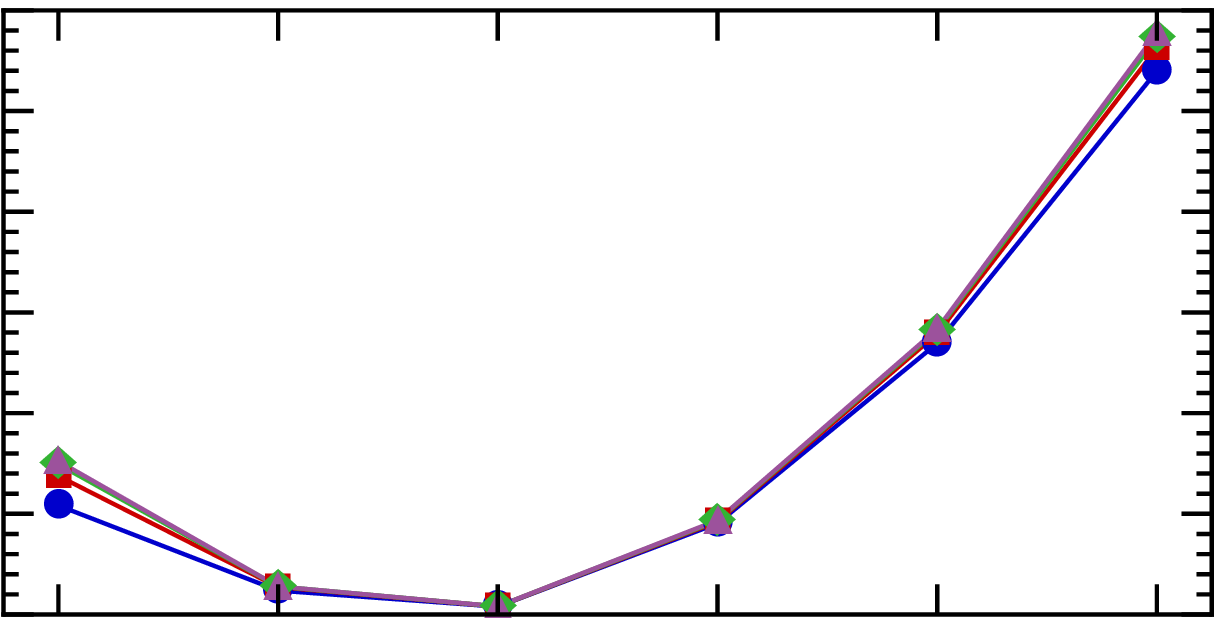}}
    \put(-0.0500,0.0000){\input{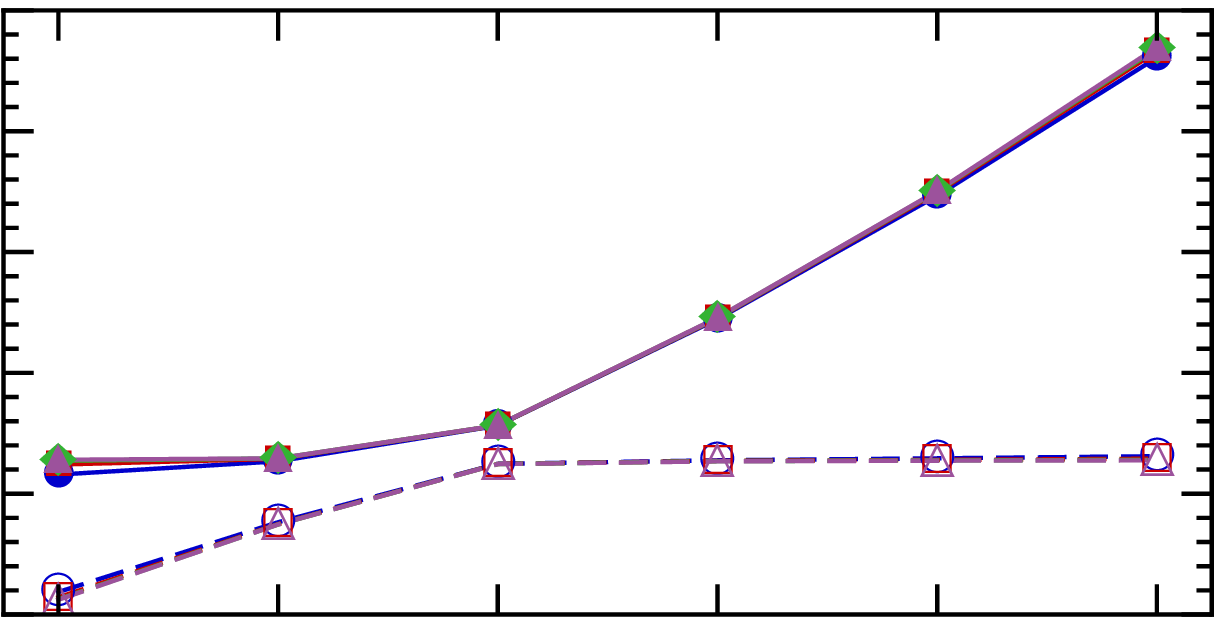}}
  \end{picture}
  \end{center}
  \vspace{-20pt}
  \caption{\label{fig:hwBar}Determination of the common center-of-mass oscillator
  frequency $\widetilde{\omega}$ (see text) for $\nuc{He}{4}$ for the chiral \NNNLO
  interaction with $\lambda=2.0\,\fmi$: Expectation value of the center-of-mass
  Hamiltonian $\Hcm(\omega)$ defined through Eq.~\eqref{eq:lawson} in the IM-SRG(2)
  wave function (upper panel), and calculated $\hwBar^\pm$ (lower panel), as functions
  of the single-particle basis size $\eMax$.}
\end{figure}

To derive $\widetilde{\omega}$, it is assumed that the many-body wave function factorizes
approximately, 
\begin{equation}\label{eq:wf_factorized}
  \ket{\Psi} = \ket{\Psi_\text{cm}(\widetilde{\omega})}\otimes\ket{\Psi_\text{int}}\,,
\end{equation}
and that the center of mass is in the ground state of a common harmonic oscillator
Hamiltonian $\Hcm(\widetilde{\omega})$. From the potential term, one can obtain the identity
\begin{equation}
  \frac{1}{2}mA\omega^2\ROV^2 = \Hcm(\omega) + \frac{3}{2} \hw - \TO_\text{cm} 
    = \frac{\omega^2}{\widetilde{\omega}^2}\left(\Hcm(\widetilde{\omega}) 
    + \frac{3}{2}\hbar\widetilde{\omega} - \TO_\text{cm}\right)\,.
\end{equation}
Taking the expectation value in the state \eqref{eq:wf_factorized}, one has
\begin{equation}
  \expect{\TO_\text{cm}} = \frac{3}{4}\hbar\widetilde{\omega}
\end{equation}
due to the virial theorem and $\expect{\Hcm(\widetilde{\omega})}\approx0$ because the center
of mass is supposed to be in the ground state. Defining
\begin{equation}
  \Ecm(\omega)\equiv \expect{\Hcm(\omega)}\neq 0
\end{equation}
one can solve for $\widetilde{\omega}$, and obtain the non-trivial solutions
\begin{equation}\label{eq:def_hwBar}
  \hwBar^\pm \equiv \hw + \frac{2}{3}\Ecm(\omega)\pm
    \sqrt{\frac{4}{9}\Ecm^2{\omega} + \frac{4}{3}\hw \Ecm(\omega)}\,.
\end{equation}

In the upper panel of Fig.~\ref{fig:hwBar}, we show the expectation values $\Ecm(\omega)$
for $\nuc{He}{4}$, calculated in IM-SRG(2) by evolving the operator $\Hcm(\omega)$ along
with the full Hamiltonian (cf.~Sec.~\ref{sec:observables}). While $\Ecm(\omega)$ is very
small in the vicinity of the basis parameter $\hw$ that minimizes the IM-SRG(2) ground-state
energy, this expectation value converges to non-zero values at other $\hw$ as the
single-particle basis grows. Applying Eq.~\eqref{eq:def_hwBar}, we obtain the oscillator
parameters $\hwBar^\pm$ shown in the lower panel of Fig.~\ref{fig:hwBar}. We find that the
parameters converge towards a value $\hwBar\approx23\,\MeV$, so that 
$\Ecm(\widetilde{\omega}) \to 0$, and the assumption of approximate factorization is
justified. Consequently, $\hwBar$ and not $\hw$ is the scale for center-of-mass excitations.

\subsubsection{\label{sec:com_localization}The Complementary View: Center-of-Mass Localization}

\begin{figure}[t]
  \setlength{\unitlength}{0.35\textwidth}
  \begin{center}
  \begin{picture}(2.8571,1.0000)
    \put(0.0000,0.0000){\input{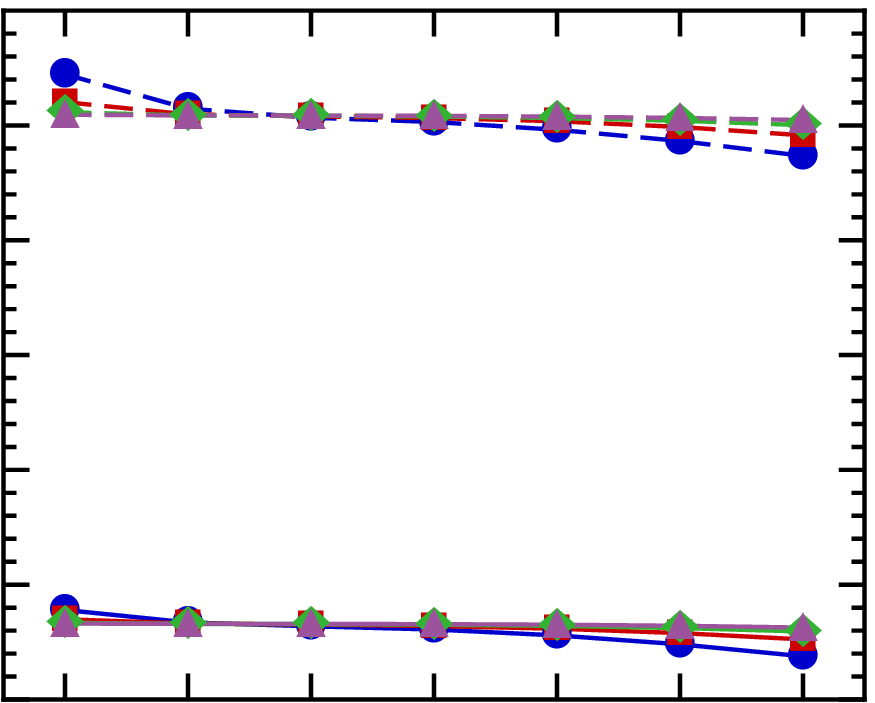}}
    \put(0.9200,0.0000){\input{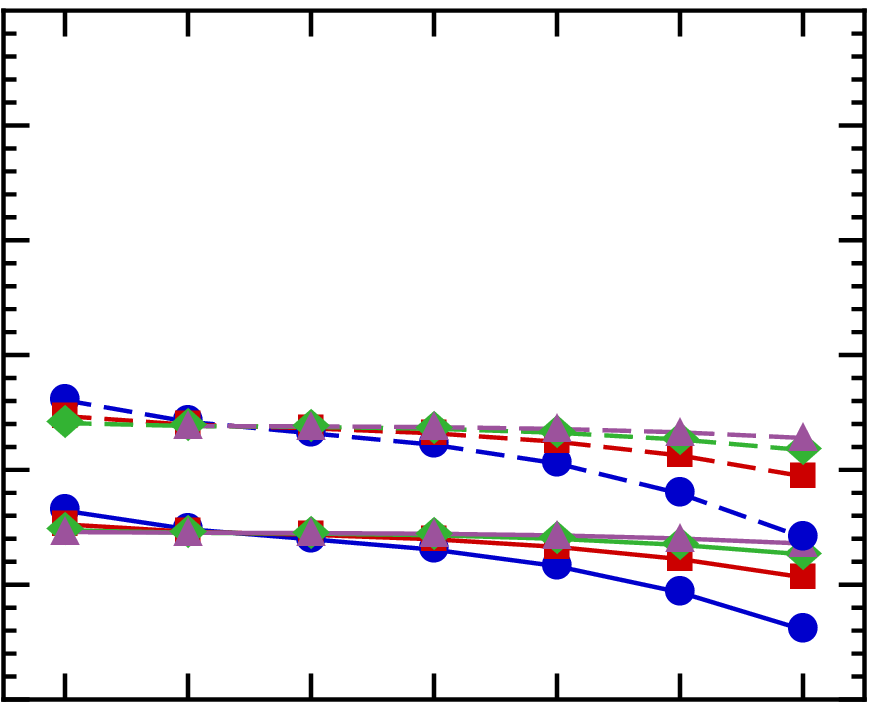}}
    \put(1.8400,0.0000){\input{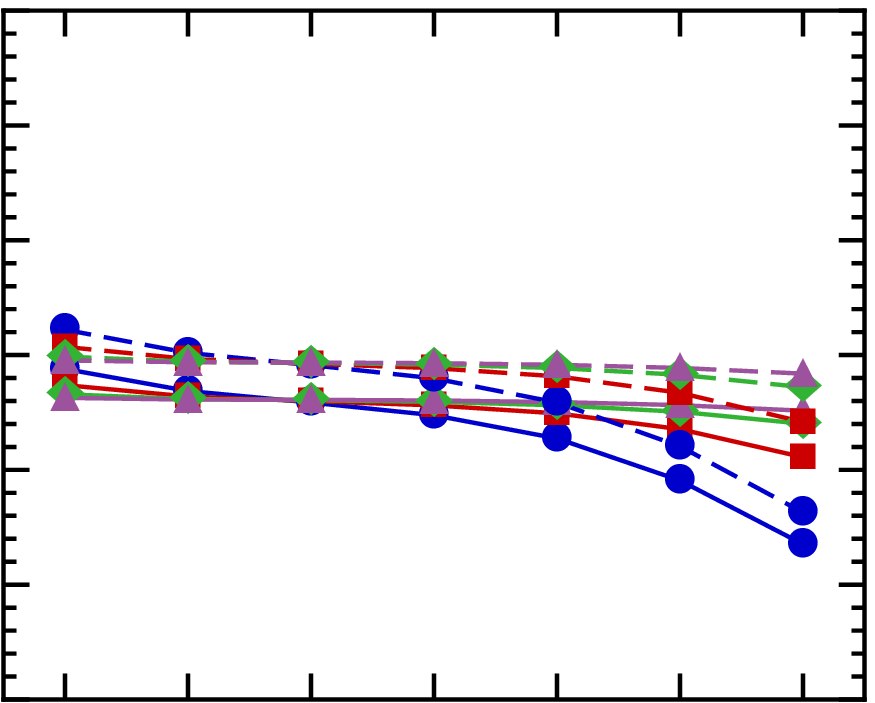}}
  \end{picture}
  \end{center}
  \vspace{-20pt}
  \caption{\label{fig:com_radii}Convergence of the total (dashed lines) and intrinsic
  mean-square radii (solid lines), as a function of the single-particle basis size
  $\eMax$, for a chiral \NNNLO interaction with $\lambda=2.0\,\fm^{-1}$. }
\end{figure}
As discussed in the introduction to this section, the choice of a truncated, localized
single-particle basis leads to an explicit breaking of the translational symmetry of 
the nuclear Hamiltonian. A translationally invariant wave function would be given by
the product of the intrinsic many-body wave function and a plane-wave state for the
center of mass,
\begin{equation}
  \ket{\Psi}=\ket{\KOV}\otimes\ket{\Psi_\text{int}}\,,\quad \braket{\ROV}{\KOV}=\frac{1}{(2\pi)^{3/2}}e^{i\KOV\cdot\ROV}\,.
\end{equation}
In this state, the expectation value of the squared center-of-mass position operator
is infinite, 
\begin{equation}
  \matrixe{\Psi}{\ROV^2}{\Psi} = \infty\,,
\end{equation}
while a localized center-of-mass wave function would clearly yield a finite value. 

In Sec.~\ref{sec:observables}, we have discussed the calculation of observables through
their evolution along with the nuclear Hamiltonian, focusing on the intrinsic radii as
a special case. From the intrinsic and total mean-square radii, we can calculate the
expectation value of $\ROV^2$, because
\begin{align}
  \ROV^2 &= \frac{1}{A^2}\sum_{ij}\rOV_i\cdot\rOV_j = \frac{1}{A^2}\sum_i\rOV^2_i 
            + \frac{2}{A^2}\sum_{i<j}\rOV_i\cdot\rOV_j\notag\\
         &= \frac{1}{A}\sum_{i}\rOV_i^2 
            - \frac{1}{A}\left(\left(1-\frac{1}{A}\right)\sum_{i}\rOV_i^2 
            - \frac{2}{A}\sum_{i<j}\rOV_i\cdot\rOV_j\right)\notag\\
         &= R_\text{ms} - R_\text{ms,int}\,.
\end{align}
In Fig.~\ref{fig:com_radii}, we show the calculated $R_\text{ms}$ and $R_\text{ms,int}$
of $\nuc{He}{4}$, $\nuc{O}{16}$, and $\nuc{Ca}{40}$ as a function of the single-particle
basis size $\eMax$. Since we used a softened \NNNLO interaction with $\lambda=2.0\,\fmi$,
the radii of the heavier nuclei are considerably smaller than observed experimentally,
but this does not matter for the present discussion. Note that the difference between total
and intrinsic mean-square radii, i.e., the correction due to the center-of-mass motion,
is largest for $\nuc{He}{4}$, and decreases rapidly as $A$ grows. We also point out that
the obtained expectation values are well-converged with respect to $\eMax$. 

\begin{figure}[t]
  \setlength{\unitlength}{0.38\textwidth}
  \begin{center}
  \begin{picture}(2.6315,1.1000)
    \put(0.0000,0.0000){\input{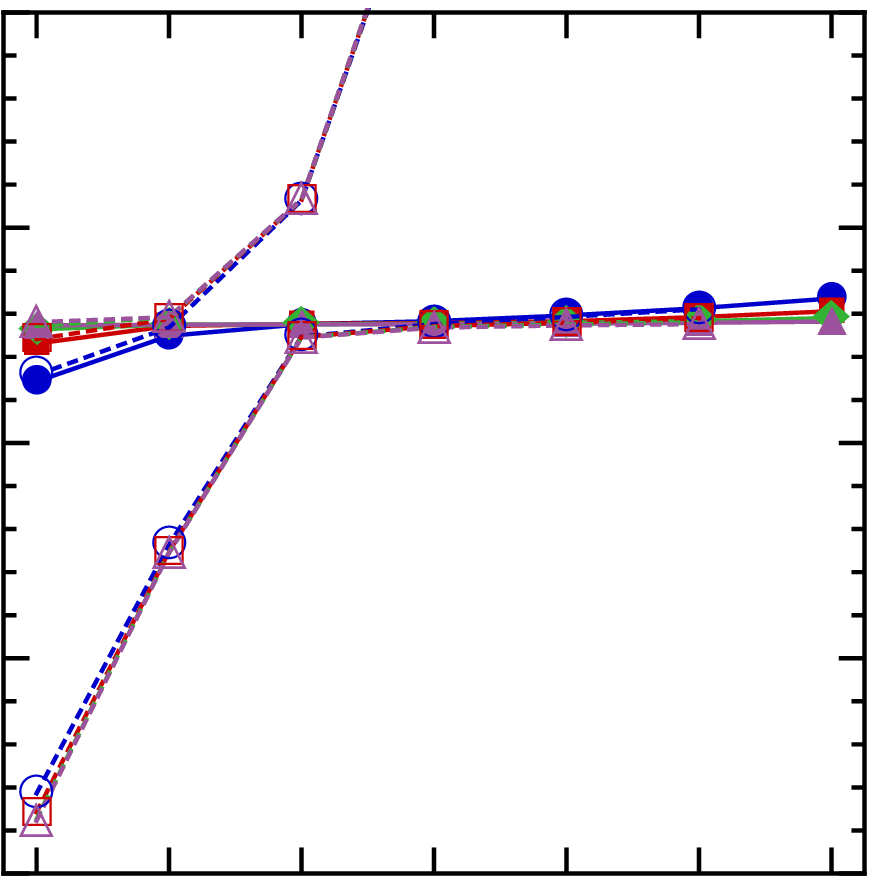}}
    \put(0.3400,0.1400){\setlength{\unitlength}{0.24\textwidth}\footnotesize\input{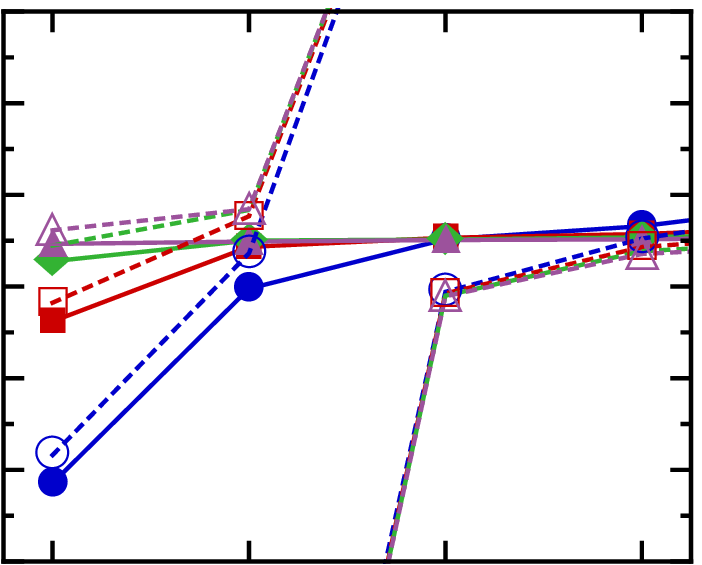}}
    \put(0.8000,0.0000){\input{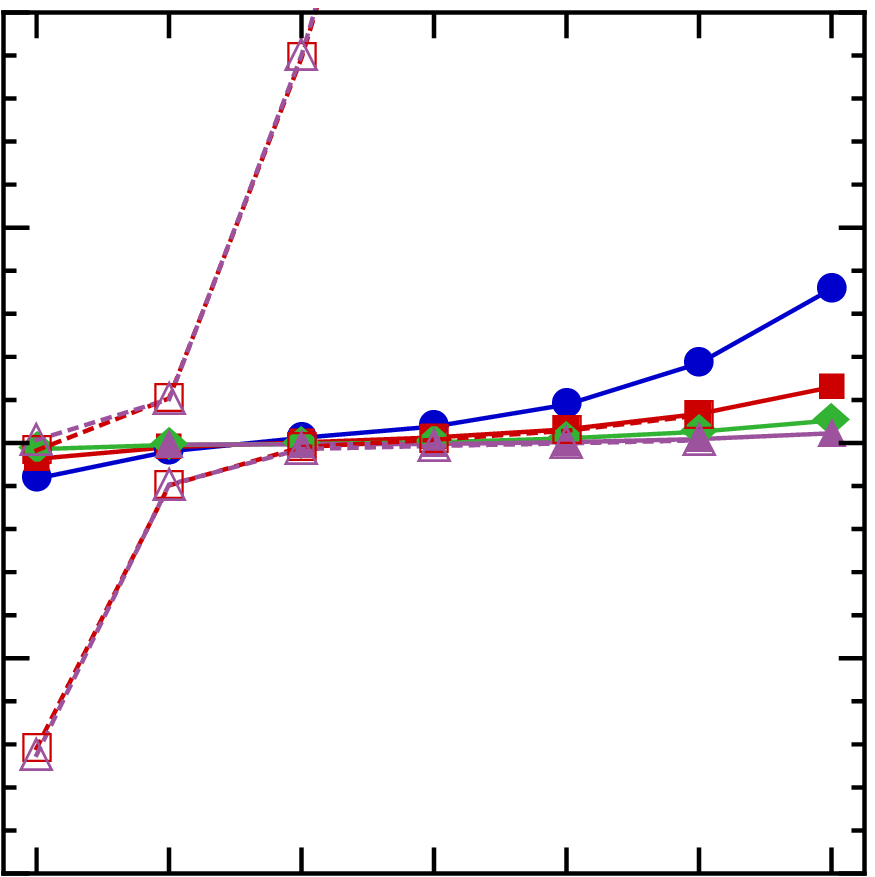}}
    \put(1.6000,0.0000){\input{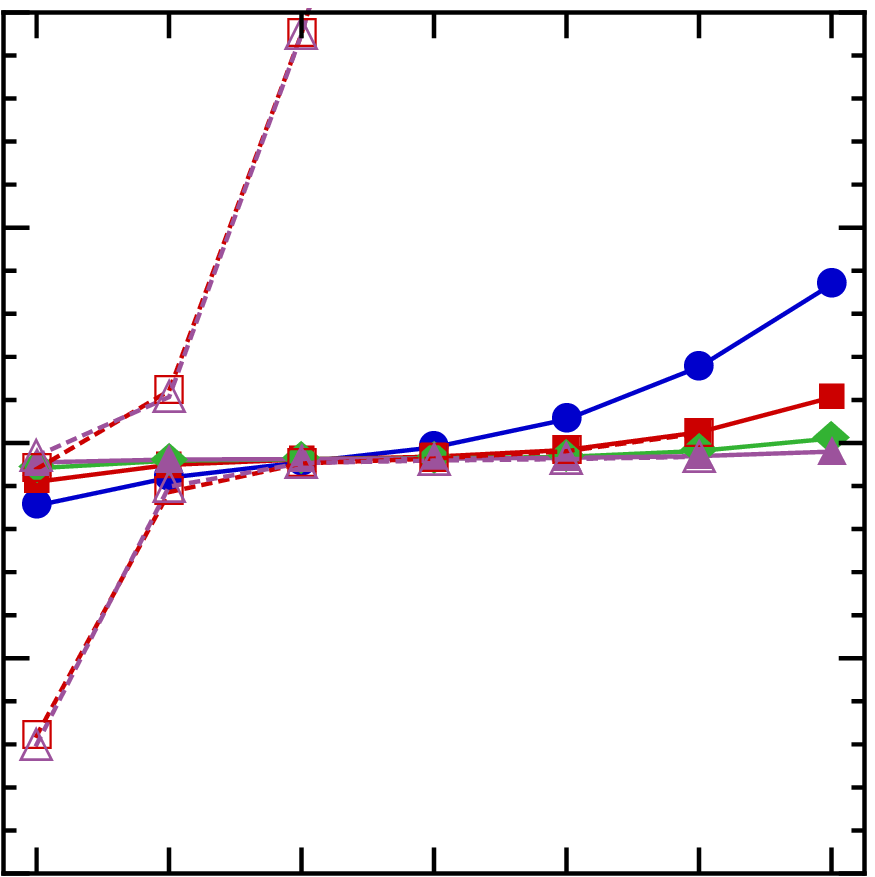}}
  \end{picture}
  \end{center}
  \vspace{-20pt}
  \caption{\label{fig:com_hwOpt}Optimal oscillator frequencies $\hwBar$ as a function of the single-particle basis size $\eMax$ (solid lines and symbols). The two branches of the construction according to Hagen et al.~\cite{Hagen:2009fk,Hagen:2010uq}, $\hwBar^{\pm}$ are indicated by dotted lines with open symbols. A chiral \NNNLO interaction with $\lambda=2.0\,\fm^{-1}$ was used for all calculations.}
\end{figure}

Let us assume now that the center-of-mass wave function is a Gaussian wave packet of
finite width $b$,
\begin{equation}
  \braket{\ROV}{\Psi_\text{loc}} = \left(\frac{1}{2\pi b^2}\right)^{3/4}e^{-\ROV^2/4b^2}\,.
\end{equation}
The origin of the center-of-mass coordinate system is chosen to coincide with that of the
localized single-particle basis. Now we have 
\begin{equation} \label{eq:gauss_R2}
  \expect{\ROV^2} = 3 b^2\,,
\end{equation}
and we can determine the widths from the calculated expectation values $\expect{\ROV^2}$.
So far, we have not made any assumptions aside from the fact that the center of mass, and
thereby the nucleus itself, are localized in space, e.g., in a laboratory experiment. Of
course, a center-of-mass potential provides a means to describe the necessary boundary
conditions for a localized wave packet in the stationary Schr\"odinger equation, and the
spherical HO potential, in particular, is known to have a Gaussian ground-state wave
function. This allows us to make a connection with the construction of the common oscillator
frequency of Hagen et al.~\cite{Hagen:2009fk} that was discussed in the previous subsection.
The relation between the width $b$ and the HO oscillator length $a$ is easily found to be
$a=\sqrt{2}b$. Combining Eq.~\eqref{eq:gauss_R2} with the relation between $a$ and $\hw$,
we obtain
\begin{equation}\label{eq:def_hwBar_rsq}
  \hwBar \equiv \frac{\hbar^2}{mA a^2} = \frac{\hbar^2}{2mA b^2} = \frac{3 \hbar^2}{2 m A \expect{\ROV^2}}\,.
\end{equation}

In Fig.~\ref{fig:com_hwOpt}, we show the values of $\hwBar$ calculated with
Eq.~\eqref{eq:def_hwBar_rsq} for $\nuc{He}{4}$, $\nuc{O}{16}$, and $\nuc{Ca}{40}$.
The dotted lines indicate the solutions $\hwBar^\pm$ of Eq.~\eqref{eq:def_hwBar}. We see that
the present construction exactly matches $\hwBar^+$ ($\hwBar^-$) for small (large) values
of the basis parameter $\hw$, and interpolates between the two solutions in the intermediate
region where they diverge.

\subsubsection{Numerical Analysis}
\begin{figure}[t]
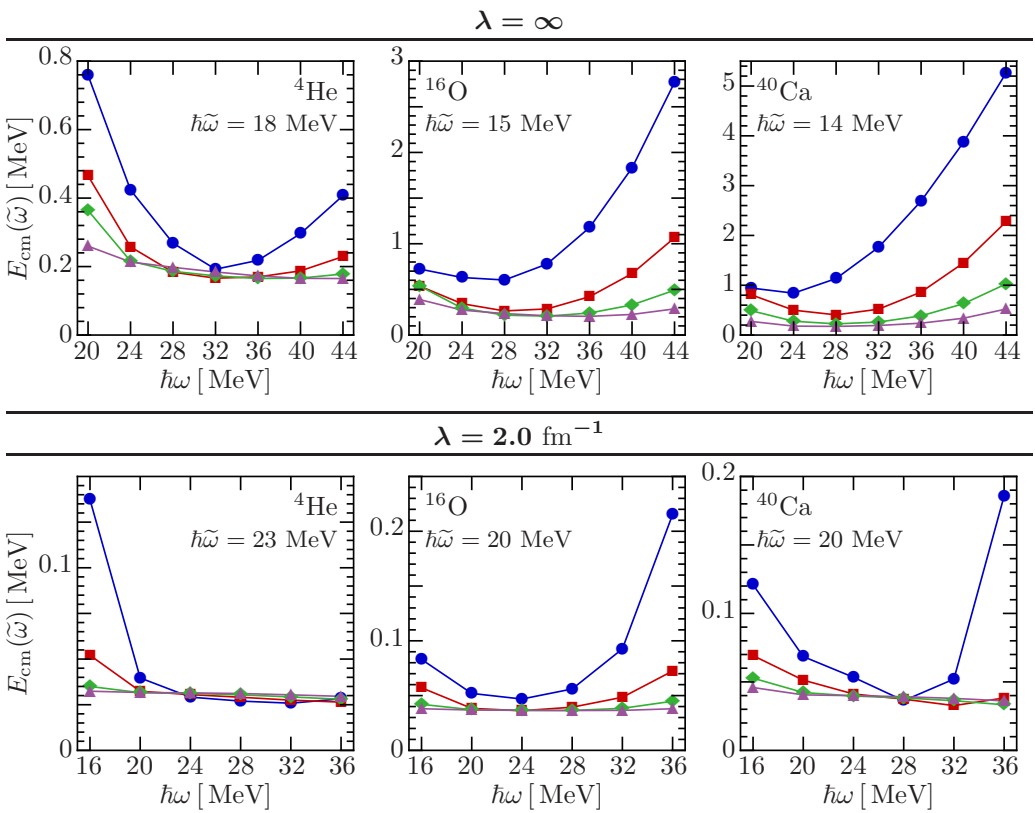

  \setlength{\unitlength}{0.34\textwidth}
  \begin{picture}(2.9411,2.4000)
    \put(0.0000,2.3000){\parbox{\textwidth}{\hrulefill\\\centering\boldmath$\lambda=\infty$\\[-8pt]\hrulefill}}
    \put(0.0000,1.2000){\input{fig/chi2b_srg0000_He4_CoM_Opt.tex}}
    \put(0.9500,1.2000){\input{fig/chi2b_srg0000_O16_CoM_Opt.tex}}
    \put(1.9000,1.2000){\input{fig/chi2b_srg0000_Ca40_CoM_Opt.tex}}
    \put(0.0000,1.1000){\parbox{\textwidth}{\hrulefill\\\centering\boldmath$\lambda=2.0\,\fmi$\\[-8pt]\hrulefill}}
    \put(0.0000,0.0000){\input{fig/chi2b_srg0625_He4_CoM_Opt.tex}}
    \put(0.9500,0.0000){\input{fig/chi2b_srg0625_O16_CoM_Opt.tex}}
    \put(1.9000,0.0000){\input{fig/chi2b_srg0625_Ca40_CoM_Opt.tex}}
  \end{picture}
  \caption{\label{fig:chi2b_srgXXXX_CoM_opt} Center-of-Mass energies 
  $E_\text{cm}(\widetilde\omega)$ of $\nuc{He}{4}$, $\nuc{O}{16}$, and $\nuc{Ca}{40}$
  as a function of the single-particle basis size $\eMax$, for the ``bare'' ($\lambda=\infty$,
  top panels) and SRG-evolved chiral \NNNLO interactions ($\lambda=2.0\,\fm^{-1}$, bottom panels). 
  }
\end{figure}

Let us now analyze the expectation value of the center-of-mass Hamiltonian with the
common frequency $\widetilde{\omega}$ that is determined by either of the methods
described in the previous subsections. In Fig.~\ref{fig:chi2b_srgXXXX_CoM_opt}, we
show these expectation values for the doubly magic nuclei $\nuc{He}{4}, \nuc{O}{16}$,
and $\nuc{Ca}{40}$. In contrast to Fig.~\ref{fig:hwBar}, we now observe that $\Ecm(\widetilde\omega)$
is converging towards a constant value instead of a parabolic function as $\eMax$ grows,
albeit in a not entirely variational manner (which is not required). 

For the unevolved chiral \NNNLO interaction, shown in the top panels, the expectation
values are in the $200-300\,\keV$ range in the largest bases. The non-zero expectation
value is caused by admixtures of center-of-mass excitations in the ground-state wave
function, which are present due to the imperfect factorization. Since the characteristic
scale of these excitations is $\hwBar$, excited states contribute multiples of $\hwBar$
to $\Ecm(\widetilde\omega)$. Considering the orthogonality of center-of-mass eigenstates
and the (semi-)positive definiteness of $\Ecm(\widetilde{\omega})$, we can estimate the
size of the excited-state admixture as $\Ecm(\widetilde\omega)/\hwBar \lesssim 2\%$.

The non-negligible size of the center-of-mass contamination is in part due to the ``slow''
convergence of the ground-state wave function for the bare \NNNLO interaction. With the
softened \NNNLO potential at $\lambda=2.0\,\fmi$, $\Ecm(\widetilde\omega)$ is an order
of magnitude smaller, ranging from merely $30$ to $40\,\keV$. Consequently, we estimate
center-of-mass admixtures of $\Ecm(\widetilde\omega)/\hwBar\lesssim 0.2\%$.

\section{Summary and Ongoing Developments}
\label{sec:sum+dev}

In this work, we have presented a comprehensive review of the IM-SRG
as a powerful \emph{ab initio} method for nuclei. The IM-SRG employs
the flow-equation formulation of the SRG to decouple an $A$-body
reference state from particle-hole excitations, thus solving the
many-body Schr\"odinger equation. Compared to other \emph{ab initio}
approaches for medium-mass nuclei, we have shown that the IM-SRG is
competitive with CC and SCGF calculations, both in terms of accuracy and of
computational efficiency.

In the IM-SRG, the Hamiltonian is normal ordered with respect to the
reference state and typically truncated at the normal-ordered two-body
level. In this way, 3N interactions can be naturally included in a
normal-ordered two-body approximation. Due to the many ongoing
developments of 3N forces, we have focused in this review on a
detailed derivation at the NN level, but discuss highlights of first
IM-SRG calculation including 3N interactions in Sec.~\ref{sec:3N}.

Going beyond the closed-shell or single-reference version, the IM-SRG
has been recently generalized to a multi-reference formulation, as
discussed in Sec.~\ref{sec:mrimsrg}. The multi-reference IM-SRG (MR-IM-SRG) is
based on a generalized version of normal-ordering and Wick's
theorem. In principle, one can use 
arbitrary reference states that are
characterized by one-, two- and three-body density matrices. In the
first multi-reference applications, particle-number-projected
Hartree-Fock-Bogoliubov states were used as reference states for
describing the even nuclei throughout semi-magic isotopic chains.

A great asset of the IM-SRG is the flexibility and simplicity of its
basic concept. Through different choices of generators and decoupling
patterns, the numerical characteristics and efficiency of the methods
can be controlled and tailored for specific applications. The
in-medium evolved Hamiltonian is directly accessible, Hermitian and readily
usable for subsequent nuclear structure calculations.

As a specific application that exploits this flexibility,
nonperturabtive derivations of valence-space Hamiltonians have been
achieved. In the IM-SRG for open-shell nuclei, states with $A_v$
particles in the valence space are additionally decoupled from those
containing non-valence admixtures. This gives the energy of the
closed-shell core (as in the standard IM-SRG), but also valence-space
single-particle energies and residual interactions. The resulting
valence-space Hamiltonian can then be diagonalized using large-scale
shell-model methods. The IM-SRG provides for the first time a
nonperturbative derivation of effective interactions in the shell
model based on nuclear forces without adjustments. This is discussed
in Sec.~\ref{sec:shell_model}.

Finally, we highlight novel technical advances of the Magnus expansion
for effective operators in Sec.~\ref{sec:magnus}. This enables a
consistent evolution of operators in the IM-SRG for applications to
nuclear structure, electroweak transitions, and for key matrix elements
needed for fundamental symmetry tests.

\subsection{Medium-Mass Nuclei with Three-Nucleon Forces}
\label{sec:3N}

\begin{figure}[t]
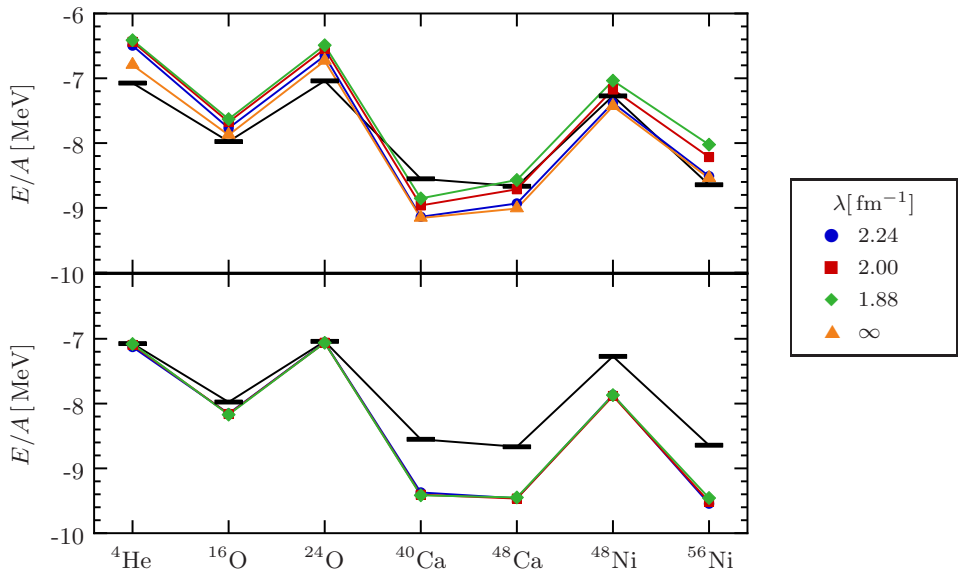

  \setlength{\unitlength}{0.8\textwidth}
  \begin{picture}(1.0000,0.8000)
    \put(-0.0500,0.3200){\small\input{fig/chi2b3bi_srgXXXX_Eint.tex}}
    \put(-0.0500,0.0000){\small\input{fig/chi2b3b400_srgXXXX_Eint.tex}}
	\put(1.0000,0.5200){\parbox{0.1600\unitlength}{
	  \framebox{
	  \begin{tabular*}{0.16\unitlength}{l@{\extracolsep\fill}l}
	  	   \multicolumn{2}{c}{\footnotesize $\lambda[\fmi]$} \\
	  	   \scriptsize\symbolcircle[FGBlue]     & \footnotesize2.24 \\
	  	   \scriptsize\symbolbox[FGRed]         & \footnotesize2.00 \\
	  	   \scriptsize\symboldiamond[FGGreen]   & \footnotesize1.88 \\
	  	   \scriptsize\symboltriangle[FGOrange] & \footnotesize$\infty$
	  \end{tabular*}
	  }
	  }
	}
  \end{picture}
  \vspace{-30pt}
  \caption{\label{fig:chi2b3bXXXX_srgXXXX_Eint}
  	IM-SRG(2) ground-state energies per nucleon for light and medium-mass closed-shell isotopes from chiral NN and NN+3N interactions at different resolution scales $\lambda$ (see text). Experimental values (black bars) taken from \cite{Wang:2012uq}. \emph{Top panel:} Bare \NNNLO at $\lambda=\infty$, and \NNNLO plus induced 3N interactions for $\lambda<\infty$. \emph{Bottom panel:} \NNNLO NN plus local \NNLO interaction with initial cutoff $\LambdaNNN=400\,\MeV$, evolved to lower resolution scale.
  }
\end{figure}

For most of the present work, we have focused on bare or softened
chiral NN interactions only for the sake of simplicity. Key aspects of
the IM-SRG, like the convergence behavior of ground-state energies as
a function of the single-particle basis size or the diagrammatic
content of the infinite-order summation, the reshuffling of many-body
correlations into the zero-body part of the flowing Hamiltonian, etc.,
are primarily governed by the resolution scale of the input
Hamiltonian. Similar features are expected (and found) regardless of
whether it contains NN, 3N, or higher many-body interactions. Despite
open issues pertaining to power counting and renormalizability, chiral
NN+3N interactions have been used in a variety of \emph{ab initio} nuclear
structure and reaction calculations, and these calculations in turn
provide new opportunities for testing the performance of chiral input
Hamiltonians.

The inclusion of 3N interactions in the IM-SRG framework is
straightforward, especially if we content ourselves with the IM-SRG(2)
truncation. In that case, any initial 3N interaction only needs to be
considered when the initial Hamiltonian is normal ordered, and its
contributions added with proper pre-factors to the normal-ordered
zero-, one-, and two-body parts of the Hamiltonian according to
Eqs.~\eqref{eq:E0}--\eqref{eq:Gamma}.

In Fig.~\ref{fig:chi2b3bXXXX_srgXXXX_Eint}, we give a summary overview
of the IM-SRG(2) ground-state energies of a range of closed-shell
nuclei up to $\nuc{Ni}{56}$, obtained with chiral NN+3N
interactions. In the top panel, we compare our usual bare \NNNLO
interaction with cutoff $\LambdaNN=500\,\MeV$ to Hamiltonians with a
lower resolution scale $\lambda$, but contrary to earlier sections
like Sec.~\ref{sec:numerics}, we include 3N interactions which are
induced by the change of $\lambda$. The g.s.~energies depend on
$\lambda$ because the free-space SRG evolution to lower resolution
scale as well as the IM-SRG evolution to solve the many-body problem
cannot be performed exactly. Thus, the spectrum of the bare
$\lambda=\infty$ NN interaction is no longer perfectly unitarily
equivalent to the spectrum of the NN+3N interaction with
$\lambda<\infty$.

Theoretical uncertainties due to the use of finite HO configuration
spaces, both for the free-space SRG evolution in Jacobi coordinates,
and the IM-SRG evolution in coupled single-particle bases, can be
controlled to much better than 1\% (see
Sec.~\ref{sec:numerics_convergence} and
Refs. \cite{Binder:2013zr,Hergert:2013mi,Roth:2014fk,Binder:2014fk}),
so the sources of the $\lambda$-dependence are the omission of induced
4N,...,$A$N interactions, the omission of the residual 3N
interaction in the initial Hamiltonian, and the use of the IM-SRG(2)
truncation. In Ref.~\cite{Hergert:2013mi}, we analyzed these
uncertainties in detail, and quantified them to be on the level of
3-4\% for the studied nuclei in the range
$\lambda=1.88,\ldots2.24\fmi$. For $\nuc{He}{4}$ and $\nuc{O}{16}$, we
benefitted from the ability to benchmark the IM-SRG(2) directly
against exact NCSM and importance-truncated NCSM results. Beyond that,
our results were consistent with CCSD and $\Lambda-$CCSD(T) results
using the same Hamiltonian.

The energy band we obtain by varying $\lambda$ spans an energy
interval that is comparable to the uncertainty given above, but we
have to be careful how we interpret this feature due to the use of a
truncated many-body method. As we have seen and discussed throughout
Sec.~\ref{sec:numerics}, methods like IM-SRG or CC benefit
significantly from the increasing perturbativeness of the nuclear
Hamiltonian as $\lambda$ is lowered. For soft interactions, more and
more ground-state energy is already recovered by low-order terms in
the MBPT series, and the IM-SRG(2) gives a more and more accurate
approximation of the exact result.

For the bare interaction, on the other hand, the error associated with
the many-body truncation is more substantial. Unfortunately, exact
calculations with methods like the (IT-)NCSM also fail to reach
convergence for all but the lightest nuclei with the bare \NNNLO
interaction, and this limited availability of benchmarks makes it hard
to quantify the size of the missing many-body contributions in the
IM-SRG(2). The size extensivity of methods like IM-SRG and CC
guarantees linear scaling of the ground-state energy with the mass
number $A$, but in order to infer ground-state energies and their
uncertainties for heavier nuclei from this property, the density has
to be (at least approximately) constant. The importance of shell and
surface effects for nuclear structure phenomena shows that this would
be an over-simplification, and thereby limits the usefulness of simple
energy scaling arguments.

Let us now consider the impact of including a chiral 3N interaction in
the initial Hamiltonian as well. Here, we have used a local \NNLO
interaction with a reduced cutoff $\LambdaNNN=400\,\MeV$, and
low-energy constants (LECs) $c_D=-0.2$ and $c_E=0.098$
\cite{Navratil:2007ve,Gazit:2009qf,Roth:2011kx}. The LECs are fixed by
fitting the triton beta decay half-life and $\nuc{He}{4}$ ground-state
energy. In Ref.~\cite{Roth:2011kx} and subsequent works,
$\LambdaNNN=400\,\MeV$ was advocated over the naively consistent
choice $\LambdaNNN=500\,\MeV$ for practical reasons: For the harder
input 3N interaction, a change of resolution scale induces strong 4N
interactions. We can see from the bottom panel of
Fig.~\ref{fig:chi2b3bXXXX_srgXXXX_Eint} that the $\lambda$ dependence
gets considerably reduced, but there are still signficant deficiencies
in reproducing the experimental ground-state energy for these NN+3N
interactions.

\subsection{The Multi-Reference IM-SRG}
\label{sec:mrimsrg}

The IM-SRG formalism and applications presented so far use a single
Slater determinant as the reference state, ideally a solution of the
Hartree-Fock equations (cf.~Sec.~\ref{sec:refstate}). Thus, the
standard IM-SRG belongs to the class of single-reference methods, such
as finite-order MBPT or CC. In nuclear physics, these approach are
only appropriate for the description of nuclei around (sub-)shell
closures.

In open-shell nuclei, correlations cause the emergence of phenomena
like nuclear superfluidity or intrinsic deformation. With
reference-state constructions, one can attempt to capture these
effects at the mean-field level to some extent, by breaking symmetries
either spontaneously or explicitly. Pairing correlations can be
treated in the Hartree-Fock-Bogoliubov (HFB) formalism, which is
formulated in terms of Slater determinants of fermionic
quasi-particles that are superpositions of particles and
holes. Intrinsic deformation will develop if the single-particle basis
is not symmetry restricted, e.g., in an $m$-scheme formalism, and
rotational symmetry breaking is energetically favored.

An $m$-scheme IM-SRG or CC calculation may be able to converge to a
solution if the excitation spectrum of the symmetry-broken reference
state has a sufficiently large gap, i.e., a single dominant
configuration. If such a solution is found, one must eventually
restore the broken symmetries through the application of projection
methods, which have a long track record in nuclear physics (see, e.g.,
\cite{Peierls:1973fk,Ring:1980bb,Egido:1982sd,Robledo:1994qf,Flocard:1997fx,Sheikh:2000xx,Dobaczewski:2007hh,Bender:2009rv,Duguet:2009ph,Lacroix:2009aj,Lacroix:2012vn,Duguet:2015ye}). At
this point, one is no longer dealing with a single-reference problem,
although the projected states retain an imprint of the original
symmetry-broken (single-)reference states that simplifies practical
implementations.

In the domain of exotic neutron-rich nuclei, the single-reference
paradigm may also break down. The complex interplay of nuclear
interactions, many-body correlations, and, in the dripline region,
continuum effects, can cause strong competition between configurations
with different intrinsic structures. This manifests in phenomena like
the erosion and emergence of shell closures
\cite{Wienholtz:2013bh,Holt:2014vn,Soma:2014eu,Hergert:2014vn}, or the
appearance of the so-called islands of inversion (see, e.g.,
\cite{Brown:2010xr}). Their description requires a true
multi-reference treatment.

\begin{figure}[t]
  \setlength{\unitlength}{0.7\textwidth}
  \begin{center}
    \hspace{-0.05\unitlength}\input{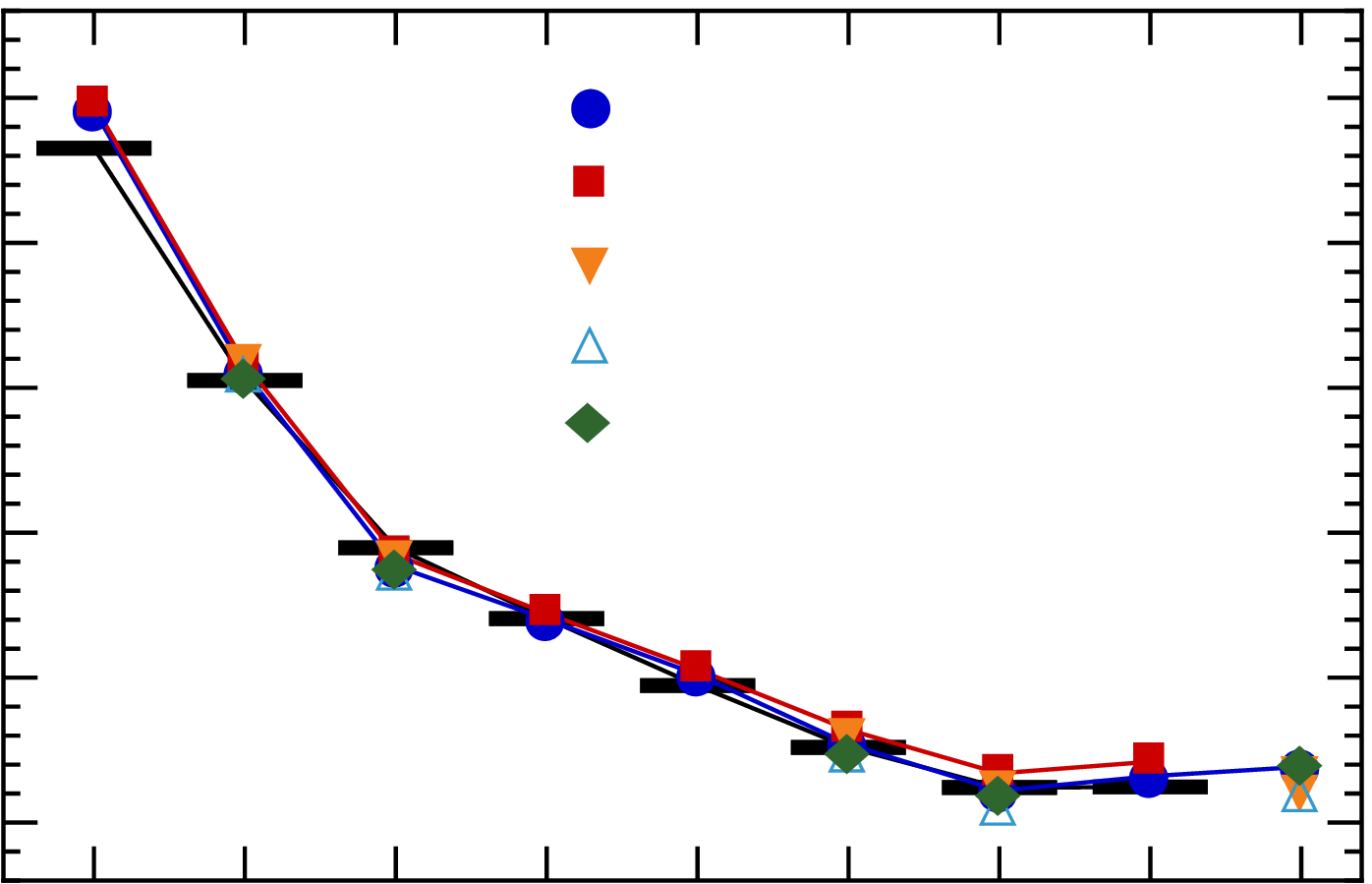}
  \end{center}
  \vspace{-30pt}
  \caption{\label{fig:mrimsrg_OXX} Ground-state energies of the oxygen isotopes from MR-IM-SRG
    and other many-body approaches, based on the NN+3N-full interaction with 
    $\LambdaNNN=400\,\MeV$, evolved to the resolution scale $\lambda=1.88\fmi$
    ($\lambda=2.0\fmi$ for the Green's Function ADC(3) results, cf.~\cite{Cipollone:2013uq}).
    Black bars indicate experimental data \cite{Wang:2012uq}.
    See Ref.~\cite{Hergert:2013ij} for additional details. 
  }
\end{figure}

The Multi-Reference IM-SRG (MR-IM-SRG) is capable of dealing with the
aforementioend scenarios
\cite{Hergert:2013ij,Hergert:2014vn,Hergert:2015qd}. It generalizes
the IM-SRG formalism discussed in this work to arbitrary correlated
reference states, using the multi-reference normal ordering and Wick's
theorem developed by Kutzelnigg and Mukherjee
\cite{Kutzelnigg:1997fk,Mukherjee:1997yg}. The idea of decoupling the
ground state from excitations readily carries over, except that
excited states are given by
\begin{equation*}
\nord{\aaO_i\aO_j\!}\!\ket{\Phi},\,\nord{\aaO_i\aaO_j\aO_l\aO_k\!}\!\ket{\Phi},\ldots\,,
\end{equation*}
and the single-particle states are no longer of pure particle or hole
character.  The flow equation formulation of the MR-IM-SRG makes it
possible to avoid complications due to the non-orthogonality and
possible linear dependency of these general excitations (see
\cite{Hergert:2015qd} for more details).

While only one-body density matrices appear in the contractions of the
standard Wick's theorem (see Eqs.~\eqref{eq:def_particle_contraction}
and \eqref{eq:def_hole_contraction}), additional contractions that
involve two- and higher-body density matrices enter that encode the
correlation content of the reference state. In the MR-IM-SRG
framework, correlations that are hard to capture as few-body
excitations of the reference state can be built directly into the
reference state.

In a first applications of the MR-IM-SRG framework, we have used
spherical, particle-number projected HFB vacua to compute the
ground-state energies of the even oxygen isotopes, starting from
chiral NN+3N forces \cite{Hergert:2013ij}. This work improved on
previous Shell Model \cite{Otsuka:2010cr,Holt:2013fk} and CC studies
\cite{Hagen:2012oq}, that included NN+3N interactions in MBPT or for
the latter with 3N forces in a more phenomenological, nuclear-matter
based normal ordering.  Based on a Hamiltonian that is entirely fixed
in the $A=3,4$ system and consistently evolved to lower resolution, we
found that MR-IM-SRG, various CC methods, and the importance-truncated
NCSM consistently predict the neutron dripline in $\nuc{O}{24}$ if
chiral 3N forces are included (see Fig.~\ref{fig:mrimsrg_OXX}), as
pointed out in the context of the Shell Model in
Ref.~\cite{Otsuka:2010cr}.

\begin{figure}[t]
  \setlength{\unitlength}{0.72\textwidth}
  \begin{center}
    \hspace{-0.07\unitlength}\input{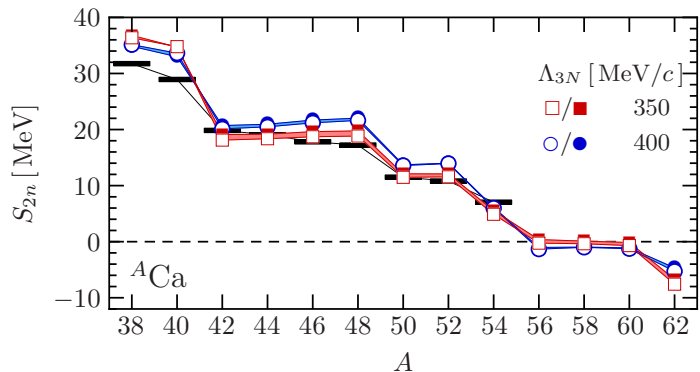}
  \end{center}
  \vspace{-30pt}
  \caption{\label{fig:mrimsrg_CaXX} 
    MR-IM-SRG results for Ca two-neutron separation energies, 
    for chiral NN+3N interactions 
    with different cutoffs in the 3N sector, and a range of resolution scales 
    from $\lambda=1.88\fmi$ (open symbols) to $2.24\fmi$ (solid symbols).
    Black bars indicate experimental data \cite{Wang:2012uq,Wienholtz:2013bh}.
    See Ref.~\cite{Hergert:2014vn} for additional details. 
    }
\end{figure}

Encouraged by this success, we moved on to the calcium and nickel
isotopic chains \cite{Hergert:2014vn}, where importance-truncated NCSM
calculations are no longer feasible. The same family of chiral NN+3N
Hamiltonians that successfully reproduce the oxygen ground-state
energies overestimate the binding energies in these isotopes by
several hundred keV per nucleon, in MR-IM-SRG and CC (also see
\cite{Roth:2012qf,Binder:2013zr,Binder:2014fk}), as well as the
second-order Gor'kov Green's Function approach \cite{Soma:2014eu}.
The revelation of these deficiencies has led to a variety of efforts
to improve on the chiral interactions
\cite{Ekstrom:2015fk,Hagen:2015ve,Epelbaum:2015fb,Epelbaum:2014sza,Entem:2015hl,Entem:2015qf,Carlsson:2015fk,Gezerlis:2014zia,Piarulli:2015rm,Lynn:2015jua}.

Contrary to the ground-state energies, chiral NN+3N forces reproduce
relative quantities like the two-neutron separation energies quite
well, aside from the exaggerated $N=20$ shell closure
(Fig.~\ref{fig:mrimsrg_CaXX}).  In particular, they show signals of
sub-shell closures in $\nuc{Ca}{52,54}$, in agreement with
Shell Model calculations based on NN+3N interactions in MBPT
\cite{Holt:2014vn,Wienholtz:2013bh}. These observations indicate which
terms in the chiral input Hamiltonian may be deficient, and this
information can be used in future optimizations.

Ongoing work focuses on three interconnected main directions:
\emph{(i)} the extension of the MR-IM-SRG to the evaluation of general
observables (see Sec.~\ref{sec:magnus}), \emph{(ii)} the description
of deformed nuclei, and \emph{(iii)} the integration of MR-IM-SRG and
Equation-of-Motion methods \cite{Rowe:1968eq} for the spectroscopy of
medium-mass and heavy neutron-rich nuclei.

\subsection{Non-Perturbative Shell-Model Interactions}
\label{sec:shell_model}

For open-shell systems, rather than solving the full $A$-body problem,
it is profitable to follow the Shell Model paradigm by constructing
and diagonalizing an effective Hamiltonian in which the active degrees
of freedom are $A_v$ valence nucleons confined to a few orbitals near
the Fermi level. Both phenomenological and microscopic
implementations of the Shell Model have been used with success to
understand and predict the evolution of shell structure, properties of
ground and excited states, and electroweak transitions
\cite{Brown:2001rg,Caurier:2005qf,Otsuka:2013vn}.

Recent microscopic Shell-Model studies have revealed the impact of 3N
forces in predicting ground- and excited-state properties in neutron-
and proton-rich nuclei
\cite{Otsuka:2010cr,Holt:2012fk,Holt:2013fk,Holt:2013hc,Holt:2013cr,Gallant:2012kx,Wienholtz:2013bh,Holt:2014vn}.
Despite the novel insights gained from these studies, they make
approximations that are difficult to benchmark. The microscopic
derivation of the effective valence-space Hamiltonian relies on MBPT
\cite{Hjorth-Jensen:1995ys}, where order-by-order convergence is unclear. Even with
efforts to calculate particular classes of diagrams nonperturbatively
\cite{Holt:2005mi}, results are sensitive to the HO frequency $\hw$ (due
to the core), and the choice of valence space
\cite{Holt:2012fk,Holt:2013fk,Holt:2013hc}. A nonperturbative method to
address these issues was developed in
\cite{Lisetskiy:2008fk,Lisetskiy:2009uq,Dikmen:2015fk}, which generates valence-space
interactions and operators by projecting their full NCSM counterparts
into a given valence space.

To overcome these limitations in heavier systems, the IM-SRG can be
extended to derive effective valence-space Hamiltonians and operators
nonperturbatively.  Calculations without initial 3N forces
\cite{Tsukiyama:2012fk} indicated that an \emph{ab initio} description
of ground and excited states for open-shell nuclei may be possible
with this approach.

The utility of the IM-SRG lies in the freedom to tailor the definition
of $H^{\rm{od}}$ to a specific problem.  For instance, to construct a
Shell Model Hamiltonian for a nucleus comprised of $A_v$ valence
nucleons outside a closed core, we define a HF reference state
$\ket{\Phi}$ for the core with $A_c$ particles, and split the
single-particle basis into hole ($h$), valence ($v$), and non-valence
($q$) particle states.  Treating all $A$ nucleons as active, i.e.,
without a core approximation, we eliminate matrix elements which
couple $\ket{\Phi}$ to excitations, just as in IM-SRG ground-state
calculations \cite{Tsukiyama:2011uq,Hergert:2013mi,Hergert:2013ij}. In
addition, we decouple states with $A_v$ particles in the valence
space, \mbox{$:\!\aaO_{v_1}\ldots\aaO_{v_{A_v}}\!\!:\!\!\ket{\Phi}$},
from states containing non-valence states.

\begin{figure}[t]
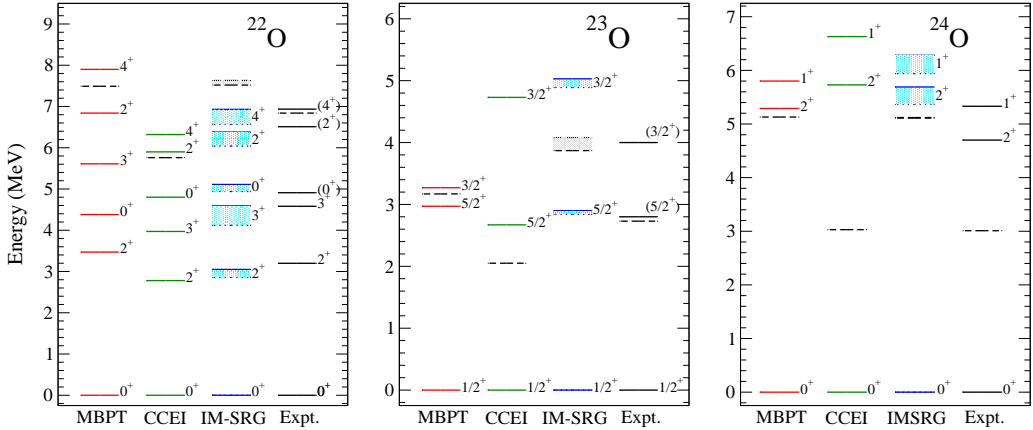

\begin{center}
\minipage{0.335\textwidth}
\includegraphics[width=\linewidth,clip=]{fig/22O_arnps}
\endminipage\hfill
\minipage{0.3\textwidth}
\includegraphics[width=\linewidth,clip=]{fig/23O_arnps}
\endminipage\hfill
\minipage{0.3\textwidth}
\includegraphics[width=\linewidth,clip=]{fig/24O_arnps}
\endminipage
\end{center}
\caption{Excited-state spectra of $^{22,23,24}$O based on chiral NN+3N
interactions and compared with experiment. Figures adapted from
Ref.~\cite{Bogner:2014tg}. The MBPT results are performed in an extended
$sdf_{7/2}p_{3/2}$ space~\cite{Holt:2013fk} based on low-momentum NN+3N
interactions, while the IM-SRG~\cite{Bogner:2014tg} and CC effective
interaction (CCEI)~\cite{Jansen:2014qf} results are in the $sd$ shell from
the SRG-evolved NN+3N-full Hamiltonian with $\hbar \omega=20$~MeV
(CCEI and dotted IM-SRG) and $\hbar \omega=24$~MeV (solid IM-SRG). The
dashed lines show the neutron separation energy.
Figure taken from Ref.~\cite{Hebeler:2015xq}.\label{fig:Ospectra}}
\end{figure}

After the IM-SRG derivation of the valence-space Hamiltonian, the
$A$-dependent Hamiltonian is diagonalized in the valence space to
obtain the ground and excited states. For the oxygen isotopes, a good
description of the experimental spectra is found
(Fig.~\ref{fig:Ospectra}).  Recently, these calculations were extended
to nearby F, Ne, and Mg isotopes showing excellent agreement with new
measurements in $^{24}$F~\cite{Caceres:2015fk} and that deformation
can emerge from these \emph{ab initio}
calculations~\cite{Stroberg:2015qr}. Future directions include
extending the valence space, which will give access to the
island-of-inversion region and potentially the full $sd$-shell (and
higher) neutron dripline.

\subsection{Technical Advances: Magnus Expansion and Effective Operators}
\label{sec:magnus}

\begin{figure}[t]
\begin{center}
\includegraphics[width=.6\columnwidth]{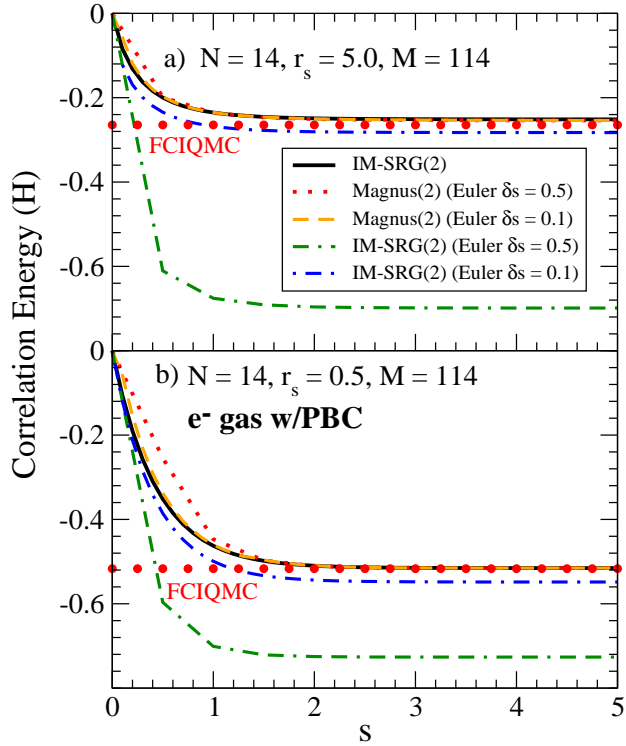}
\caption{\label{fig:timestep_HEG}
(Color online) Flowing IM-SRG(2) and Magnus(2) correlation energy for 
the electron gas, $E_0(s)-E_{\rm HF}$, for Wigner-Seitz radii of a)
$r_s=5.0$ and b) $r_s=0.5$. The solid black line corresponds to the
IM-SRG(2) results using an adaptive solver based on the
Adams-Bashforth method, while the other lines correspond to Magnus(2)
and IM-SRG(2) results using different Euler step sizes. The red
circles denote the quasi-exact FCIQMC results of
Ref.~\cite{Shepherd:2012hl}.}
\end{center}
\end{figure}

Despite modest computational scaling and the flexibility to tailor the
generator to different systems, IM-SRG calculations based on the
direct integration of Eqs.~(\ref{eq:opflow}) and~(\ref{eq:obsflow}) are
limited by the memory demands of the ODE solver. The use of a
high-order solver is essential, as the accumulation of time-step
errors destroys the unitary equivalence between $H(s)$ and $H(0)$ even
if no truncations are made in the flow equations. State-of-the-art
solvers can require the storage of 15-20 copies of the solution vector
in memory, which is the main computational bottleneck of the method.
What is worse, the dimensionality of the flow equations roughly
doubles for each additional observable one wishes to calculate, and
each operator can evolve with rather different timescales than the
Hamiltonian, increasing the likelihood of the ODEs becoming
stiff.

\begin{figure}[t]
\begin{center}
\includegraphics[width=.6\columnwidth]{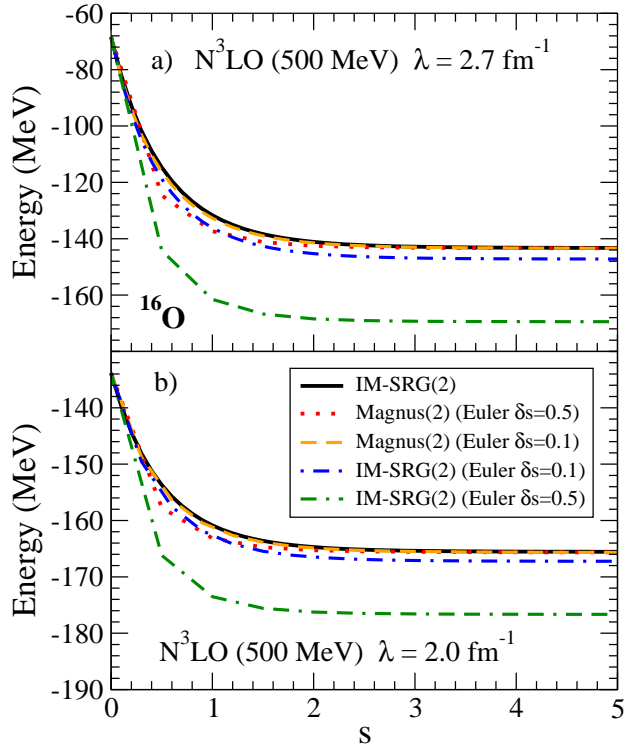}
\caption{\label{fig:timestep_O16}
(Color online) Flowing IM-SRG(2) and Magnus(2) ground-state energy, 
$E_0(s)$, for $^{16}$O starting from the N$^3$LO NN interaction of
Entem and Machleidt~\cite{Entem:2003th,Machleidt:2011bh} evolved by
the free-space SRG to a) $\lambda=2.7$ fm$^{-1}$ and $\lambda = 2.0$
fm$^{-1}$. The solid black line corresponds to IM-SRG(2) results using
an adaptive solver based on the Adams-Bashforth method, while the
other lines correspond to Magnus(2) and IM-SRG(2) results using
different Euler step sizes. The calculations were done in an
$e_{max}=8$ model space, with $\hbar\omega = 24$~MeV for the
underlying harmonic-oscillator basis.}
\end{center}
\end{figure}

To bypass these limitations, an improved formulation of the IM-SRG was
proposed in Ref.~\cite{Morris:2015ve} that utilizes the Magnus
expansion from the theory of matrix differential
equations~\cite{Magnus:1954xy,Blanes:2009fk}.  In essence, the problem
is recast so that rather than solving flow equations for the
Hamiltonian and other operators of interest, one solves flow equations
for the anti-Hermitian operator $\Omega(s)$, where $U(s) =
e^{\Omega(s)}$. The unitary operator $U(s)$ is then used to transform
the Hamiltonian and any other operators of interest via the
Baker-Cambell-Hausdorff (BCH) formula. The advantage of the Magnus
formulation stems from the fact that the flow equations for
$\Omega(s)$ can be solved using a simple first-order Euler step method
without any loss of accuracy, resulting in substantial memory savings
and a modest reduction in CPU time.  In the conventional approach,
time-step errors accumulate directly in the evolved $H(s)$,
necessitating the use of a high-order solver to preserve an acceptable
level of accuracy.  In the Magnus formulation, even though sizable
time-step errors accumulate in $\Omega(s)$ with a first-order method, 
upon exponentiation the transformation is still unitary, and
the transformed $H(s)=\UO(s)H\UUO(s)$ is unitarily equivalent to
the initial Hamiltonian modulo any truncations made in evaluating the BCH formula.  For
further details on the implementation of the Magnus formulation, see
Ref.~\cite{Morris:2015ve}.

The insensitivity to time-step errors is illustrated in
Figs.~\ref{fig:timestep_HEG} and~\ref{fig:timestep_O16}, which show
the 0-body part of the flowing Hamiltonian $H(s)$ versus the flow
parameter for the electron gas, where we plot $E_0(s)-E_\text{HF}$ as an
approximation of the correlation energy at large $s$, and for
$^{16}$O, respectively. The black solid lines denote the results of a
standard IM-SRG(2) calculation using the predictor-corrector solver of
Shampine and Gordon, while the other curves denote IM-SRG(2) and
Magnus(2) calculations using a first-order Euler method with different
step sizes $\delta s$. For the electron gas, the exact full
configuration quantum Monte Carlo (FCIQMC)
results~\cite{Shepherd:2012hl} are shown for
reference. Unsurprisingly, the IM-SRG(2) calculations using a
first-order Euler method are very poor, with the various step sizes
converging to different large-$s$ limits. The Magnus(2) calculations,
on the other hand, converge to the same large-$s$ limit in excellent
agreement with the standard IM-SRG(2) and the FCIQMC results.

The evaluation of general operators poses considerable computational
challenges in the conventional formulation of the IM-SRG. In the
Magnus expansion formulation, the evolution of additional operators is
relatively straightforward since the dimensionality of the flow
equations is fixed, regardless of how many additional operators are
being evolved. Proof-of-principle operator evolutions have been
carried out for for the momentum distribution in the electron gas, and
the generalized center-of-mass Hamiltonian in $^{16}$O with encouraging
results~\cite{Morris:2015ve}.  The relative ease of performing
operator evolution is especially encouraging for shell model
applications, as it opens up the exciting possibility for consistent,
nonperturbative calculations of both Shell-Model Hamiltonians and
effective electroweak operators (e.g., the $0\nu\beta\beta$ matrix
element, quenching of Gamow-Teller strength, etc.) relevant for
studies of fundamental symmetries in nuclei.

\section*{Acknowledgments}
\addcontentsline{toc}{section}{\protect\numberline{}Acknowledgments}%

We thank C.~Barbieri, S.~Binder, A.~Calci, T.~Duguet, J.~Engel,
F.~Evangelista, R.~J.~Furnstahl, G.~Hagen, K.~Hebeler,
M.~Hjorth-Jensen, J.~D.~Holt, J.~Langhammer, J.~Men\'endez, T.~Otsuka,
T.~Papenbrock, R.~Roth, J.~Simonis, V.~Som\`a and S.~R.~Stroberg for
useful discussions on the topics of this review. This work was
supported in part by the NUCLEI SciDAC Collaboration under the
U.S.~Department of Energy Grants No.~DE-SC0008533 and DE-SC0008511,
the National Science Foundation under Grants No.~PHY-1002478,
PHY-1306250, PHY-1068648 and PHY-1404159, the European Research
Council Grant No.~307986 STRONGINT, the BMBF under Contracts
No.~06DA70471 and 05P15RDFN1, and the DFG through Grant SFB 634.

\section*{Appendices}
\appendix
\renewcommand*{\thesection}{\Alph{section}}

\section{\label{app:commutators}Fundamental Commutators}
For convenience, we collect the expressions for the fundamental commutators which are required for the derivation of the IM-SRG flow equations and Wegner-type generators. We write one-, two-, and three-body operators as
\begin{align}
	A^{(1)}	&= \sum_{ij} A_{ij} \nord{\aaO_i\aO_j}\,,\\
 	A^{(2)} &= \frac{1}{(2!)^2} \sum_{ijkl} A_{ijkl}\nord{\aaO_i\aaO_j\aO_l\aO_k}\,,\\
 	A^{(3)} &= \frac{1}{(3!)^2}\sum_{ijklmn}A_{ijklmn} \nord{\aaO_i\aaO_j\aaO_k\aO_n\aO_m\aO_l}\,, 
\end{align}
where the two- and three-body matrix elements are assumed to be fully anti-symmetrized. 
Single-particle indices refer to natural orbitals, so that occupation numbers are $n_i =0,1$, and we use the notation $\nn_a=1-n_a$. We also recall that the commutator of two operators of rank $M$ and $N$ can only have contributions of rank $|M-N|,\ldots,M+N-1$,
\begin{equation}
  \comm{\AO^{(M)}}{\BO^{(M)}} = \sum_{k=|M-N|}^{M+N-1}\CO^{(k)}\,.
\end{equation}

\subsection{\texorpdfstring{\boldmath$\comm{\AO^{(1)}}{\circ}$}{[A(1),]}}
\begin{align}
 	\comm{\AO^{(1)}}{\BO^{(1)}}^{(1)} &= \sum_{ij}\sum_a\nord{\aaO_i\aO_j}
 	\left(A_{ia}B_{aj}-B_{ia}A_{aj}\right) \\
 	\comm{\AO^{(1)}}{\BO^{(1)}}^{(0)} &= \sum_{ij}A_{ij}B_{ji}(n_i-n_j)
\end{align}

\begin{align}
	\comm{\AO^{(1)}}{\BO^{(2)}}^{(2)} &= \frac{1}{4}\sum_{ijkl}\sum_{a}\nord{\aaO_i\aaO_j\aO_l\aO_k}
 	\left\{(1-P_{ij}) A_{ia}B_{ajkl}-(1-P_{kl})A_{ak}B_{ijal} \right\} \\
 	\comm{\AO^{(1)}}{B^{(2)}}^{(1)} &= \sum_{ij}\sum_{ab}\nord{\aaO_i\aO_j}
 	\left\{(n_a-n_b)A_{ab}B_{biaj}\right\} 
\end{align}

\begin{align}
	\comm{\AO^{(1)}}{\BO^{(3)}}^{(3)} &= \frac{1}{36}\sum_{ijklmn}\sum_a\nord{\aaO_i\aaO_j\aaO_k \aO_n \aO_m \aO_l} \notag\\
	&\qquad\times\left\{(1-P_{ij}-P_{ik})A_{ia}B_{ajklmn}-(1-P_{lm}-P_{ln})A_{al}B_{ijkamn}  \right\}\\[1mm]
 	\comm{\AO^{(1)}}{\BO^{(3)}}^{(2)} &= \sum_{ijkl}\sum_{ab}\nord{\aaO_i\aaO_j\aO_l\aO_k}
 	\left(n_a-n_b\right)A_{ab}B_{bijakl} 
\end{align}

\subsection{\texorpdfstring{\boldmath$\comm{\AO^{(2)}}{\circ}$}{[A(2),]}}
\begin{align}
	\comm{\AO^{(2)}}{\BO^{(2)}}^{(3)} &= \frac{1}{36}\sum_{ijklmn}\sum_a 
	\nord{\aaO_i\aaO_j\aaO_k \aO_n\aO_m\aO_l}\notag\\
	 &\qquad\qquad\qquad\times P(ij/k)P(l/mn)\left(A_{ijla}B_{akmn}-B_{ijla}A_{akmn}\right)\\[1mm]
	\comm{\AO^{(2)}}{\BO^{(2)}}^{(2)} &= \frac{1}{4}\sum_{ijkl}\sum_{ab}\nord{\aaO_i\aaO_j\aO_l\aO_k}
	\biggl\{ \frac{1}{2}(A_{ijab}B_{abkl}-B_{ijab}A_{abkl})(1-n_a-n_b) \notag\\
	&\qquad\qquad\qquad\qquad\qquad\quad+(n_a-n_b)(1-P_{ij}-P_{kl}+P_{ij}P_{kl})A_{aibk}B_{bjal} \biggr\} \\
 	\comm{\AO^{(2)}}{\BO^{(2)}}^{(1)} &= \frac{1}{2}\sum_{ij}\sum_{abc} \nord{\aaO_i \aaO_j}
 	\left(A_{ciab}B_{abcj}-B_{ciab}A_{abcj}\right)\left(\bar{n}_a\bar{n}_bn_c+n_a n_b \bar{n}_c\right)\\
 	\comm{\AO^{(2)}}{\BO^{(2)}}^{(0)} &= \frac{1}{4}\sum_{ijkl}n_in_j\bar{n}_k\bar{n}_l
	\left(A_{ijkl}B_{klij}-B_{ijkl}A_{klij}\right)
\end{align}

\begin{align}
 	\comm{\AO^{(2)}}{\BO^{(3)}}^{(3)} &= \frac{1}{72}\sum_{ijklmn}\sum_{ab}
 	\nord{\aaO_i\aaO_j\aaO_k\aO_n\aO_m\aO_l} (1-n_a-n_b)\notag\\
	&\qquad\qquad\qquad\times\left(P(ij/k)A_{ijab}B_{abklmn}-P(l/mn)A_{abmn}B_{ijklab}\right)\\[1mm]
	\comm{\AO^{(2)}}{\BO^{(3)}}^{(2)} &= -\frac{1}{8}\sum_{ijkl}\sum_{abc}
	\nord{\aaO_i\aaO_j\aO_l\aO_k}
   	(n_a\bar{n}_b\bar{n}_c+\bar{n}_an_bn_c)\notag\\
   	&\qquad\qquad\qquad\times\left(1-P_{ij}P_{ik}P_{jl}-P_{kl}+P_{ik}P_{jl}\right)A_{bcak}B_{aijbcl}\\[1mm]
	\comm{\AO^{(2)}}{\BO^{(3)}}^{(1)} &= -\frac{1}{4}\sum_{ij}\sum_{abcd}\nord{\aaO_i\aO_j}
	(n_an_b \bar{n}_c\bar{n}_d-\bar{n}_a\bar{n}_bn_cn_d)A_{cdab}B_{abijcd}
\end{align}

\subsection{\texorpdfstring{\boldmath$\comm{\AO^{(3)}}{\circ}$}{[A(3),]}}
\begin{align}
	&\comm{\AO^{(3)}}{\BO^{(3)}}^{(3)}\notag\\ 
	&= \frac{1}{36}\sum_{ijklmn}\sum_{abc}
	\nord{\aaO_i\aaO_j\aaO_k\aO_n\aO_m\aO_l}\notag\\ 
	&\qquad\times\left\{
	\frac{1}{6}(n_an_bn_c + \bar{n}_a\bar{n}_b\bar{n}_c)
	(A_{ijkabc}B_{abclmn}-B_{ijkabc}A_{abclmn}) \right. \notag \\
	&\qquad\qquad
 	\left. + \frac{1}{2}(n_an_b\bar{n}_c+\bar{n}_a\bar{n}_bn_c)P(ij/k)P(l/mn)
	(A_{abkcmn}B_{cijabl}-A_{cjkabn}B_{iablmc})
	\right\}\\
	&\comm{\AO^{(3)}}{\BO^{(3)}}^{(2)} \notag\\
	&= \frac{1}{4}\sum_{ijkl}\sum_{abcd}
	\nord{\aaO_i\aaO_j\aO_l\aO_k} \notag\\
	&\qquad\times\left\{\frac{1}{6}(n_a\bar{n}_b\bar{n}_c\bar{n}_d-\bar{n}_a n_bn_cn_d) 
	(A_{aijbcd}B_{bcdakl}-A_{bcdakl}B_{aijbcd}) \right. \notag \\
	&\qquad \qquad 
	\left. +\frac{1}{4}(\bar{n}_a\bar{n}_b n_c n_d-n_an_b\bar{n}_c\bar{n}_d)
	(1-P_{ij})(1-P_{kl})A_{abicdl}B_{cdjabk}
	\right\}\\[1mm]
 	&\comm{\AO^{(3)}}{\BO^{(3)}}^{(1)} \notag\\ 
 	&= \frac{1}{12} \sum_{ij}\sum_{acde} \nord{\aaO_i\aO_j}
	(n_an_b \bar{n}_c\bar{n}_d\bar{n}_e + \bar{n}_a\bar{n}_b n_cn_dn_e)
	(A_{abicde}B_{cdeabj}-B_{abicde}A_{cdeabj})\\
	&\comm{\AO^{(3)}}{\BO^{(3)}}^{(0)} \notag \\
	&= \frac{1}{36} \sum_{ijklmn}
	(n_in_jn_k\bar{n}_l\bar{n}_m\bar{n}_n-\bar{n}_i\bar{n}_j\bar{n}_k n_l n_m n_n)
	A_{ijklmn}B_{lmnijk}
\end{align}

\section{\label{app:imsrg3}IM-SRG(3) Flow Equations}
The IM-SRG(3) flow equations can be derived using the fundamental commutators from Appendix \ref{app:commutators}. The permuation symbols $P_{ij}, P(ij/k),$ and $P(i/jk)$ have been defined in Eqs.~\eqref{eq:def_Pij}, \eqref{eq:def_Pijk1}, and \eqref{eq:def_Pijk2}. The normal-ordered Hamiltonian is given by
\begin{equation}
  \HO(s) \approx E(s) + f(s) + \Gamma(s) + W(s)\,.
\end{equation}
The particle ranks of the individual contributions of $\HO$ and the generator $\etaO$ are obvious from the indices of the associated matrix elements. 
\begin{align}
	\frac{d}{ds}E &= \sum_{ab}(n_a-n_b)\eta_{ab} f_{ba} 
		+ \frac{1}{2} \sum_{abcd}\eta_{abcd}\Gamma_{cdab} n_a n_b\bar{n}_c\bar{n}_d \notag\\
		&\quad
		+ \frac{1}{18}\sum_{abcdef}\eta_{abcdef}W_{defabc}  n_a n_b n_c \bar{n}_d\bar{n}_e \bar{n}_f  
		\label{eq:imsrg3_0b}
\end{align}

\begin{align}
	\frac{d}{ds}f_{ij} &= 
	\sum_{a}(1+P_{ij})\eta_{ia}f_{aj} +\sum_{ab}(n_a-n_b)(\eta_{ab}\Gamma_{biaj}-f_{ab}\eta_{biaj}) \notag\\ 
	&\quad +\frac{1}{2}\sum_{abc}(n_an_b\bar{n}_c+\bar{n}_a\bar{n}_bn_c) (1+P_{ij})\eta_{ciab}\Gamma_{abcj}\notag \\
    &\quad +\frac{1}{4}\sum_{abcd}(n_an_b\bar{n}_c\bar{n}_d)(\eta_{abicdj}\Gamma_{cdab}-W_{abicdj}\eta_{cdab})\nonumber\\
	&\quad + \frac{1}{12}\sum_{abcde}(n_an_b \bar{n}_c\bar{n}_d\bar{n}_e + \bar{n}_a\bar{n}_b n_cn_dn_e)
		(\eta_{abicde}W_{cdeabj}-W_{abicde}\eta_{cdeabj})
	\label{eq:imsrg3_1b} \\
	\frac{d}{ds}\Gamma_{ijkl}&= 
	\sum_{a}\Bigl\{ 
		(1-P_{ij})(\eta_{ia}\Gamma_{ajkl}-f_{ia}\eta_{ajkl} )
		-(1-P_{kl})(\eta_{ak}\Gamma_{ijal}-f_{ak}\eta_{ijal} )
		\Bigr\}\notag \\
	&\quad+ \frac{1}{2}\sum_{ab}(1-n_a-n_b)(\eta_{ijab}\Gamma_{abkl}-\Gamma_{ijab}\eta_{abkl})\notag\\
	&\quad -\sum_{ab}(n_a-n_b) (1-P_{ij})(1-P_{kl})\eta_{bjal}\Gamma_{ai bk}
		\notag \\
	&\quad+\sum_{ab}(n_a-n_b)
		\left(
			\eta_{aijbkl}f_{ba}-W_{aijbkl}\eta_{ba}
		\right)
		\notag \\
	&\quad+\frac{1}{2}\sum_{abc}(n_a \bar{n}_b \bar{n}_c+ \bar{n}_an_bn_c)
		(1-P_{ik}P_{jl}P_{ij}-P_{kl}+P_{ik}P_{jl})\notag\\
 	&\qquad\qquad\times
		(\eta_{aijbcl}\Gamma_{bcak}-W_{aijbcl}\eta_{bcak})
		\notag \\
	&\quad + \frac{1}{6}\sum_{abcd}
		(n_a\bar{n}_b\bar{n}_c\bar{n}_d-\bar{n}_an_bn_cn_d)
		(\eta_{aijbcd}W_{bcdakl}-\eta_{bcdakl}W_{aijbcd})  
		\notag \\
	&\quad + \frac{1}{4}\sum_{abcd}
		(\bar{n}_a\bar{n}_b n_c n_d-n_an_b\bar{n}_c\bar{n}_d)
		(1-P_{ij})(1-P_{kl})\eta_{abicdl}W_{cdjabk}
		\label{eq:imsrg3_2b}
\end{align}

\begin{align}
 	\frac{d}{ds}W_{ijklmn}& =\sum_{a}
 		\Bigl\{
			P(i/jk)\eta_{ia}W_{ajklmn}-P(l/mn)\eta_{al}W_{ijkamn}
		\Bigr\}
		\notag \\
	&\quad -\sum_{a}
		\Bigl\{
			P(i/jk)f_{ia}\eta_{ajklmn}-P(l/mn)f_{al}\eta_{ijkamn}
		\Bigr\} 
		\notag \\
	&\quad +\sum_{a}
		P(ij/k)P(l/mn)(\eta_{ijla}\Gamma_{akmn}-\Gamma_{ijla}\eta_{akmn})
		\notag \\
	&\quad +\frac{1}{2}\sum_{ab}
		(1-n_a-n_b)(P(i/jk))(\eta_{ijab}W_{abklmn}-\Gamma_{ijab}\eta_{abklmn})
		\notag \\
	&\quad -\frac{1}{2}\sum_{ab}
		(1-n_a-n_b)(P(lm/n))(\eta_{ablm}W_{ijkabn}-\Gamma_{ablm}\eta_{ijkabn})
		\notag \\
	&\quad -\sum_{ab}
		(n_a-n_b)P(i/jk)p(l/mn)(\eta_{bial}W_{ajkbmn}-\Gamma_{bial}\eta_{ajkbmn})
		\notag \\
	&\quad +\frac{1}{6}\sum_{abc}
		(n_an_bn_c + \bar{n}_a\bar{n}_b\bar{n}_c)
		(\eta_{ijkabc}W_{abclmn}-W_{ijkabc}\eta_{abclmn})  
		\notag \\
	&\quad +\frac{1}{2}\sum_{abc}
		(n_an_b\bar{n}_c+\bar{n}_a\bar{n}_bn_c)P(ij/k)P(l/mn)\notag\\
	&\qquad\qquad\times
		(\eta_{abkcmn}W_{cijabl}-\eta_{cjkabn}W_{iablmc})
	 \label{eq:imsrg3_3b}
\end{align}

\section{\label{app:diagram}Diagram Rules}
\newcommand{\skelOB}{
	\setlength{\unitlength}{0.12\textwidth}
	\begin{picture}(0.4600,1.4000)
	  \put(0.0000,0.2000){\includegraphics[height=0.9\unitlength]{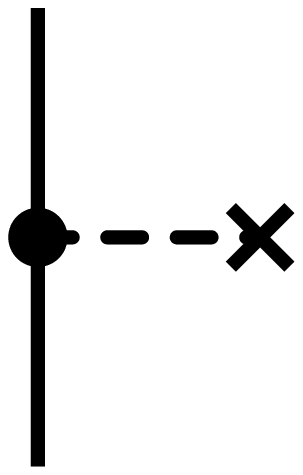}}
	  \put(0.0200,0.0000){$j$}
	  \put(0.0200,1.2000){$i$}
	\end{picture}
}
\newcommand{\skelTB}{
	\setlength{\unitlength}{0.12\textwidth}
	\begin{picture}(0.6000,1.4000)
	  \put(0.0000,0.2000){\includegraphics[height=0.9\unitlength]{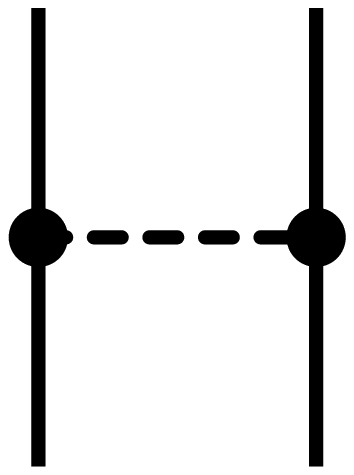}}
	  \put(0.0200,0.0000){$k$}
	  \put(0.5400,0.0000){$l$}
	  \put(0.0200,1.2000){$i$}
	  \put(0.5400,1.2000){$j$}
	\end{picture}
}
\newcommand{\skelTHB}{
	\setlength{\unitlength}{0.12\textwidth}
	\begin{picture}(1.0000,1.40000)
	  \put(0.0000,0.2000){\includegraphics[height=0.9\unitlength]{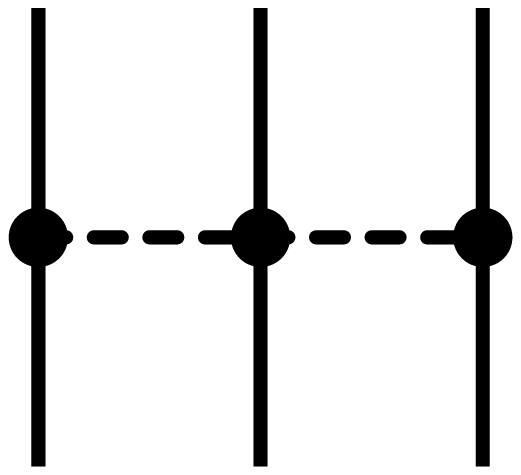}}
	  \put(0.0200,0.0000){$l$}
	  \put(0.3800,0.0000){$m$}
	  \put(0.8600,0.0000){$n$}
	  \put(0.0200,1.2000){$i$}
	  \put(0.4200,1.2000){$j$}
	  \put(0.8600,1.2000){$k$}
	\end{picture}
}

\newcommand{\skelHOB}{
	\setlength{\unitlength}{0.12\textwidth}
	\begin{picture}(0.4600,1.4000)
	  \put(0.0000,0.2000){\includegraphics[height=0.9\unitlength]{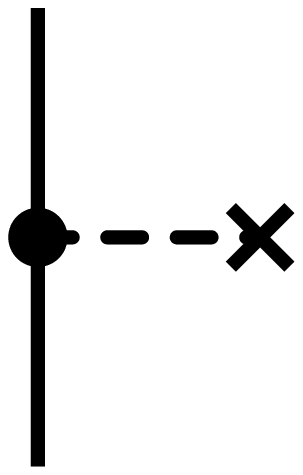}}
	  \put(0.0200,0.0000){$j$}
	  \put(0.0200,1.2000){$i$}
	\end{picture}
}
\newcommand{\skelHTB}{
	\setlength{\unitlength}{0.12\textwidth}
	\begin{picture}(0.6000,1.4000)
	  \put(0.0600,0.2000){\includegraphics[height=0.9\unitlength]{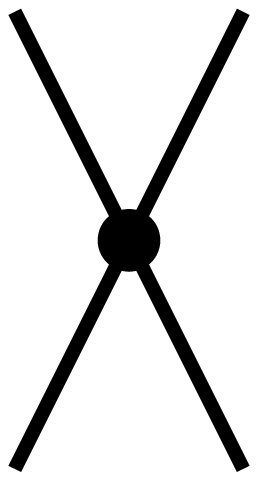}}
	  \put(0.0000,0.0000){$k$}
	  \put(0.5200,0.0000){$l$}
	  \put(0.0000,1.2000){$i$}
	  \put(0.5200,1.2000){$j$}
	\end{picture}
}
\newcommand{\skelHTHB}{
	\setlength{\unitlength}{0.12\textwidth}
	\begin{picture}(1.0000,1.40000)
	  \put(0.0800,0.2000){\includegraphics[height=0.9\unitlength]{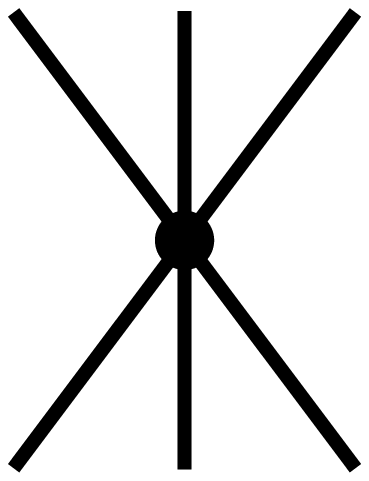}}
	  \put(0.0200,0.0000){$l$}
	  \put(0.3600,0.0000){$m$}
	  \put(0.7400,0.0000){$n$}
	  \put(0.0200,1.2000){$i$}
	  \put(0.4000,1.2000){$j$}
	  \put(0.7400,1.2000){$k$}
	\end{picture}
}

\newcommand{\thirdpp}{
  \setlength{\unitlength}{0.135\columnwidth}
  \begin{picture}(1.3000,1.4000)
    \put(0.1500,0.0000){\includegraphics[width=\unitlength]{fig/hugenholtz_3rd_2B_pp.eps}}
    \put(0.3100,0.3400){\footnotesize$p_1$}
    \put(0.8500,0.3400){\footnotesize$p_2$}
    \put(0.3100,0.8600){\footnotesize$p_3$}
    \put(0.8500,0.8600){\footnotesize$p_4$}
    \put(0.0000,0.6000){\footnotesize$h_1$}
    \put(1.1500,0.6000){\footnotesize$h_2$}
  \end{picture}
}
\newcommand{\thirdph}{
  \setlength{\unitlength}{0.135\columnwidth}
  \begin{picture}(1.3000,1.4000)
    \put(0.1500,0.0000){\includegraphics[width=\unitlength]{fig/hugenholtz_3rd_2B_ph.eps}}
    \put(0.3000,0.3400){\footnotesize$h_1$}
    \put(0.8500,0.3400){\footnotesize$p_1$}
    \put(0.3100,0.8600){\footnotesize$h_3$}
    \put(0.8500,0.8600){\footnotesize$p_3$}
    \put(0.0000,0.6000){\footnotesize$p_1$}
    \put(1.1500,0.6000){\footnotesize$h_2$}
  \end{picture}
}

\newcommand{\fdiagram}{
  \setlength{\unitlength}{0.135\columnwidth}
  \begin{picture}(1.2000,1.2000)
    \put(0.0000,0.0000){\includegraphics[height=\unitlength]{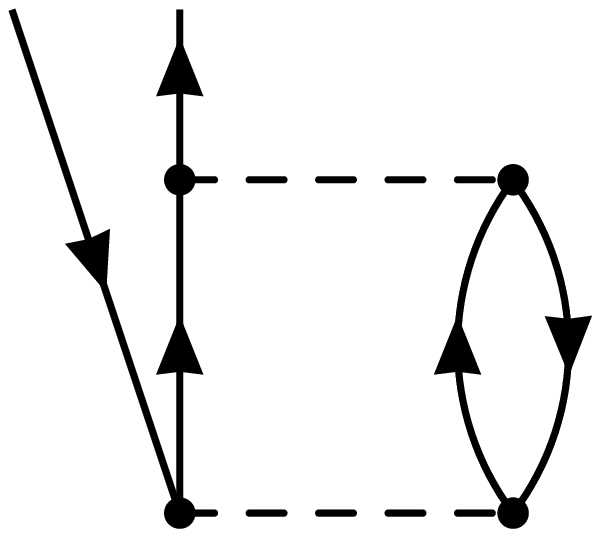}}
    \put(0.0000,1.0600){\parbox{0.100\unitlength}{\footnotesize\raggedright$h\quad\; p$}}
    \put(0.4000,0.3000){\footnotesize$p'$}
    \put(0.6600,0.3000){\footnotesize$p''$}
    \put(1.1500,0.3000){\footnotesize$h'$}
  \end{picture}	
}

\newcommand{\Gammadiagram}{
  \setlength{\unitlength}{0.2\columnwidth}
	\begin{picture}(1.1000,1.2000)
	    \put(0.1000,0.0000){\includegraphics[height=\unitlength]{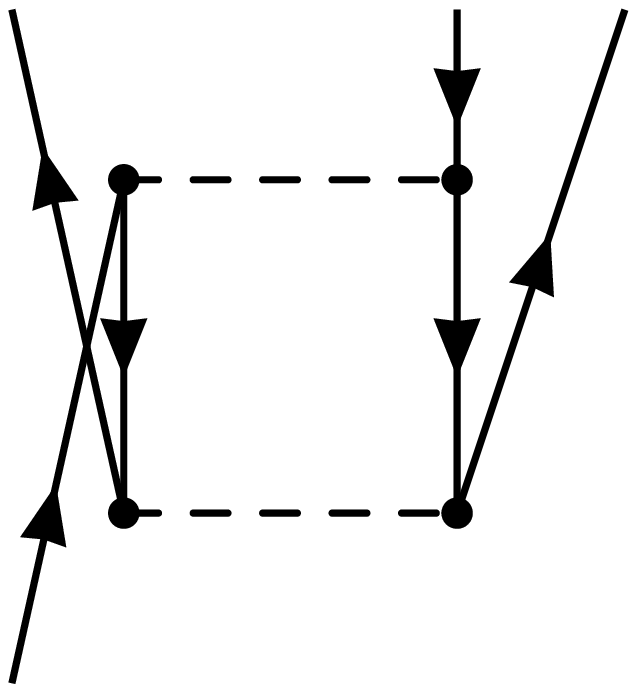}}
	    \put(0.0000,0.0500){\footnotesize\raggedright$p_3$}
	    \put(0.0000,0.9600){\footnotesize\raggedright$p_1$}
	    \put(0.6600,0.9600){\footnotesize\raggedright$h$}
	    \put(0.3200,0.5000){\footnotesize\raggedright$h'$}
	    \put(0.6200,0.5000){\footnotesize\raggedright$h''$}
	    \put(0.8900,0.9600){\footnotesize\raggedright$p_2$}
	\end{picture}
}

\newcommand{\Wdiagram}{
  \setlength{\unitlength}{0.2\textwidth}
  \begin{picture}(1.4000,0.700)
    \put(0.0500,0.0000){\includegraphics[height=0.6\unitlength]{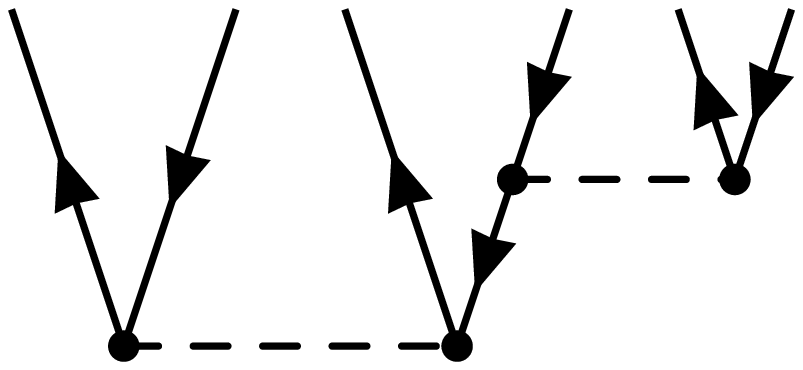}}
	\put(0.0000,0.6200){\footnotesize$p_1$}
	\put(0.3800,0.6200){\footnotesize$h_1$}
	\put(0.5800,0.6200){\footnotesize$p_2$}
	\put(0.9600,0.6200){\footnotesize$h_2$}
	\put(1.1200,0.6200){\footnotesize$p_3$}
	\put(1.3000,0.6200){\footnotesize$h_3$}
	\put(0.9200,0.1200){\footnotesize$h'$}
  \end{picture}
}


For convenience, we briefy summarize the rules for interpreting the antisymmetrized Goldstone and Hugenholtz diagrams that appear in the perturbative discussion of the IM-SRG in Sec.~\ref{sec:mbpt}. Detailed derivations can be found in standard texts on many-body theory, e.g., in Ref.~\cite{Shavitt:2009,Negele:1998ve,Fetter:2003ve}, as well as in Refs.~\cite{Kuo:1990zr,Hjorth-Jensen:1995ys,Kuo:1981fk}, which are particularly useful for diagrammtic treatments of effective nuclear Hamiltonians. 

\begin{enumerate}
\item Solid lines represent single-particle states (indices), with up- and downward pointing arrows indicating particle ($\varepsilon>\varepsilon_F$) and hole states ($\varepsilon\leq\varepsilon_F$), respectively.
\item Interaction vertices are represented as dots in Hugenholtz diagrams, 
  \begin{equation}
    \matrixe{i}{f}{j} = \vcenter{\hbox{\skelHOB}} \;\,,\;\;\;
    \matrixe{ij}{\Gamma}{kl} = \!\!\!\vcenter{\hbox{\skelHTB}} \,,\;\;\;
    \matrixe{ijk}{W}{lmn} =\hspace{-0.8em}\vcenter{\hbox{\skelHTHB}}\,,
  \end{equation}
where the two- and three-body matrix elements are fully antisymmetrized. Throughout this work, we will also use the short-hand notation $f_{ij}=\matrixe{i}{f}{j}\,,\Gamma_{ijkl}=\matrixe{ij}{\Gamma}{kl},$ etc. 

For the discussion of the effective one- and two-body Hamiltonians, we switch from Hugenholtz diagrams to antisymmetrized Goldstone diagrams for clarity (see, e.g., Ref.~\cite{Shavitt:2009}). To this end, the Hugenholtz point vertices are stretched into dashed interaction lines,
  \begin{equation}
    \matrixe{i}{f}{j} = \vcenter{\hbox{\skelOB}} \;\,,\;\;\;
    \matrixe{ij}{\Gamma}{kl} = \vcenter{\hbox{\skelTB}} \,,\;\;\;
    \matrixe{ijk}{W}{lmn} =\vcenter{\hbox{\skelTHB}}\,.
  \end{equation}
Note that the matrix elements are still antisymmetrized: Each of the diagrams shown here represents all allowed exchanges of single-particle lines/indices in the bra and ket. This is reflected in the rules for prefactors that we adopt in the following~\cite{Shavitt:2009}.
\item Assign a factor $1/2^{n_d}$ for $n_d$ equivalent pairs, i.e., pairs of particle or hole lines that start at the same interaction vertex and end at the same interaction vertex. Likewise, assign $1/6^{n_t}$ for $n_t$ equivalent triples connecting the same interaction vertices.
\item Assign a phase factor $(-1)^{n_l+n_h+n_c+n_{exh}}$ to each diagram, where $n_l$ is the number of closed fermion loops, $n_h$ the total number of hole lines, $n_c$ is the number of crossings of distinct external lines, and $n_{exh}$ the number of hole lines continuously passing through the whole diagram (i.e., $n_{exh}=0$ for energy diagrams).
\item For each interval between interactions with particle lines $p_1,\ldots, p_M$ and hole lines $h_1,\ldots, h_N$ multiply the expression with the energy denominator 
\begin{equation}
  \frac{1}{\Omega + \sum_{i=1}^N \varepsilon_{h_i} - \sum_{i=1}^M \varepsilon_{p_i}}\,,
\end{equation}
where $\Omega$ is the unperturbed energy of the state entering the diagram relative to the reference state, reading from bottom to top (e.g., $\Omega = 0$ for energy diagrams). Throughout this work, the energies are given by the diagonal matrix elements of the one-body part of the Hamiltonian $\varepsilon_{i}=f_{ii}$; for Hartree-Fock reference states, $f$ is diagonal, of course. The sum over intermediate particle and hole lines in the denominator is the unperturbed energy of the excited $M$p$N$h state in a M{\o}ller-Plesset type perturbation theory with respect to the reference state. In the Epstein-Nesbet case, it is replaced with the diagonal matrix element of the Hamiltonian in the same state, i.e., 
\begin{align}
&\matrixe{\Phi}{\nord{\aaO_{h_N}\ldots\aaO_{h_1}\aO_{p_M}\ldots\aO_{p_1}}H\nord{\aaO_{p_1}\ldots\aaO_{p_M}\aO_{h_1}\ldots\aO_{h_N}}}{\Phi} - E_0\notag\\
&\qquad\qquad=\sum_{i=1}^M\varepsilon_{p_i} - \sum_{i=1}^N\varepsilon_{h_i} + \text{additional terms}\,,
\end{align}
where $E_0=\matrixe{\Phi}{\HO}{\Phi}$.
\item Sum freely over all internal single-particle indices.
\end{enumerate}

Let us demonstrate the use of the diagram rules for a few examples. For the third-order particle-ladder diagram,
\begin{equation}
  \vcenter{\hbox{\thirdpp}}=\frac{1}{8}\sum_{\substack{p_1p_2p_3p_4\\h_1h_2}}\frac{\Gamma_{h_1h_2p_3p_4}\Gamma_{p_3p_4p_1p_2}\Gamma_{p_1p_2h_1h_2}}{(\varepsilon_{h_1}+\varepsilon_{h_2}-\varepsilon_{p_1}-\varepsilon_{p_2})(\varepsilon_{h_1}+\varepsilon_{h_2}-\varepsilon_{p_3}-\varepsilon_{p_4})}\,.
\end{equation}
Here $n_c=n_{exh}=0$, $n_h=2$, and the number of closed fermion loops is $n_l=2$, namely $p_1 \to p_3 \to h_1 \to p_1$ and $p_2 \to p_4 \to h_2 \to p_2$. For the particle-hole diagram, we have
\begin{equation}
  \vcenter{\hbox{\thirdph}}=-\sum_{\substack{p_1p_2p_3\\h_1h_2h_3}}\frac{\Gamma_{h_3h_2p_1p_3}\Gamma_{h_1p_3h_3p_2}\Gamma_{p_1p_2h_1h_2}}{(\varepsilon_{h_1}+\varepsilon_{h_2}-\varepsilon_{p_1}-\varepsilon_{p_2})(\varepsilon_{h_2}+\varepsilon_{h_3}-\varepsilon_{p_1}-\varepsilon_{p_3})}\,,
\end{equation}
with $n_c=n_{exh}=0$, $n_h=3$, and two closed loops ($n_l=2$), $p_1\to h_3\to h_1\to p_1$ and $p_2\to p_3 \to h_2$, giving a negative sign. Since the interaction vertices are connected by one particle and one hole line each, $n_d=0$, and the pre-factor is 1. 

For the second-order effective Hamiltonian, diagram $f_4$ in Fig.~\ref{fig:mbpt2_f} translates into
\begin{equation}
  \vcenter{\hbox{\fdiagram}}=\frac{1}{2}\sum_{p'p''h'}\frac{\Gamma_{ph'p'p''}\Gamma_{p'p''hh'}}{\varepsilon_{h}+\varepsilon_{h'}-\varepsilon_{p'}-\varepsilon_{p''}}\,.
\end{equation}
Reading from bottom to top, we have $\Omega=0$ just like in an energy diagram. To determine the phase, we note that there is one fermion loop ($p''\to h' \to p''$), there are two hole lines, one of which is external and passing through the diagram via $h \to p' \to p$. Thus $n_l=1, n_h=2, n_{exh}=1$, and $n_c=0$, so the phase factor is $+1$. There is one pair of equivalent particle lines, $n_d=1$, giving rise to the pre-factor $\tfrac{1}{2}$.

As an example for a second-order two-body interaction, we consider diagram $\Gamma_3$ in Fig.~\ref{fig:mbpt2_Gamma}:
\begin{equation}
  \vcenter{\hbox{\Gammadiagram}}=\frac{1}{2}\sum_{h'h''}\frac{\Gamma_{p_1p_2h'h''}\Gamma_{h'h''p_3h}}{\varepsilon_{h'}+\varepsilon_{h''}-\varepsilon_{p_1}-\varepsilon_{p_2}}\,,
\end{equation}
where $n_l=n_c=0, n_h=3$, and there is one external hole line ($n_{exh}=1$) passing through the diagram, $h\to h'' \to p$, giving a phase factor $+1$. There is one pair of equivalent hole lines ($n_d=1$), and the starting energy is $\Omega=p_3$, which explains the symmetry pre-factor and energy denominator, respectively.

Our final example is an induced three-body interaction, diagram $W_3$ in Fig.~\ref{fig:mbpt2_W}. The expression is
\begin{equation}
  \vcenter{\hbox{\Wdiagram}}=-\sum_{h'}\frac{\Gamma_{p_1p_2h_1h'}\Gamma_{h'p_3h_3h_3}}{\varepsilon_{h_1}+\varepsilon_{h'}-\varepsilon_{p_1}-\varepsilon_{p_2}}\,,
\end{equation}
where $\Omega=0$, the phase factor is $-1$ because $n_l=n_c=0, n_h=4, n_{exh}=3$. Due to the lack of equivalent lines, the overall pre-factor of the diagram is $1$.

\renewcommand{\bibname}{References}
\section*{\bibname}
\addcontentsline{toc}{section}{\protect\numberline{}\bibname}%
\bibliography{2013_imsrg_long}

\end{document}